\documentclass[fleqn, usenatbib]{mnras}

\usepackage{newtxtext,newtxmath}
\usepackage[T1]{fontenc}

\usepackage{amsmath, bm, dcolumn, epsfig, graphicx, hyperref, 
subfigure, mathrsfs, soul, newtxmath, newtxtext}

\usepackage{flushend}
\usepackage{cuted}

\usepackage[dvipsnames]{xcolor}

\title[Distinguishing gravitational and emission physics in black-hole
  imaging]{Distinguishing gravitational and emission physics in
  black-hole imaging: spherical symmetry}

\author{
Prashant Kocherlakota$^1$
and
Luciano Rezzolla$^{1,2,3}$
\\
$^1$Institut f{\"u}r Theoretische Physik,Goethe-Universit{\"a}t, 
Max-von-Laue-Str. 1, 60438 Frankfurt, Germany\\
$^2$Frankfurt Institute for Advanced Studies, Ruth-Moufang-Str. 1, 
60438 Frankfurt, Germany\\
$^3$School of Mathematics, Trinity College, Dublin 2, Ireland
}

\date{Accepted XXX. Received YYY; in original form ZZZ}

\pubyear{2021}

\begin{document}
\label{firstpage}
\pagerange{\pageref{firstpage}--\pageref{lastpage}}
\maketitle

\begin{abstract}
Imaging a supermassive black hole and extracting physical information
requires good knowledge of both the gravitational and the astrophysical
conditions near the black hole. When the geometrical properties of the
black hole are well understood, extracting information on the emission
properties is possible. Similarly, when the emission properties are well
understood, extracting information on the black-hole geometry is
possible. At present however, uncertainties are present both in the
geometry and in the emission, and this inevitably leads to degeneracies
in the interpretation of the observations. We explore here the impact of
varying geometry and emission coefficient when modelling the imaging of a
spherically-accreting black hole. Adopting the Rezzolla-Zhidenko
parametric metric to model arbitrary static black-holes, we first
demonstrate how shadow-size measurements leave degeneracies in the
multidimensional space of metric-deviation parameters, even in the limit
of infinite-precision measurements. Then, at finite precision, we show
that these degenerate regions can be constrained when multiple pieces of
information, such as the shadow-size and the peak image intensity
contrast, are combined. Such degeneracies can potentially be eliminated
with measurements at increased angular-resolution and flux-sensitivity.
While our approach is restricted to spherical symmetry and hence
idealised, we expect our results to hold also when more complex
geometries and emission processes are considered.
\end{abstract}



\section{Introduction}
The characteristic central brightness depression in the first Event
Horizon Telescope (EHT) image of M87$^*$ provides strong evidence of
its nature as a supermassive black hole (BH; \citealt{EHT_M87_PaperI,
  EHT_M87_PaperII, EHT_M87_PaperIII, EHT_M87_PaperIV, EHT_M87_PaperV,
  EHT_M87_PaperVI}). The size of the narrow, bright ring in the observed
image sets an upper bound on the size of the central dark compact
object. By comparing the observed image against a large library of
synthetic images obtained from general-relativistic radiative-transfer
(GRRT) computations coupled to general-relativistic magnetohydrodynamics
(GRMHD) simulations of Kerr BHs being orbited by a turbulent, hot,
magnetized disk in general relativity (GR; \citealt{EHT_M87_PaperIV,
  EHT_M87_PaperV}), the angular gravitational radius $\theta_{\text{g}} =
GM/c^2D$ of M87$^*$ can be ascertained, with $M$ its mass and $D$ the
distance to it. The consistency of this value obtained by the EHT with
the one obtained previously from stellar-dynamics measurements
\citep{Gebhardt11} demonstrates that GR passes the null-hypothesis test
\citep{EHT_M87_PaperVI}.

The width of the posterior on the fractional difference between these two
measurements of $\theta_{\text{g}}$ can be used to infer crucial, albeit
approximate, bounds on the size of the shadow of M87$^*$. Furthermore,
while all BHs cast shadows, there exist other compact objects (or
black-hole mimickers), e.g., horizonless compact objects, both with and
without singularities, wormholes, etc., that can be used to model the
observed image of M87$^*$ (see, e.g.,
\citealt{Kocherlakota2020}). Besides testing what type of exotic object
M87$^*$ is, there are also theory-specific possibilities that one can
test for indirectly by considering whether the EHT image is consistent
with the BH solutions from those theories. The latter include looking for
the existence of additional matter fields, conspicuous in the strong
gravitational fields around BHs (such as the axion in \citealt{Sen1992}),
anomalous couplings between gravity and/or fields, possibly leading to
violations of the Einstein equivalence principle \citep{Gibbons1988},
etc. Such a comparison, using the EHT shadow-size constraints,
considering various well-known BH and non-BH solutions from GR and
alternative theories has recently been reported in
\cite{Kocherlakota2021}. An accurate quantification of the size of
deviation of these solutions from the Schwarzschild solution may be found
in \cite{Kocherlakota2020}.

While such analyses \citep{Psaltis2020_EHT, Kocherlakota2021, Volkel2021}
make use solely of the shadow-size constraints, the measurements in
reality contain additional potentially useful information. Barring a few
recent studies \citep{Broderick2014, Johannsen2016c, Mizuno2018,
  Olivares2020}, the dearth of simulations of accreting non-Kerr BH and
non-BH spacetimes, and of the associated analyses that would identify and
extract novel features that they may present, limits the present ability
to distinguish BHs of various theories. Notwithstanding the computational
costs and the conceptual hurdles involved in expanding the presently
available body of GRRT and GRMHD codes to accommodate non-Maxwell and
non-Einstein-Hilbert theories respectively, it is presently unclear
whether this is the most effective route at the moment.

Thus, proposing and developing semi-analytical accretion and emission
models \citep{Bambi2013, Younsi2016, Pu2018, Shaikh2019, 
Gimeno-Soler2019, Shaikh2019b, Narayan2019, Gan2021, Yang2021}, and studying
the variation of image features with these models and with varying
spacetime geometry, as a path-finding step, could already yield important
insights. Such analyses require fewer expensive operations such as
integrations, and greatly reduce the overall computational cost, enabling
a broader parameter survey. This is the logic followed in this work,
where we adopt a generic but rapidly converging parameterisation of
spherically symmetric BHs via the Rezzolla-Zhidenko (RZ) expansion
\citep{Rezzolla2014}, and a series of simple but effective spherically
symmetric emission models. Furthermore, in addition to the shadow-size
bounds, we also use (a) the scaling parameter $\alpha = d/
\theta_{\text{g}}$ of the angular diameter $d$ of the observed ring
feature in the M87$^*$ image with its angular gravitational radius, (b) 
the lower bound on the maximum contrast in the image 
$\mathcal{C}_{\text{peak, EHT-min}}$, and (c) the fractional-width of 
the ring $f_{\text{w}}$, inferred by the EHT (see Table 1 of 
\citealt{EHT_M87_PaperI}).

Adopting this approach, it then becomes possible to address, albeit
within the present context of spherical models, some of the concerns that
have recently been raised, e.g., in \cite{Gralla2021}, about the location
of peak of emission near M87$^*$, when modelled as a Schwarzschild BH,
namely whether it is closer to the location of the innermost stable
circular orbit or the photon sphere. We can also identify certain
characteristic features in the observed image that are reasonably robust
against varying emission prescriptions, such as the location of the peak
contrast in the observed image and some that are not (value of peak
contrast, fractional ring width, etc.). Indeed, it can be argued that
this latter set of observables can be used to gain insight into the
emission physics near accreting supermassive BHs. In this sense, recent
work has considered the use of principal-component analyses to use all of
the information in the image (see, e.g., \citealt{Medeiros2020,
  Lara2021}).

More importantly, the work reported here can be used to determine which
degeneracies may emerge when distinguishing gravitational and emission
physics in BH imaging. It is clear, in fact, that extracting physical
information from a supermassive BH image requires a good knowledge of
both the gravitational and astrophysical conditions near the black
hole. The uncertainties associated with the spacetime geometry and with
the emission mechanisms, and which are presently still rather large,
inevitably lead to degeneracies in the interpretation of the
observations. Indeed, while the EHT shadow-size measurements already
impose non-trivial constraints on the relevant parameter space of any
family of BHs, it is in general extremely difficult to distinguish --
even in the limit of a perfectly precise measurement -- BHs which belong
to a two-parameter family that have differing horizon-sizes but which
predict identical shadow-sizes (see the top row of Fig.
\ref{fig:RZ_eps_a0_a1_two_param}). This limitation, however, is simply
due to the use of a single observable: the shadow-size (see, e.g., 
\citealt{Volkel2021}). Hence, complementing the latter with the location 
of the observed peak of emission (characterized by $\alpha$) drastically 
changes the scenario, and it is therefore possible, in principle, to 
isolate the relevant metric parameters (see the bottom row of Fig. 
\ref{fig:RZ_eps_a0_a1_two_param}). We also illustrate that ever more
detailed descriptions of the spacetime geometry, which emerge when
considering BHs described by three independent parameters, bring in
additional degeneracies. These, however, can be further resolved by
exploiting other observables such as precise measurements of the
fractional-width of the observed bright ring or with future observables
such as the sizes of photon subrings that the next-generation EHT (ngEHT) 
hopes to observe (\citealt{Gralla2019, Johnson2020, Broderick2021}).

The structure of the paper is as follows. In
Sec.~\ref{sec:Accretion_Emission} we summarise the simple radially freely
infalling accretion model that we will use exclusively here, and discuss
also a prescription for the associated emission that naturally follows
(see \citealt{Narayan2019}). In Sec.~\ref{sec:Schw_Emissivity_Results},
keeping the spacetime geometry fixed to that of a Schwarzschild BH, we
compare the observed contrast profiles obtained by applying the
aforementioned accretion-emission model against those obtained from other
emission models that are currently in use. We also introduce an
artificial inner cut-off for the emitting region as a simplistic proxy
for a peaked emission profile and study how this impacts the final
image. In Sec.~\ref{sec:Ring_Size_RZBHs}, we vary the spacetime metric
when using two characteristic emission prescriptions, and explore the
constraints imposed by the EHT measurements on the space of parameters
that induce deviations from the Schwarzschild geometry.  In
Sec.~\ref{sec:Fractional_Width}, we characterise the ring widths of the
emission rings around various RZ BHs for different emission models and
show that the shadow boundary always forms the inner edge of the emission
ring, independently of whether or not there is an inner cutoff in the
emission zone outside the horizon.

The purpose of the present work is to demonstrate, using
spherically symmetric geometry and emission, that is possible to test the
nature of strong gravitational fields near supermassive compact objects
with the EHT. We expect that much of what discussed here will continue to
hold also when more realistic emission models, such as those considered
in \cite{EHT_M87_PaperV} and \cite{EHT_M87_PaperVI}, would be used in
conjunction with axisymmetric spacetimes \citep{Konoplya2016a}.

\section{Spherical Accretion and Emission Models}
\label{sec:Accretion_Emission}

In what follows, we will assume that photons and accreting material 
moves on geodesics of the spacetime metric. Further, the symbol $\nu$ 
and the index $\nu$ will be used to denote frequencies and specific 
quantities; Thus, e.g., $I_\nu$ will be the specific intensity and is 
a scalar quantity.

The line-element describing the exterior geometry of an arbitrary 
static BH can be written, in areal-polar coordinates $x^\alpha = 
(t, r, \theta, \phi)$, as,
\begin{equation}
  \label{eq:Static_Spacetime} 
ds^2 = g_{\alpha\beta}dx^\alpha dx^\beta = -N^2(r)dt^2 + \frac{B^2(r)}
{N^2(r)}dr^2 + r^2 d\Omega_2^2\,,
\end{equation} 
where $d\Omega_2^2 := d\theta^2 + \sin^2\theta~d\phi^2$ is the 
standard line-element on the two-sphere. 

Due to the existence of the Killing vectors $\partial_t$ and 
$\partial_\phi$ in such a spacetime (Eq. \ref{eq:Static_Spacetime}), there 
exist a pair of associated conserved quantities, $E$ and $L$, along 
every geodesic, $x^\alpha(\lambda)$. These can be obtained 
straightforwardly from the Lagrangian for geodesic motion, 
$2\mathscr{L} := g_{\alpha\beta}\dot{x}^\alpha\dot{x}^\beta$, as,
\begin{align}
- E :=&\ p_t = \partial_{\dot{t}} \mathscr{L} = -N^2 \dot{t}\,, \\
L :=&\ p_\phi = \partial_{\dot{\phi}} \mathscr{L} = r^2\sin^2\theta
\dot{\phi}\,, \nonumber
\end{align}
where the overdot denotes a derivative with respect to the affine parameter
$\lambda$ along the geodesic. The Lagrangian $\mathscr{L}$ is itself also
a conserved quantity along a geodesic ($2\mathscr{L} = 0, -1$ for null
and timelike geodesics), and when rewritten in terms of the other
conserved quantities, as,
\begin{equation}
2\mathscr{L} = -\frac{E^2}{N^2} + \frac{B^2}{N^2}\dot{r}^2 + 
r^2\dot{\theta}^2 + \frac{L^2}{r^2\sin^2\theta}\,,
\end{equation}
yields a separable equation for the radial and the polar velocity
components,
\begin{equation}
0 = \left(\frac{r^2B^2}{N^2}\frac{\dot{r}^2}{E^2} - \frac{r^2}{N^2} - 
\frac{2\mathscr{L}r^2}{E^2}\right) + 
\left(r^4\frac{\dot{\theta}^2}{E^2} + \frac{L^2}{E^2}\csc^2\theta
\right)\,. \nonumber
\end{equation}
The existence of a fourth conserved quantity, called the Carter constant
$C$, associated with the existence of an additional ``hidden'' symmetry
and characterized by the Killing-Yano tensor (see, e.g.,
\citealt{Hioki2009,Abdujabbarov2015}) allows for us to write, with $\xi
:= L/E$ and $\mathscr{I} := C/E^2 + \xi^2$,
\begin{align}
\frac{\dot{r}^2}{E^2} =&\ \frac{1 + 2\mathscr{L} E^{-2} N^2 - 
\mathscr{I} r^{-2} N^2}{B^2} := V_{\text{r}}(r)\,, \\
\frac{\dot{\theta}^2}{E^2} =&\ \frac{\mathscr{I} - 
\xi^2\csc^2\theta}{r^4}\,. \nonumber
\end{align}
Thus the four-velocity along an arbitrary geodesic $x^\alpha(\lambda)$ 
with conserved quantities $\left\{E, \xi, \mathscr{L}, 
\mathscr{I}\right\}$ is given as,
\begin{align} \label{eq:Four_Vel_General_Geodesics}
\frac{\dot{x}_{\pm,\pm}^\alpha}{E} =&\ \left(\frac{1}{N^2}, \pm
\frac{\left[1 + 2 \mathscr{L}E^{-2}N^2 - \mathscr{I} r^{-2} N^2
\right]^{1/2}}{B} \right. \,, \\ 
&\ \left. \pm\frac{\sqrt{\mathscr{I} - \xi^2\csc^2\theta}}{r^2}, 
\frac{\xi}{r^2\sin^2\theta}\right)\,, \nonumber
\end{align}
where the indices $\pm, \pm$ above correspond to the sign of the 
radial and polar velocities respectively.

\subsection{Spherical Accretion Model and Observed Intensity Profile}
\label{sec:Spherical_Accretion}
The spherical accretion model of interest here is that of a fluid in
radial free-fall towards the central compact object. The four-velocity
$u^\alpha_{\text{e}}$ along such a radially freely infalling ($d\Omega_2
= 0$) timelike emitter is [see Eq.
  \eqref{eq:Four_Vel_General_Geodesics}],
\begin{equation} \label{eq:Emitter_Four_Vel}
u_{\text{e}}^\alpha = \bar{E}\left(\frac{1}{N^2}, 
-\frac{\sqrt{1 - \bar{E}^{-2}N^2}}{B}, 0, 0\right)\,, 
\end{equation}
where $\bar{E}$ is the energy per unit mass of a fluid element at 
infinity. 

As is reasonable, we will restrict to the case when the fluid elements 
have negligible (radial) kinetic energy at infinity (i.e., 
$\lim_{r\rightarrow\infty}u^r_{\text{e}} \simeq 0$ or $\bar{E} \simeq 
1$). Further, the four-velocity $u^\alpha_{\text{o}}$ of an asymptotic 
static observer ($d\Omega_2 = dr = 0$) is,
\begin{equation} \label{eq:Observer_Four_Vel}
u_{\text{o}}^\alpha = \lim_{r\rightarrow\infty}\left(N^{-1}, 0, 0,
 0\right) = (1, 0, 0, 0)\,. 
\end{equation}
A photon that is emitted at a radius $r$ with an emitter-frame frequency
of $\nu_{\text{e}}$ appears in the frame of the observer to be of
frequency $\nu_{\text{o}}$, due to gravitational and Doppler
redshift. This redshift factor $\gamma_\pm$ for radially-outgoing (+) and
-ingoing (-) photons is then given as,
\begin{equation} \label{eq:Redshifts}
\gamma_\pm(r) = \frac{\nu_{\text{o}}}{\nu_{\text{e}}} = \frac{\left(
k_{\pm,\pm}\right)_\alpha u^\alpha_{\text{o}}}{\left(k_{\pm,\pm}
\right)_\beta u^\beta_{\text{e}}} 
\simeq \frac{N^2}{1 \pm \sqrt{\left(1 - \mathscr{I} r^{-2} N^2\right)
\left(1 - N^2\right)}}\,,
\end{equation}
where in writing the last relation above we have used $\bar{E} 
\approx 1$, and $k^\mu_{\pm, \pm}$ denotes the four-velocity of a 
photon%
\footnote{With dual, $\left(k_{\pm,\pm}\right)_\mu = E\left(-1,
\pm\frac{B\sqrt{1 - \mathscr{I} r^{-2} N^2}}{N^2}, \pm\sqrt{\mathscr{I} -
  \xi^2\csc^2\theta}, \xi\right)\,.$} [see
  Eq.~\eqref{eq:Four_Vel_General_Geodesics}],
\begin{equation} \label{eq:Photon_Four_Vel}
k_{\pm,\pm}^\mu = E\left(\frac{1}{N^2}, \pm\frac{\sqrt{1 - \mathscr{I} 
r^{-2} N^2}}{B}, \pm\frac{\sqrt{\mathscr{I} - \xi^2\csc^2\theta}}{r^2}, 
\frac{\xi}{r^2\sin^2\theta}\right)\,, \nonumber
\end{equation}
Notice that due to the purely radial motion of the emitter [see
  Eq.~\eqref{eq:Emitter_Four_Vel}] the redshift experienced by a photon
is independent of its polar $(k_{\pm,\pm}^\theta)$ and azimuthal
$(k^\phi)$ velocities, thus removing all $\theta$-dependence.

In general, in the presence of a fluid that both emits and absorbs 
radiation, the covariant radiative-transfer equation, written in the 
local fluid-frame, reads (see, e.g., \citealt{Younsi2012}),
\begin{equation}
\frac{dI_\nu}{d\lambda} = -\left(k_{\pm, \pm}\right)_\beta 
u_{\text{e}}^\beta\left(- q_\nu I_\nu + \frac{j_\nu}{\nu^3}\right)\,,
\end{equation}
where $I_\nu$ is the specific intensity of a ray at frequency $\nu$, 
and $q_\nu$ and $j_\nu$ are the absorption and emission coefficients 
at that frequency respectively. Here however, as is relevant for the 
observing frequency of the EHT, we will consider the scenario of an 
optically-transparent fluid ($q_\nu = 0$) that emits isotropically and 
monochromatically in its rest frame, $j_\nu = \delta(\nu - 
\nu_\star)j$. For this scenario, the 
total intensity $I$ at a location $(\alpha, \beta)$ on the 
observer's celestial plane is given as (see, e.g., Sec. 2.3 of 
\citealt{Jaroszynski1997}; See also \citealt{Bambi2013a, Shaikh2019}),
\begin{equation} \label{eq:Total_Intensity}
I(\alpha, \beta) = -\fint\int\gamma_\pm^4\left(k_{\pm, \pm}
\right)_\beta u_{\text{e}}^\beta j_\nu~\text{d}\lambda~\text{d}\nu 
= \fint\gamma_\pm^3 j\frac{\text{d}r}{k_{\pm, \pm}^r} \,,
\end{equation}
where the slash indicates that the integral is evaluated along the full
inextensible worldline, through the emitting fluid, of the photons that
appear in the observer's sky at $(\alpha, \beta)$, and the sign of the
redshift factor is chosen depending on whether the photon velocity is
locally directed radially-outward ($+$) or -inward ($-$). The Cartesian
celestial coordinates $(\alpha, \beta)$ on the sky of an asymptotic
static observer, inclined at an angle $\theta_{\text{o}}$ with respect to
the BH's (positive) $z$-axis, that were introduced above can be related
to the conserved quantities of the photons that appear there via
\citep{Bardeen1974, Hioki2009},
\begin{equation}
\left(\alpha, \beta_{\pm}\right) := \left(-\xi
\csc{\theta_{\text{o}}}\,, \pm\sqrt{\mathscr{I} -\xi^2\csc^2{
\theta_{\text{o}}}}\right)\,,
\end{equation}
with the sign determined by the polarity of its polar velocity. 

From Eq.~\eqref{eq:Total_Intensity} we see that the inclination angle of
the observer $\theta_{\text{o}}$ does not impact image formation in this
simplified spherical accretion and emission setup; i.e., the only
combination of $(\alpha, \beta)$ in the above is $\alpha^2 + \beta^2 =
\mathscr{I}$, and the final image is circularly symmetric on the
observer's celestial plane. These properties greatly simplify the image
construction, and we can simply compute the integral above (Eq. 
\ref{eq:Total_Intensity}) along the the $x$-axis of the image plane for
an equatorial asymptotic observer w.l.g., i.e., for $\theta_{\text{o}} =
\pi/2$ and $\beta = 0$, and simply rotate $\mathit{I}(\alpha\!=\!\xi,
\beta\!=\!0) := I(\xi)$ to obtain the full 2-d intensity profile,
$\mathit{I}(\alpha, \beta)$. Therefore, the integral we will be concerned
with is,
\begin{align} \label{eq:Intensity_Integral_Inner_Boundary}
\mathit{I}(\xi) =
\begin{cases}
\int_{r_{\text{cut}}}^\infty \gamma_+^3~j~\Lambda~\text{d}r\,, 
& |\xi| < \xi_{\text{ps}} \\
\int_{\max{\{r_{\text{cut}}, r^+_{\text{ps}}\}}}^\infty 
\gamma_+^3~j~\Lambda~\text{d}r\,, 
& |\xi| = \xi_{\text{ps}} \\
\int_{\max{\{r_{\text{cut}}, r_{\text{tp}}(\xi)\}}}^\infty 
\left(\gamma_-^3 + \gamma_+^3\right)~j~\Lambda~\text{d}r\,,
& |\xi| > \xi_{\text{ps}}\,,
\end{cases}
\end{align}
where $r = r_{\text{cut}} \geq r_0$ is the location of the inner edge of
the emission zone, where $r_0$ denotes the location of the outermost
Killing horizon, $\Lambda := |1/k_{\pm, \pm}^r|$, $\xi = \xi_{\text{ps}}$
is the impact factor of a photon moving on the photon sphere which is
located at $r = r_{\text{ps}}$ (the superscript $+$ denotes an
infinitesimally larger value), and $r = r_{\text{tp}}(\xi)$ is the
location of the (radial) turning point of a photon with an impact factor
of $\xi$ (see Sec. \ref{sec:rtp_xips} below).

Although the existence of a such a sharp inner edge of emission is highly
unrealistic, we employ $r_{\text{cut}}$ here as a proxy to study the
impact of severe emission suppression close to the event horizon on the
final image. Alternatively, and possibly more reasonably,
$r=r_{\text{cut}}$ can be thought of as representing the location of the
surface of a hypothetical exotic compact object such as a gravastar 
(see, e.g., \citealt{Mazur2004}), whose size we can then potentially 
constrain.

\subsection{Radial Turning Points of Null Geodesics and the Photon 
Sphere} 
\label{sec:rtp_xips}
The (radial) turning point of an equatorial photon ($\theta = \pi/2$;
$\mathscr{I} = \xi^2$) is defined as the location where its radial
velocity changes sign, i.e., $\dot{r} = 0$, which implicitly determines
its radial coordinate $r = r_{\text{tp}}(\xi)$ as a function of its
impact factor $\xi$,
\begin{equation} \label{eq:rtp}
r_{\text{tp}}(\xi) = \xi N(r_{\text{tp}}(\xi))\,.
\end{equation}
Equivalently, we can write the impact factor of a photon that has a
radial turning point at a coordinate $r$ from the inverse relation,
$\xi_{\text{tp}}(r) := r/N(r)$.

For the special case of a photon on an equatorial circular null 
geodesic (CNG), $\dot{r} = \ddot{r} = 0$, we require additionally that,
\begin{equation}
\label{eq:CNG_Eq}
r \partial_r N - N = 0\,. 
\end{equation}
If the equation above has a single root, and the CNG is unstable to
radial perturbations ($\partial^2_r V_{\text{r}} < 0$), then it locates
the photon sphere\footnote{In general, there exist non-equatorial photons
that also satisfy $\dot{r} = \ddot{r} = 0$, with $\mathscr{I} =
\xi_{\text{ps}}^2$, which move on a \textit{sphere} of radius
$r_{\text{ps}}$.}, $r = r_{\text{ps}}$, in the spacetime. The associated
impact factor $\xi = \xi_{\text{ps}}$ of a photon on such an orbit
corresponds to the size of the shadow as seen by an asymptotic observer
[Eq.~\eqref{eq:rtp}],
\begin{equation} \label{eq:xips}
\xi_{\text{ps}} = r_{\text{ps}}/N\left(r_{\text{ps}}\right)\,.
\end{equation}
Therefore, in writing the total intensity integral above (Eq.
\ref{eq:Intensity_Integral_Inner_Boundary}) for photons with $|\xi| >
\xi_{\text{ps}}$, we have just explicitly split the integral into the
parts corresponding to the ingoing and outgoing pieces of the photon
orbit respectively. Furthermore, the piecewise definition of the integral
above (Eq. \ref{eq:Intensity_Integral_Inner_Boundary}) becomes clear when
remembering that photons with $|\xi| < \xi_{\text{ps}}$ do not turn:
these are either ingoing and are captured by the BH at the photon sphere,
or are outgoing and reach asymptotic infinity. Of course, the latter set
of photons can be sourced only outside the outermost Killing horizon (see
also Appendix \ref{app:Schw_Images_Addl} for a graphical representation
of the radial orbits of null geodesics in the Schwarzschild BH spacetime)

\begin{figure*}
\includegraphics[width=2.\columnwidth]{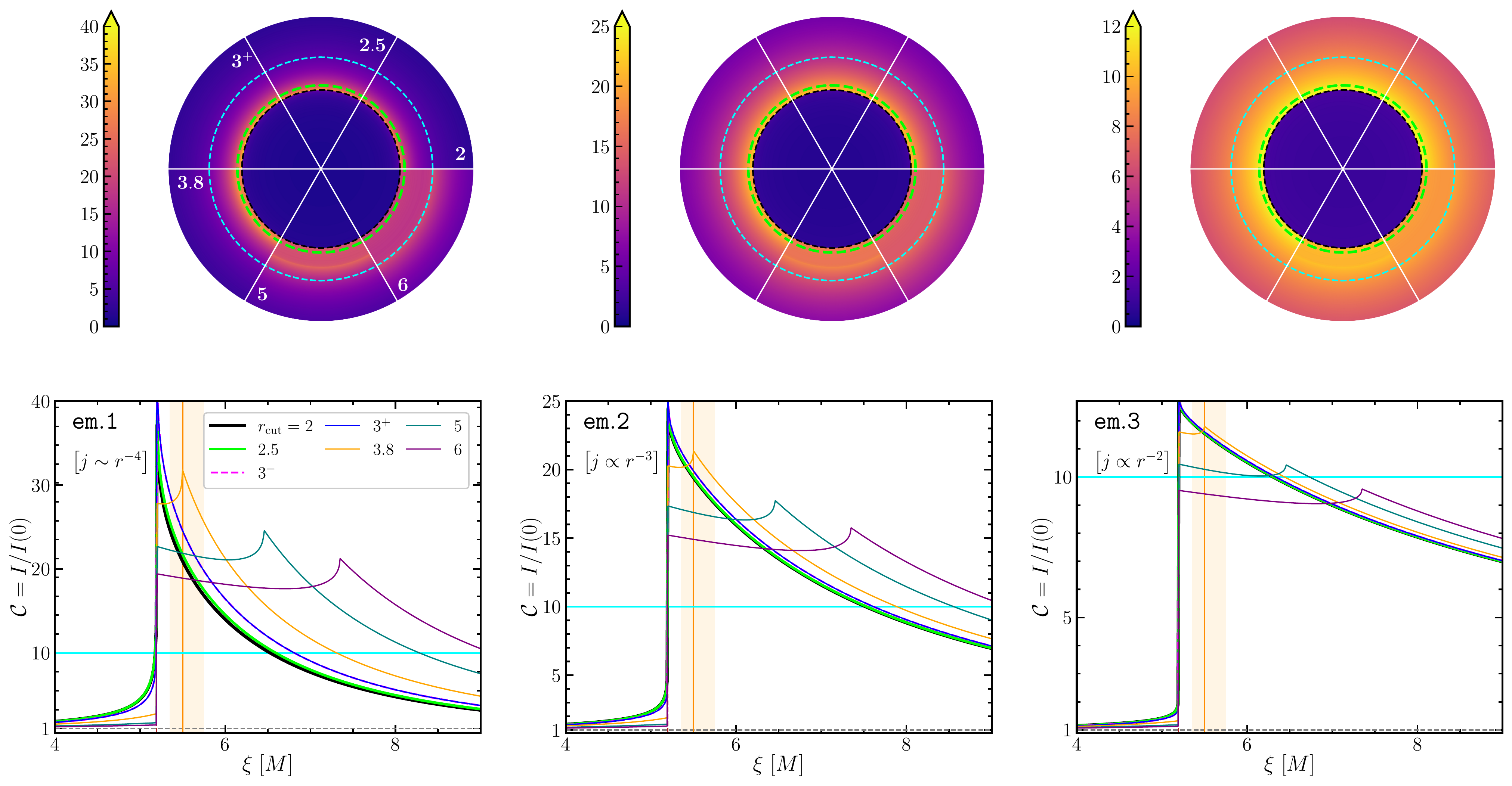}
\caption{In the top row we show the variation, on the celestial plane of
  an asymptotic observer, of the contrast in the images of Schwarzschild
  BHs that are undergoing spherical accretion. In each panel, we keep the
  isotropic, monochromatic emission prescription $j$ unchanged but vary
  the inner boundary of emitting region $r_{\text{cut}}$ across sextants,
  as indicated in the top left panel.  We also mark the impact factors of
  photons that ``turn'' at the photon sphere $r=3 M$ (black circle),
  $r=3.8 M$ (lime; corresponding to an impact factor of $\xi_{\text{peak;
      EHT}} \approx 5.5 M$), and at the ISCO $r=6M$ (cyan), for
  reference. The characteristic central contrast depression, seen in all
  the images as a violet disk, is bounded precisely by the black
  circle. In the bottom row, we show a one-dimensional cut of all of the
  sextants from the figure above in different colours as indicated in the
  bottom left panel. While the horizontal cyan line marks the lower bound
  of the maximum-contrast as measured by the EHT for M87$^*$
  $C_{\text{peak, EHT-min}}$, the vertical orange line marks the
  approximate location of the peak of emission $\xi_{\text{peak; EHT}}$,
  with the orange band around it representing the error in the
  same. Finally, across columns, we run over the emission models tabled
  in Table \ref{table:Emission_Models}, to study the impact of $j$ on the
  observed image. All radii in the figure are reported in units of the BH
  ADM mass $M$. Notice also that these figures plots are independent of
  the value of the constant mass accretion rate $\dot{m}_\infty$ since
  the contrast is a ratio of intensities.}
\label{fig:Schwarzschild_Emissivity}
\end{figure*}

\subsection{Emission Coefficient}
\label{sec:Emission_Coeff}
Thus far we have left the spacetime and the radial variation of the 
emission coefficient unfixed. Here we show how adopting a constant 
mass accretion rate $\dot{m}_\infty$ for the spherical accretion 
process described above automatically yields one prescription for the 
emission coefficient $j$, as in  \cite{Narayan2019}. 

An infinitesimal fluid element of mass $dm$ with an energy per unit mass
at infinity $\bar{E} = 1$ (i.e., at rest at infinity) that free-falls
radially inwards converting gradually its potential energy into kinetic
energy. The fluid element then comes to rest at a radius $r$, where it
loses its kinetic energy releasing a maximum energy in radiation
$\bar{E}_{\text{emit}}$ in radiation corresponding to the difference in
its binding energy [see Eq.~\eqref{eq:Observer_Four_Vel}]
\begin{equation}
  \label{eq:Binding_Energy_Diff}
E_{\text{emit}}(r) = dm\left(\bar{E} - \bar{E}(r)\right) = dm\left(1 - 
N(r)\right)\,.
\end{equation}

Thus, the net emitted luminosity $L$ ($\text{erg}~\text{s}^{-1}$) in this
scenario, as measured at infinity, is given as, $L(r) = \dot{m}_\infty
E_{\text{emit}}$, from which we can infer the isotropic specific emission
coefficient $j_\nu$ ($\text{erg}~\text{s}^{-1}
\text{cm}^{-3}\text{ster}^{-1}\text{Hz}^{-1}$) in the proper frame of the
fluid using the relation,
\begin{equation}	
L(r) = \int_r^\infty \int_0^{2\pi} \int_0^\pi \int_\nu \sqrt{-|g|}
\sqrt{-g_{tt}}4\pi~j_\nu~\text{d}r^\prime~\text{d}\phi~\text{d}\theta~
\text{d}\nu\,, \nonumber
\end{equation}	
where $|g|$ is the determinant of the metric. For monochromatic, 
isotropic emission in the rest frame,
\begin{align} \label{eq:Emission_Coeff}
j(r) =\ -\frac{\partial_r L}{16\pi^2 r^2 B N}
= \frac{\dot{m}_\infty\partial_r N}{16\pi^2 r^2 B N}\,.
\end{align} 
For such emission around a Schwarzschild BH, we can write,
\begin{equation} \label{eq:Emission_Coeff_Schw}
j_{\text{Schw}}(r) = \frac{\dot{m}_\infty}{16\pi^2 r^3 (r - 2 M)}\,.
\end{equation}
The large-$r$ ($r \gtrsim 3M$) approximation for the maximum emitted 
energy in a Schwarzschild spacetime reads, from Eq. 
\ref{eq:Binding_Energy_Diff},
\begin{equation}
E_{\text{emit}} = dm\left(1 - \sqrt{1 - \frac{2M}{r}}\right) \approx 
\frac{dm~M}{r}\,.
\end{equation}
This is equivalent to Eq. (12) of \cite{Narayan2019}, from which we can
also obtain their emission coefficient, given in Eq. (11) of the same
paper.

\section{Variation of Observed Ring Size with Emission Model: 
Schwarzschild black hole}
\label{sec:Schw_Emissivity_Results}

Here we fix the spacetime metric to that of a Schwarzschild BH 
$(N^2 = 1 - 2M/r; B^2 = 1)$, and study the impact of varying the 
emission coefficient $j$ and the inner cut-off of the emission region, 
$r_{\text{cut}}$, on the final image. We summarise the various 
emission prescriptions that we will use in this section in Table 
\ref{table:Emission_Models}, and report the observed contrast profile, 
$\mathcal{C}(\xi) = \mathit{I}(\xi)/\mathit{I}(0)$, for each model in 
Fig. \ref{fig:Schwarzschild_Emissivity}. 

\begin{table}
\centering
\caption{We compile below the different models we use to construct the 
images of a Schwarzschild BH undergoing spherical accretion, displayed in 
Fig. \ref{fig:Schwarzschild_Emissivity}. We have used $k_3$ and $k_2$ to 
represent arbitrary constants.}
\label{table:Emission_Models}
\begin{tabular}{lll}
\hline
Model & Emissivity model & Reference\\
\hline
$\texttt{em.1}$ & $j = {\dot{m}_\infty}/{16\pi^2 r^3 (r - 2 M)}$ & Eq.~\eqref{eq:Emission_Coeff_Schw} \\
$\texttt{em.2}$ & $j = k_3 r^{-3}$         			& \cite{Jusufi2021} \\
$\texttt{em.3}$ & $j = k_2 r^{-2}$ 				& \cite{Bambi2013a} \\
\hline
\end{tabular}
\end{table}

Independently of the emission model $j$ and of an inner emission region
cutoff at $r = r_{\text{cut}}$, we find an invariably sharp drop in
contrast at the shadow boundary, $\xi = \xi_{\text{ps}} = 3\sqrt{3}M$,
already evident from Eq. \ref{eq:Intensity_Integral_Inner_Boundary}. The
primary role of the emission coefficient is in determining the rate of
decay of the intensity with increasing $\xi$, and thus determines the
width of the observed emission ring, defined using the locations of the
maximum and the half-maximum intensities. Imposing a cut-off in the
emission inside the photon sphere $r_0 < r_{\text{cut}} \leq
r_{\text{ps}}$ modifies the intensity inside the shadow boundary $|\xi| <
\xi_{\text{ps}}$, and thus the overall contrast profile
slightly. However, the peak in the observed contrast profile remains at
the shadow boundary $\xi_{\text{peak}} = \xi_{\text{ps}}$, with its value
remaining divergent $\mathcal{C}_{\text{peak}} := \mathcal{C}
(\xi_{\text{peak}}) = \infty$. On imposing an artificial cut-off at a
larger radius $r_{\text{cut}} > r_{\text{ps}}$ we find that the peak of
the observed contrast profile shifts to $\xi_{\text{peak}} =
\xi_{\text{tp}}(r_{\text{cut}})$, i.e., to the impact factor of the
photon that ``turns'' at $r = r_{\text{cut}}$ (see below
Eq. \ref{eq:rtp}), as is expected from
Eq.~\eqref{eq:Intensity_Integral_Inner_Boundary}. We also find a drop in
the maximum contrast in the image, which becomes finite, and decreases
with increasing $r_{\text{cut}}$. Furthermore, the value of the contrast
at a fixed radius on the observer's sky, e.g., at $\xi = \xi_0$ increases
with increasing $r_{\text{cut}}$ until $r_{\text{cut}} =
r_{\text{tp}}(\xi_0)$, after which it starts decreasing (see
Sec.~\ref{sec:Fractional_Width} for further details).  Finally, we note
that the observed location and the magnitude of peak of the contrast in
the image can be compared against EHT data to obtain constraints on the
emission models used here. While we discuss the results one may derive
from such an exercise below for demonstrative purposes, it is important
to note here that (a) the properties of the accretion flow around
astrophysical BHs, such as M87$^*$ and Sgr~A$^*$, are expected to be
qualitatively different, best modelled by radiatively inefficient
accretion flows (RIAFs; \citealt{Yuan2014}), and (b) the
$\alpha$-measurement we use below depends critically on the measurement
of $\theta_{\text{g}}$, via a calibration of EHT data against the EHT
synthetic image library, the latter having been constructed from
simulations of RIAFs around the Kerr BHs of GR.

Indeed, the EHT has reported a measurement for the scaling $\alpha$ of
the angular diameter of the observed bright ring $d$ in the image of
M87$^*$ in terms of its angular gravitational radius
$\theta_{\text{g}}$, as $\alpha = d/\theta_{\text{g}} = 11^{+0.5}_{-0.3}$
(Table 1 of \citealt{EHT_M87_PaperI}), along with a lower-bound on the
maximum observed contrast $C_{\text{peak, EHT-min}} \gtrsim 10$. We use
the former measurement, i.e., $\xi_{{\text{peak, }}{\text{\tiny{EHT}}}}/M := \alpha/2
\approx 5.5$ to find that the location of the peak of emission in the
spacetime must be approximately at $r_{\text{cut}} \approx 3.77 M$, for
the Schwarzschild BH in the present accretion-emission setup. Using the
$1\!-\!\sigma$ values leads to a $10\%$ variation at the most. As can be
seen from Eq. \ref{eq:Intensity_Integral_Inner_Boundary} and Fig.
\ref{fig:Schwarzschild_Emissivity}, this is independent of the emission
models we use here. At the same time, it is important to note that the
maximum contrast in the image $\mathcal{C}_{\text{peak}} :=
\mathcal{C}(\xi = \xi_{\text{peak}})$ varies with $j$; In particular, if
we write it as $j = k_n r^{-n}$ ($k_n$ representing a constant), then, as
expected, steeper emission coefficients (larger $n$) yield larger
$\mathcal{C}_{\text{peak}}$. We refer the reader to see Sec.
\ref{sec:Fractional_Width} for further details. Thus, we are in broad
agreement with the results of the EHT, despite the accretion model being
drastically different, that the peak in emission occurs quite close to
the photon sphere. To compare, the impact parameter of a photon that has
a turning point at the location of the Schwarzschild innermost stable
circular orbit (ISCO) is $\xi_{\text{tp}} (r_{\text{ISCO}} = 6M) =
\sqrt{54}M$, which implies that if the peak of emission in the spacetime
were located instead at the ISCO, the radius of the observed ring in the
image would be about $34\%$ larger than the fiducial value of
$\xi_{{\text{peak, }}{\text{\tiny{EHT}}}}$ that we have adopted here.

\begin{figure*}
\centering
\includegraphics[width=.67\columnwidth]{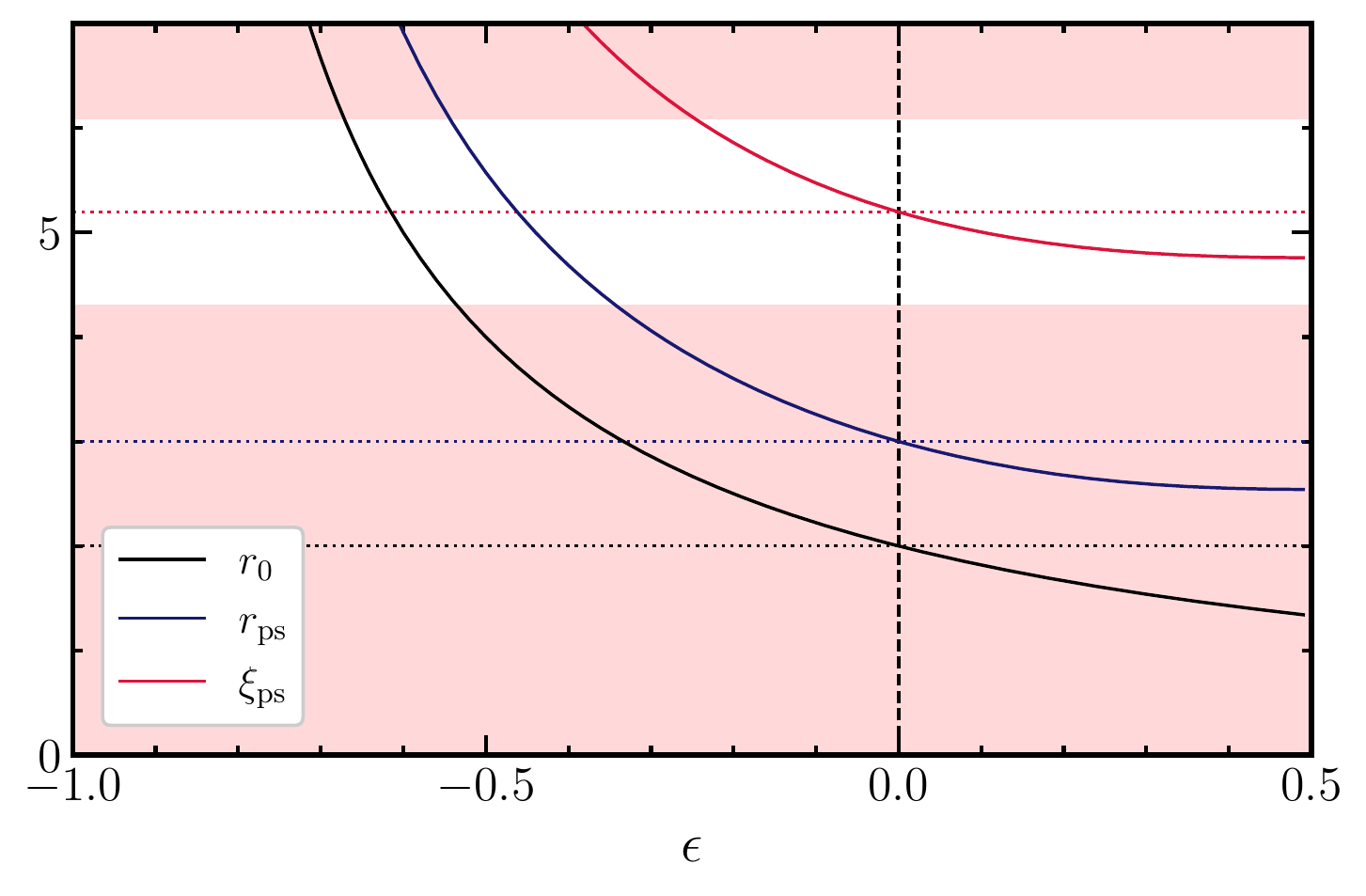}
\includegraphics[width=.67\columnwidth]{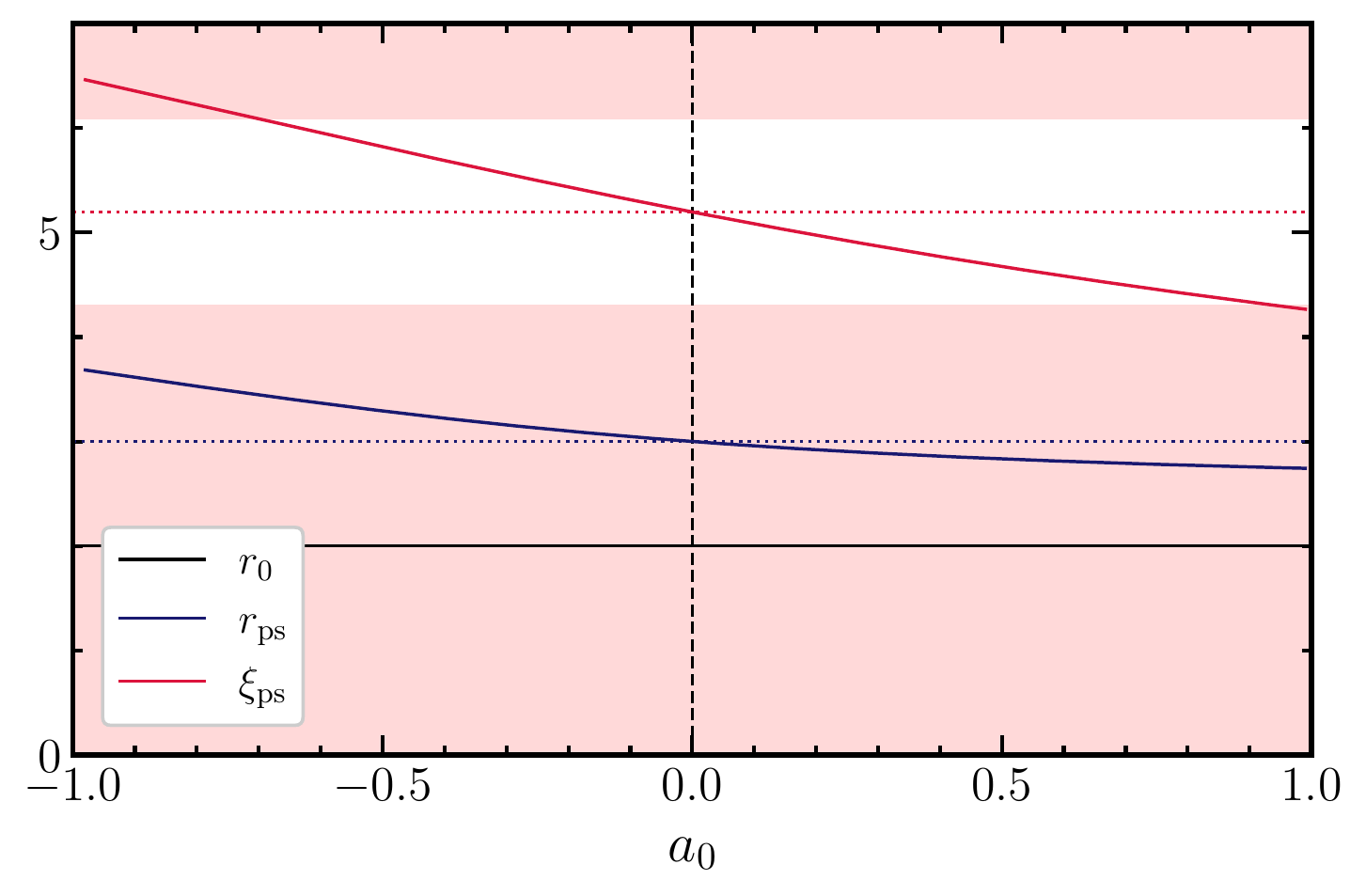}
\includegraphics[width=.67\columnwidth]{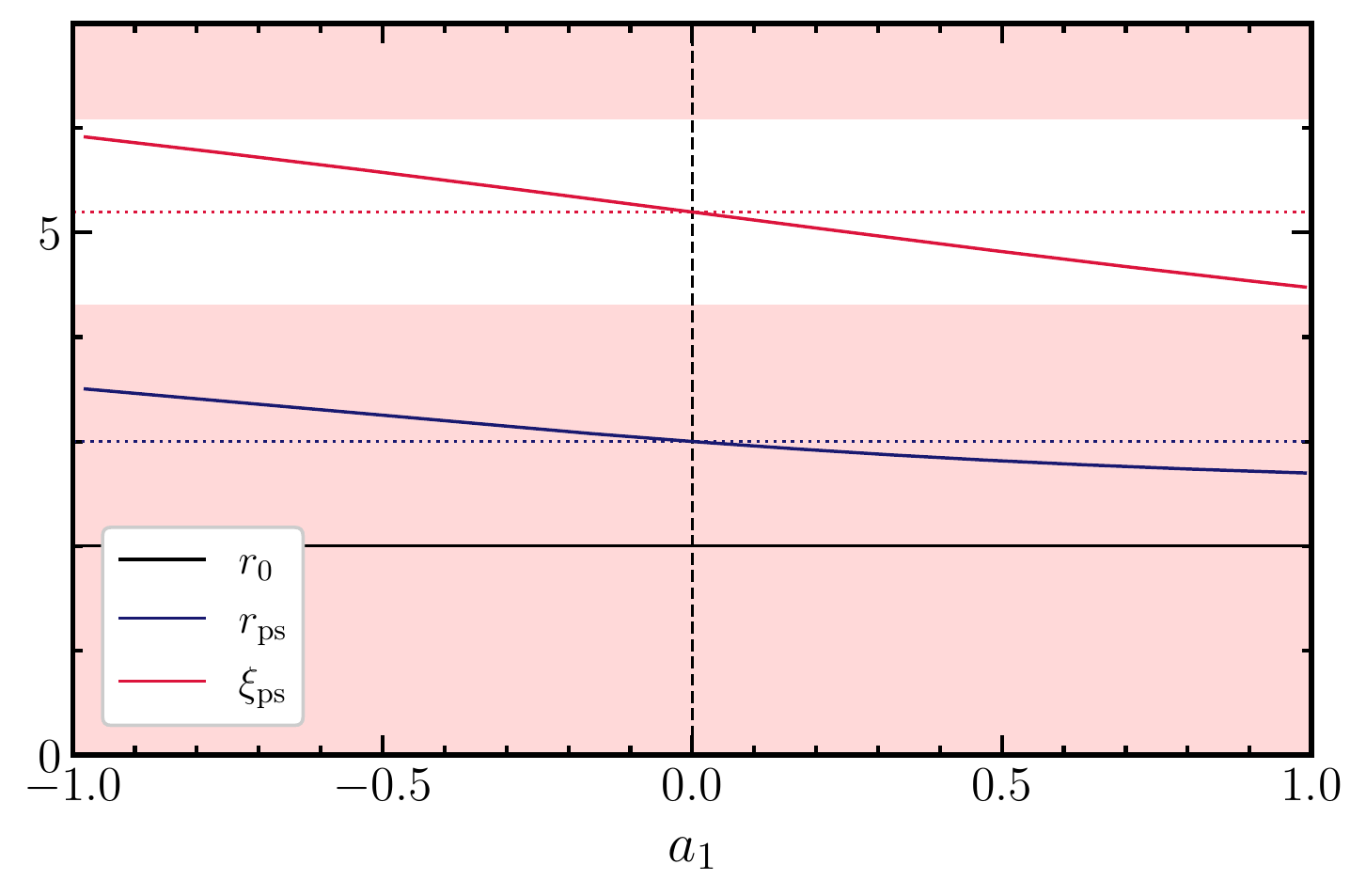}
\includegraphics[width=.68\columnwidth]{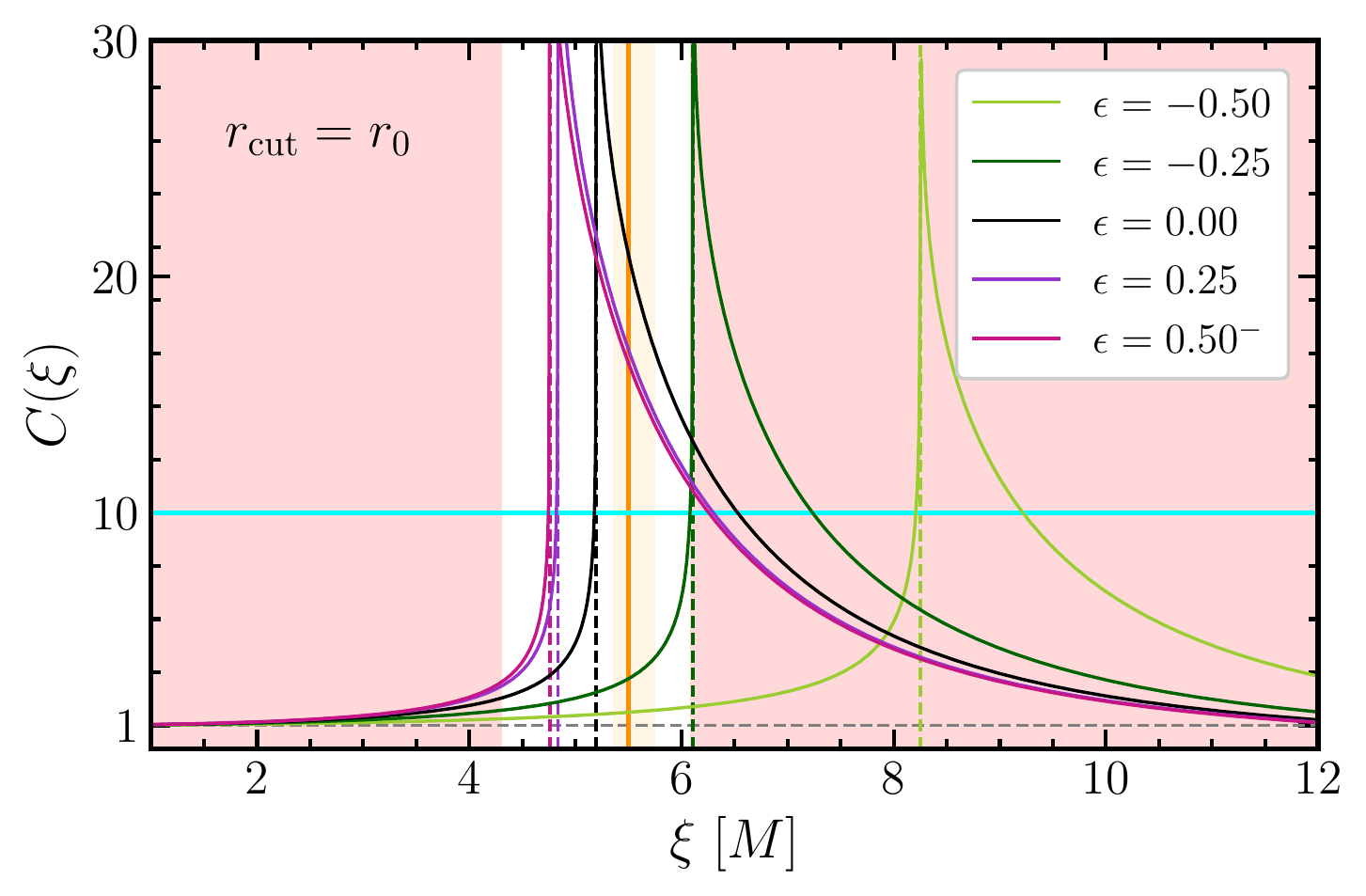}
\includegraphics[width=.68\columnwidth]{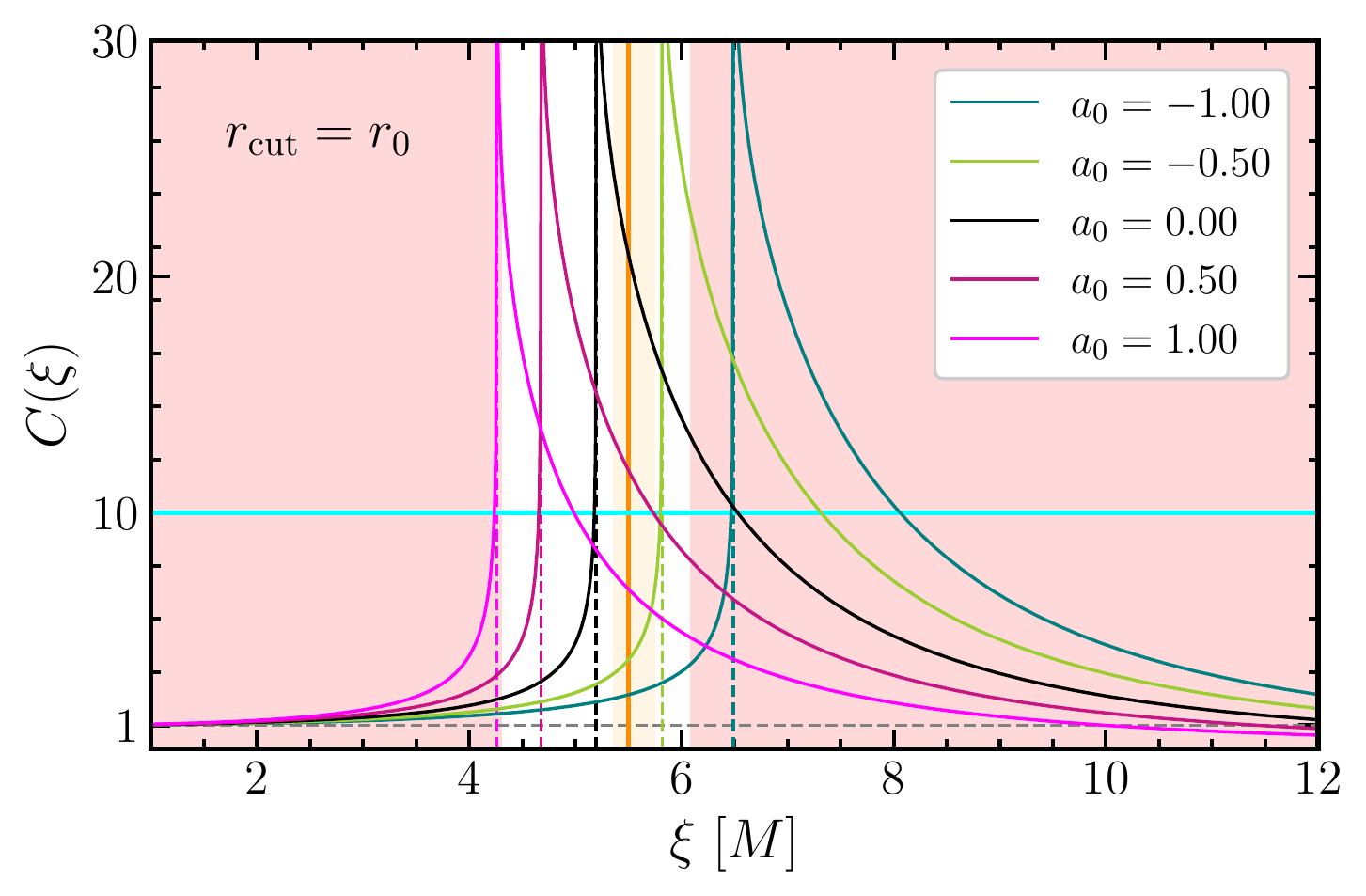}
\includegraphics[width=.68\columnwidth]{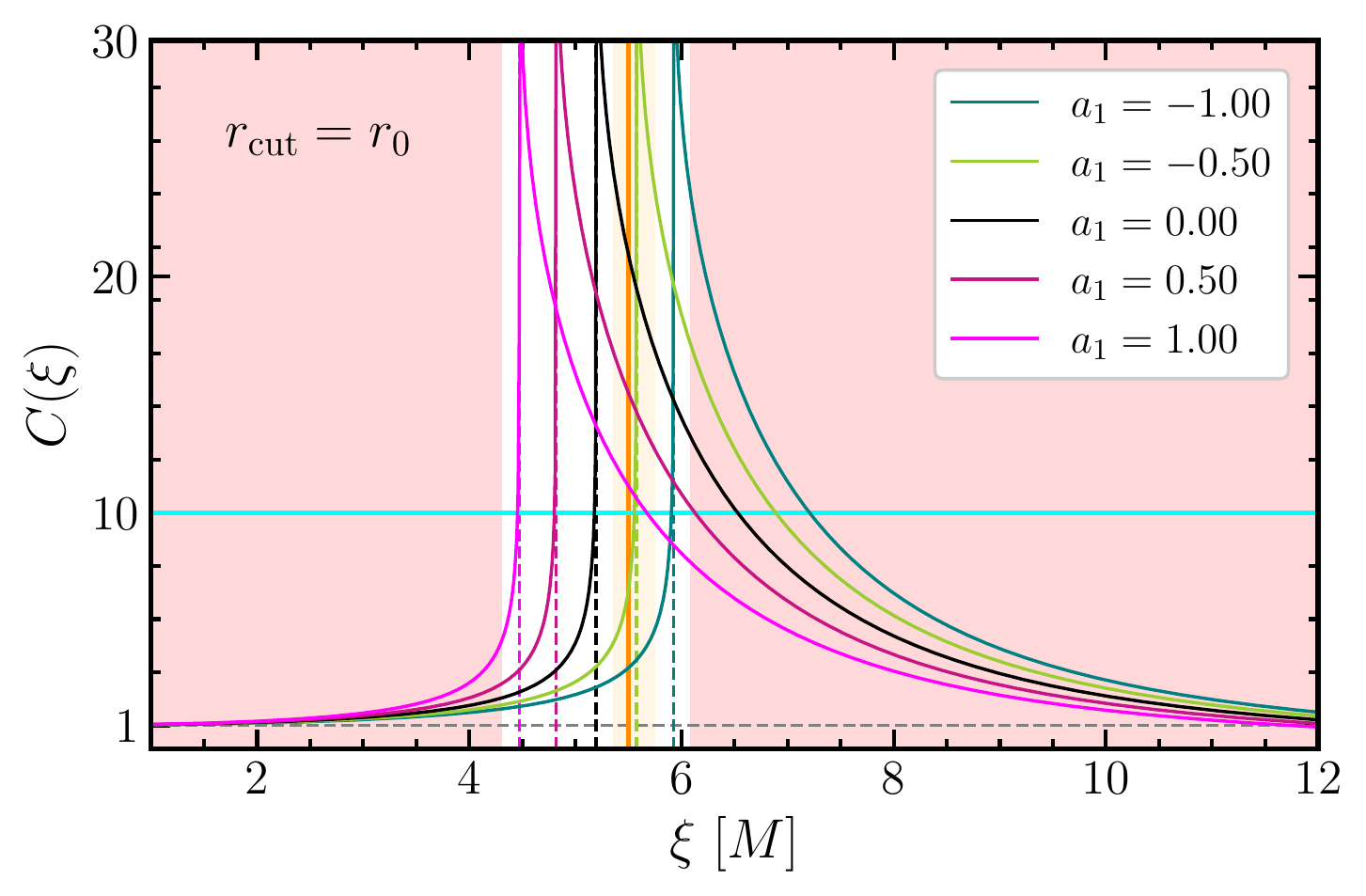}
\includegraphics[width=.68\columnwidth]{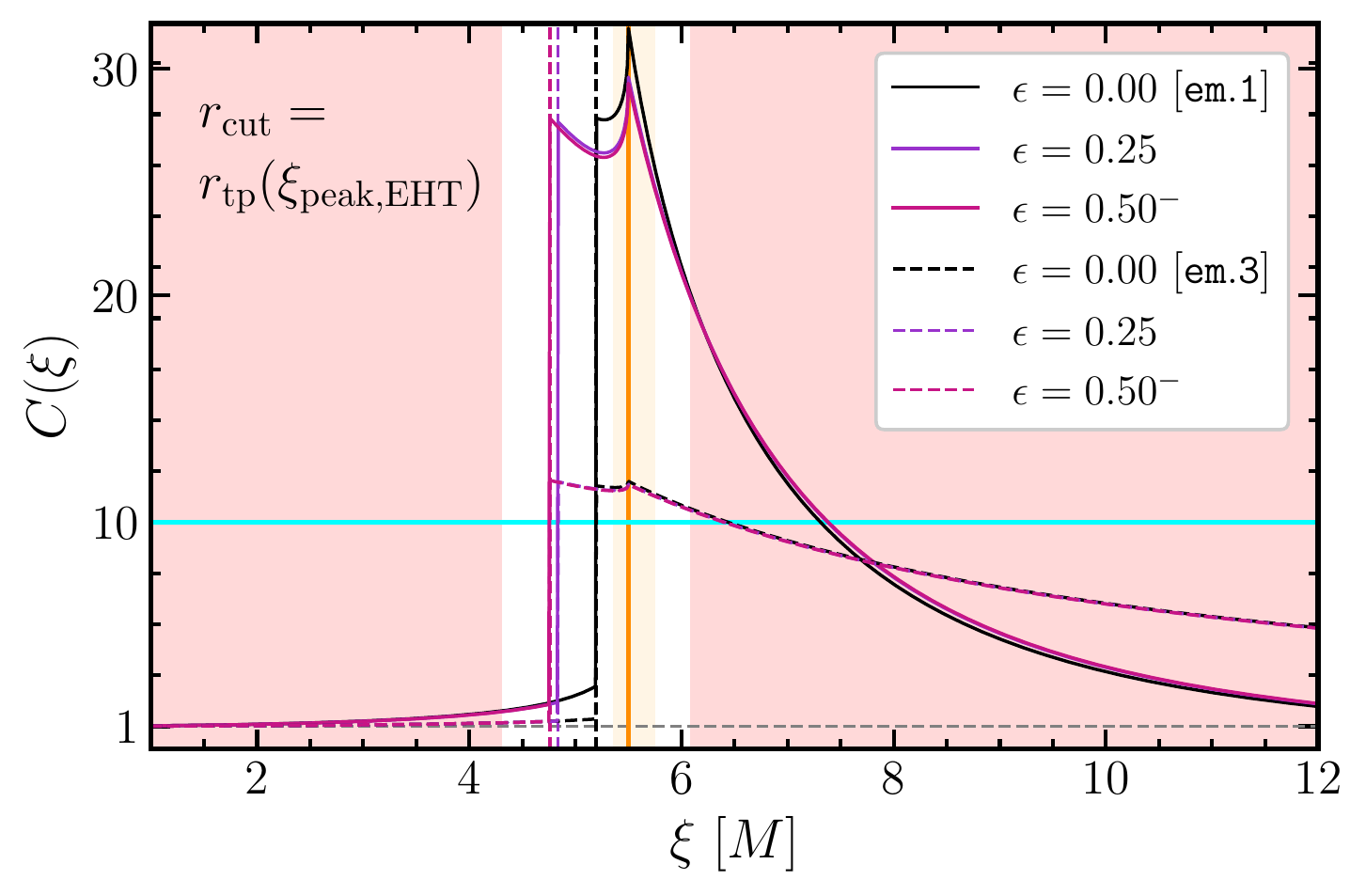}
\includegraphics[width=.68\columnwidth]{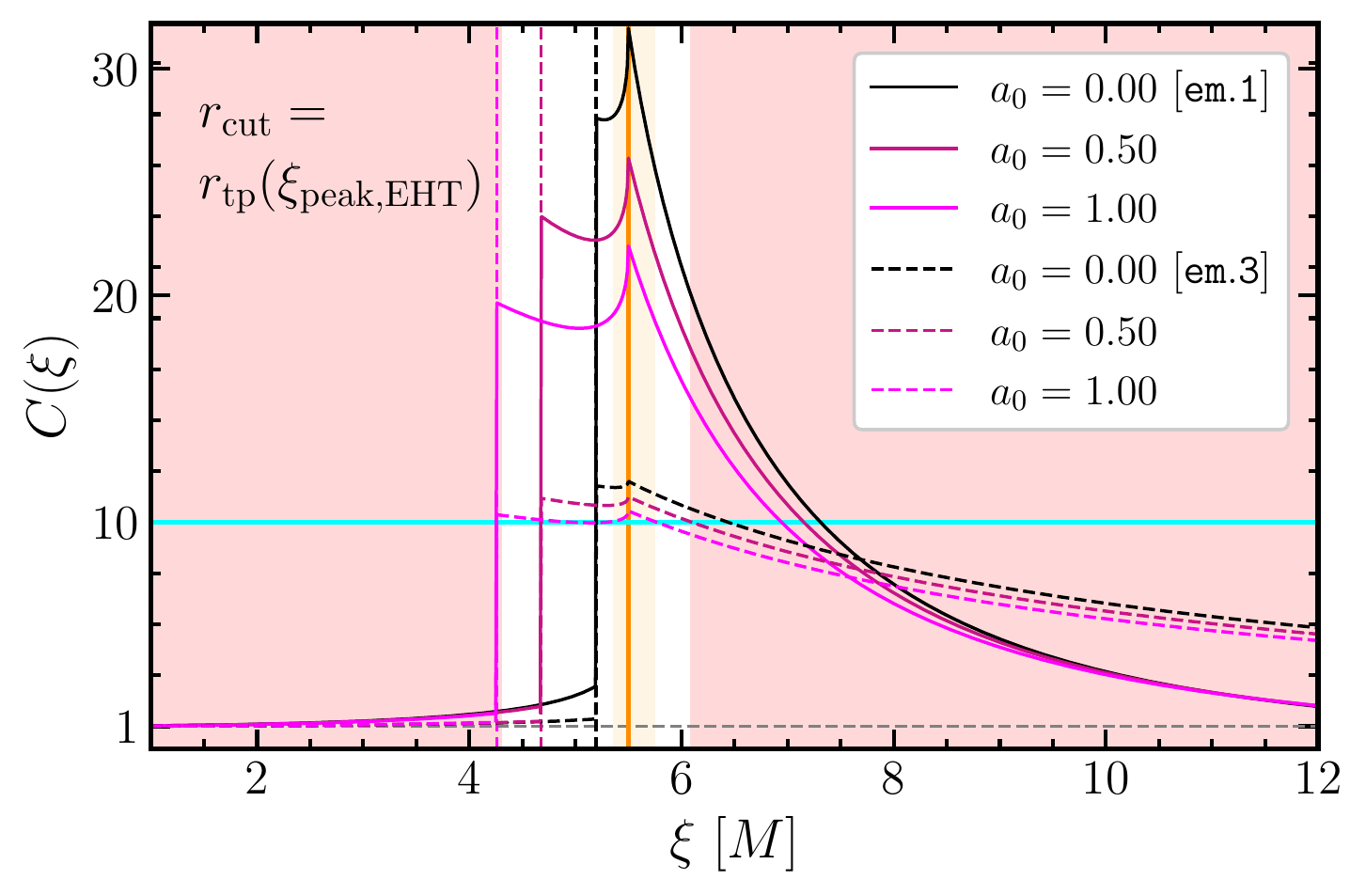}
\includegraphics[width=.68\columnwidth]{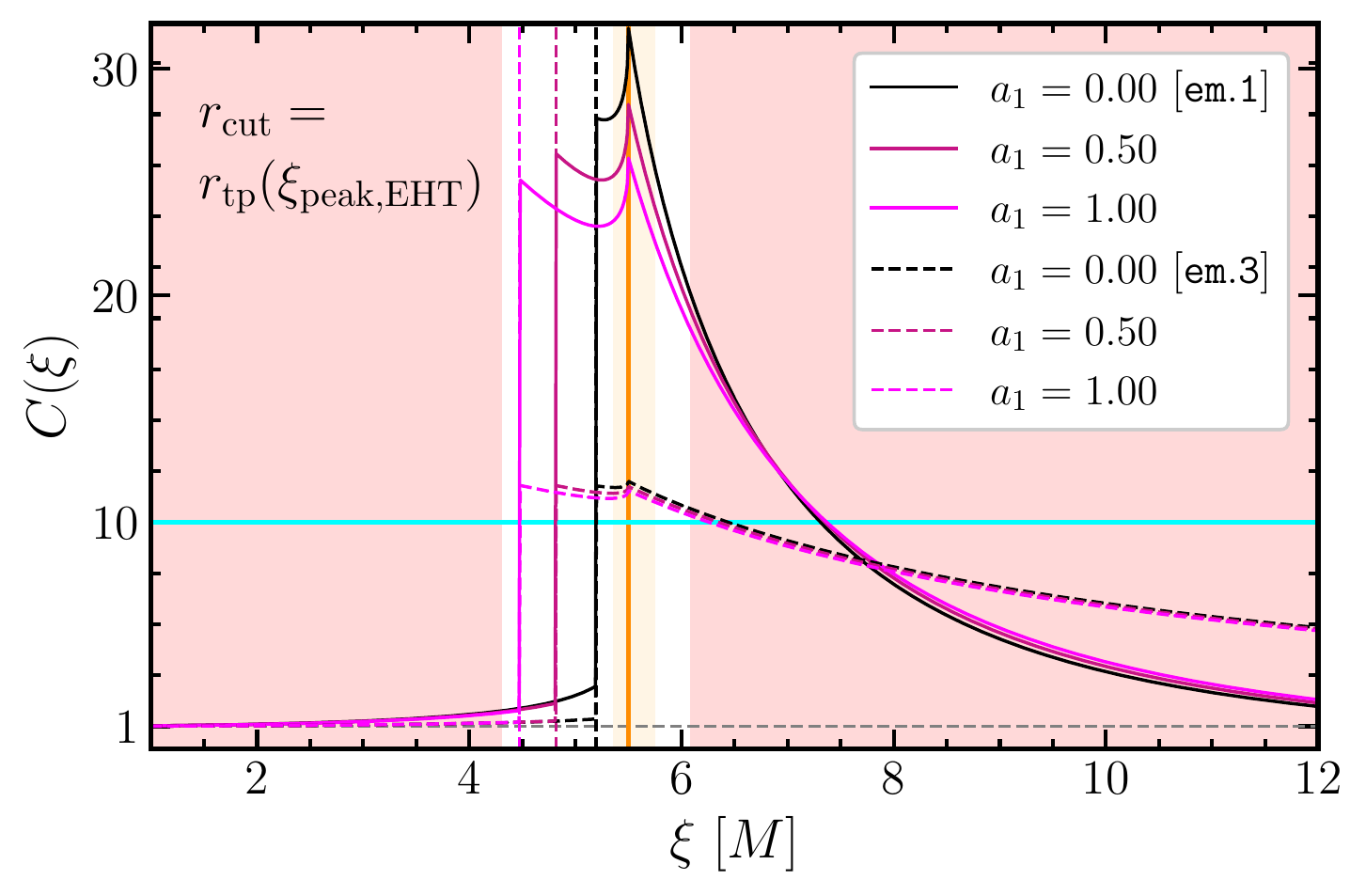}
\caption{In the top row, we show the variation in the sizes of the event
  horizon $r_0$, the photon sphere $r_{\text{ps}}$, and the shadow
  $\xi_{\text{ps}}$, with the relevant metric parameter for the
  single-parameter RZ BHs we use here. In the middle row, we display the
  contrast profile $\mathcal{C}(\xi)$, as seen on the observer's sky,
  when the emission region extends all the way down to the BH event
  horizon.  The bottom row shows the contrast profiles of the set of BHs
  that have $\xi_{\text{ps}} < \xi_{{\text{peak, }}{\text{\tiny{EHT}}}}$ with an imposed
  inner cut-off for the emission region at $r_{\text{cut}} =
  r_{\text{tp}}( \xi_{{\text{peak, }}{\text{\tiny{EHT}}}})$, so that the location of the
  peak in the contrast profile is at the EHT reported value. The red
  region in all panels shows the range of parameter disallowed by the
  EHT-reported shadow-size constraints. The vertical orange line
  corresponds to $\xi = \xi_{{\text{peak, }}{\text{\tiny{EHT}}}}$, and the orange band
  shows the associated $1\!-\!\sigma$ range. The horizontal cyan line is
  the lower bound for $\mathcal{C}_{\text{peak, EHT}}$ reported by the
  EHT.}
\label{fig:RZ_eps_a0_a1_one_param}
\end{figure*}

\begin{table*}
\centering
\caption{
We show here the theoretically-allowed parameter ranges for various 
sections of the $\mathscr{M}(\epsilon; a_0; a_1)$ RZ BHs we consider 
here. We also show which PN parameters vanish in general in each case, 
and when the horizon size is identical to that of a Schwarzschild BH, 
$r_0 = 2M$.}
\label{table:RZ_Parameter_Space_Constraints_PPN}
\begin{tabular}{||l||l|l|l||r|r|r||c||}
\hline
Model & \multicolumn{3}{l||}{RZ Parameters: Theoretically-Allowed Range} 
& \multicolumn{3}{l||}{PN Parameters} & $r_0 = 2M$ \\
\hline
& $\epsilon$ & $a_0$ & $a_1$ & $\zeta_1$ & $\zeta_2$ & $\zeta_3$ &  \\
\hline
$\mathscr{M}(\epsilon)$ & $-1 < \epsilon \leq 1/2$ & 0 & 0 & 0 & 
$\neq 0$ & 0 & $\times$ \\
\hline
$\mathscr{M}(a_0)$ & 0 & $a_0 \geq -1$  & 0 & $\neq 0$ & $\neq 0$ & 0 
& $\checkmark$ \\
\hline
$\mathscr{M}(a_1)$ & 0 & 0 & $a_1 \geq -1$ & 0 & $\neq 0$ & $\neq 0$ 
& $\checkmark$ \\
\hline
$\mathscr{M}(a_0; a_1)$ & 0 & $a_0 \leq -3$ & $ a_1 \geq 
\frac{2\sqrt{-a_0^3}}{3\sqrt{3}}$ & $\neq 0$ & $\neq 0$ & $\neq 0$
& $\checkmark$ \\
& & $a_0 > -3$ & $a_1 \geq -1 - a_0$ & & & & \\
\hline
$\mathscr{M}(\epsilon; a_1)$ & $-1 < \epsilon \leq 1$ & 0 & $a_1 
\geq -1 + 2\epsilon$ & 0 & $\neq 0$ & $\neq 0$ & $\times$ \\
& $\epsilon > 1$ & & \(\displaystyle a_1 \geq \frac{1}{27}
\left[9\epsilon^2 + 2\epsilon^3 + 2\sqrt{\epsilon^3(3+\epsilon)^3}
\right]\) & & & & \\
\hline
$\mathscr{M}(\epsilon; a_0)$ & $-1 < \epsilon \leq 2 $ & $a_0 \geq -1 + 
2\epsilon$ & 0 & $\neq 0$ & $\neq 0$ & 0 & $\times$ \\
& $\epsilon > 2 $ & \(\displaystyle a_0 \geq \frac{4\epsilon + 
\epsilon^2}{4}\) & & & & & \\
\hline
\end{tabular}
\end{table*}

\begin{figure}
\centering
\includegraphics[width=0.45\textwidth]{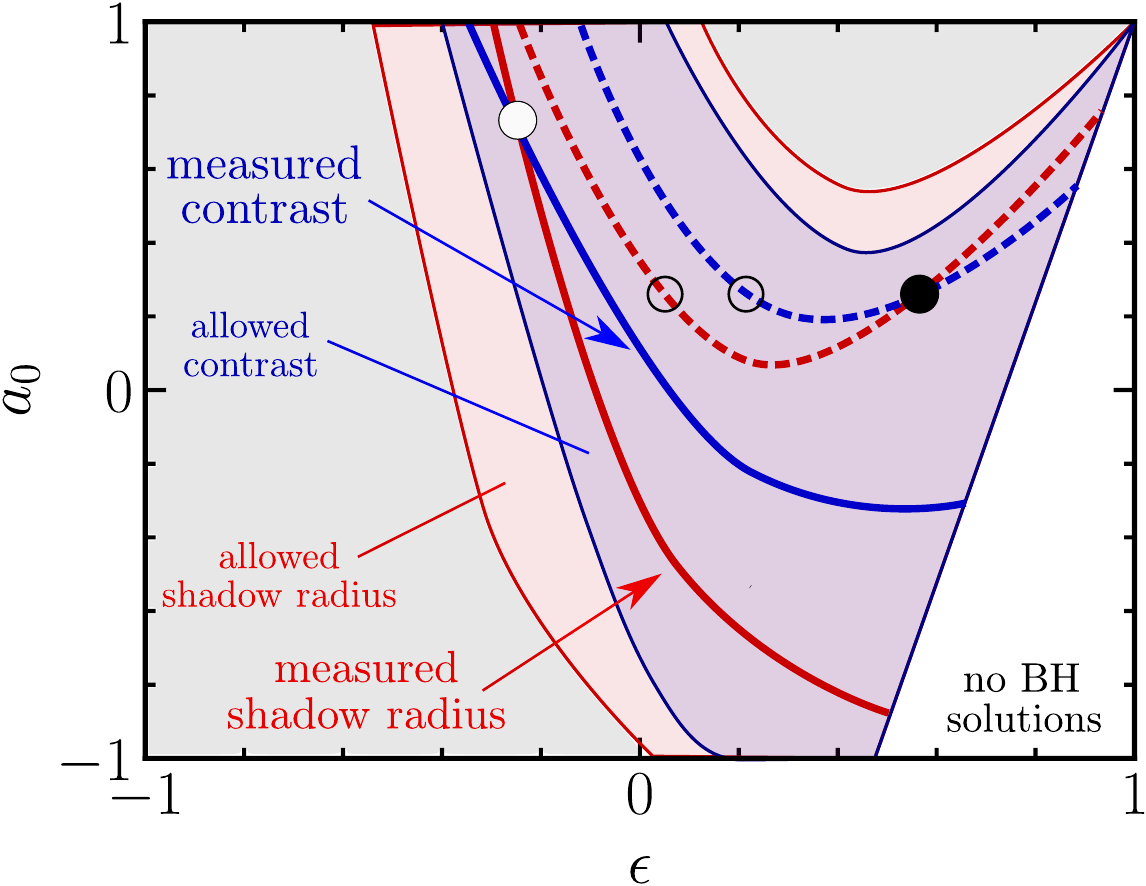}
\caption{Schematic representation of the combined use of the shadow
  radius and of the intensity contrast to obtain constraints on the
  values of the coefficients of the RZ-metric. Shown respectively as
  shaded red and blue areas are the allowed regions for the RZ
  coefficients for the shadow radius and contrast,
  respectively. Indicated instead with solid lines are the constraints on
  the coefficients once a shadow radius (red line) and the contrast (blue
  line) have been measured with very high precision. Their intersection
  can then be used to set precise constraints on the coefficients (filled
  white circle). Note that the measurements of the shadow radius and
  contrast when considered alone can be degenerate (dashed red and blue
  lines), namely, multiple values of the coefficient $\epsilon$ can be
  used for the same value of the coefficient $a_1$ (empty
  circle). However, the intersection of the two measurements leads to a
  precise constraint of the coefficients (filled black circle). The panel
  refers to the RZ coefficients $a_0$ and $\epsilon$, but a similar logic
  can be applied when considering the coefficients $a_1$ and
  $\epsilon$. Figures \ref{fig:RZ_eps_a0_a1_two_param} and
  \ref{fig:RZ_eps_a0_a1_two_param_rcut} provide more quantitative
  examples of this schematic diagram.}
\label{fig:cartoon}
\end{figure}
\begin{figure*}
\centering
\includegraphics[width=.68\columnwidth]{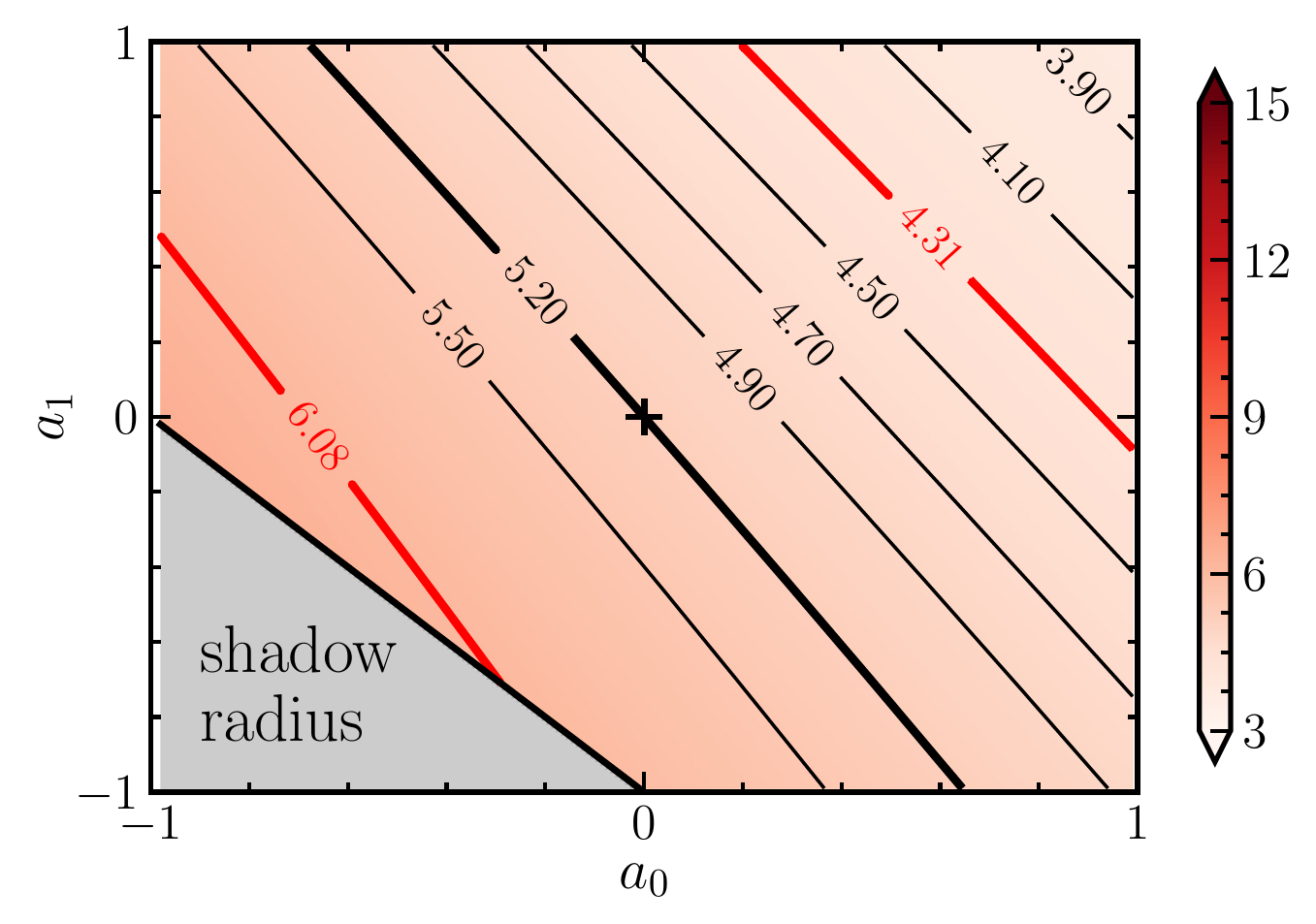}
\includegraphics[width=.68\columnwidth]{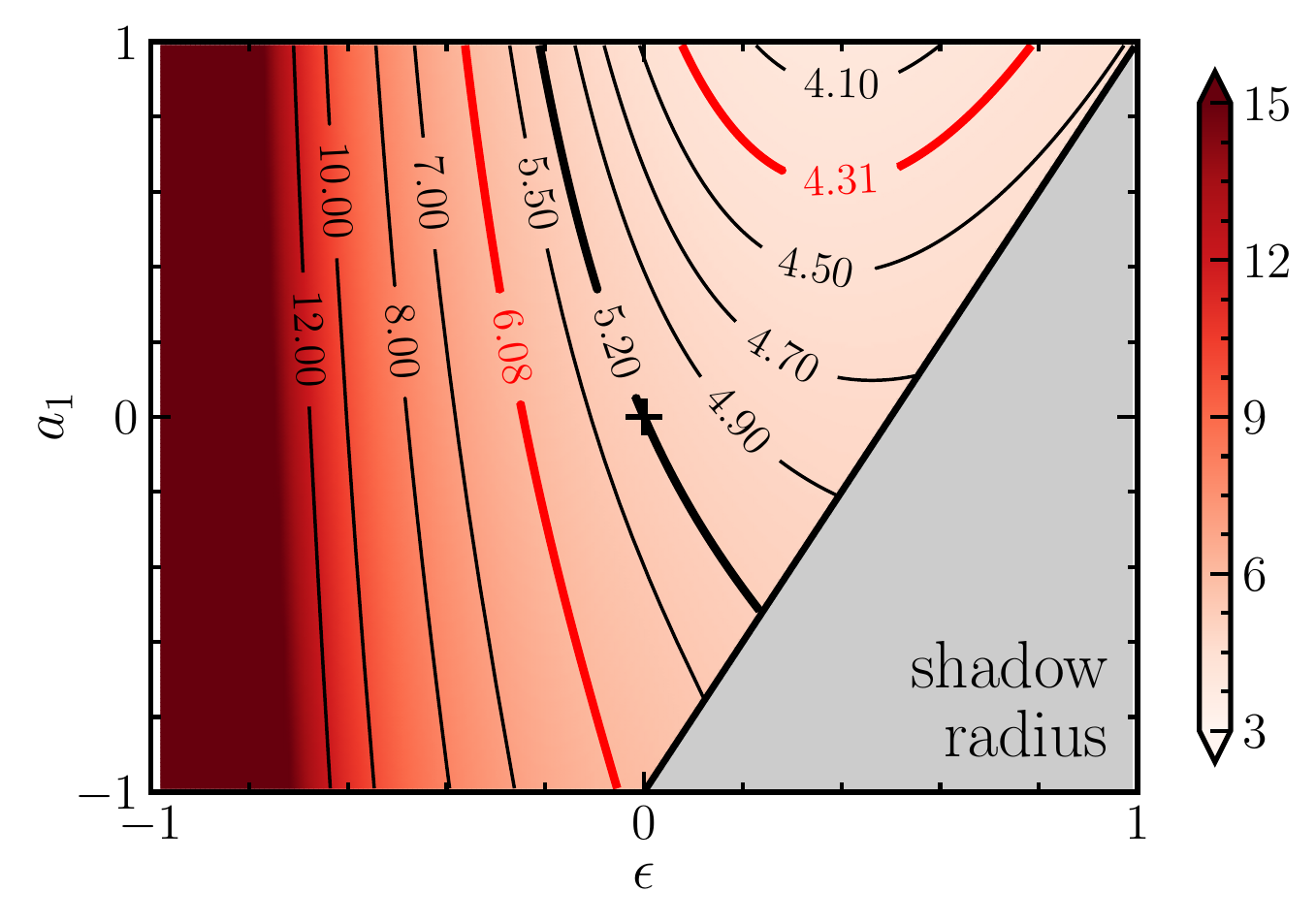}
\includegraphics[width=.68\columnwidth]{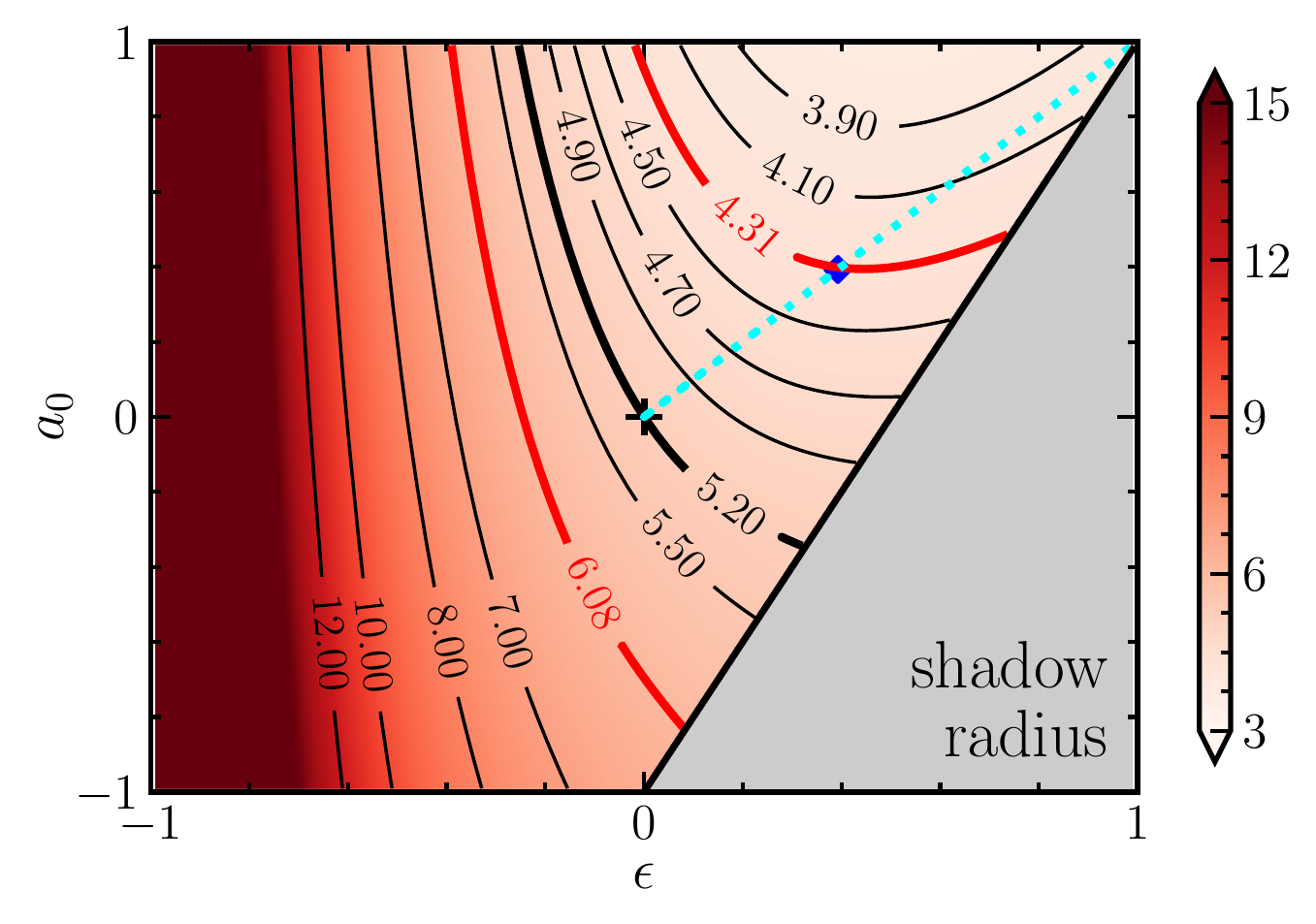}
\includegraphics[width=.68\columnwidth]{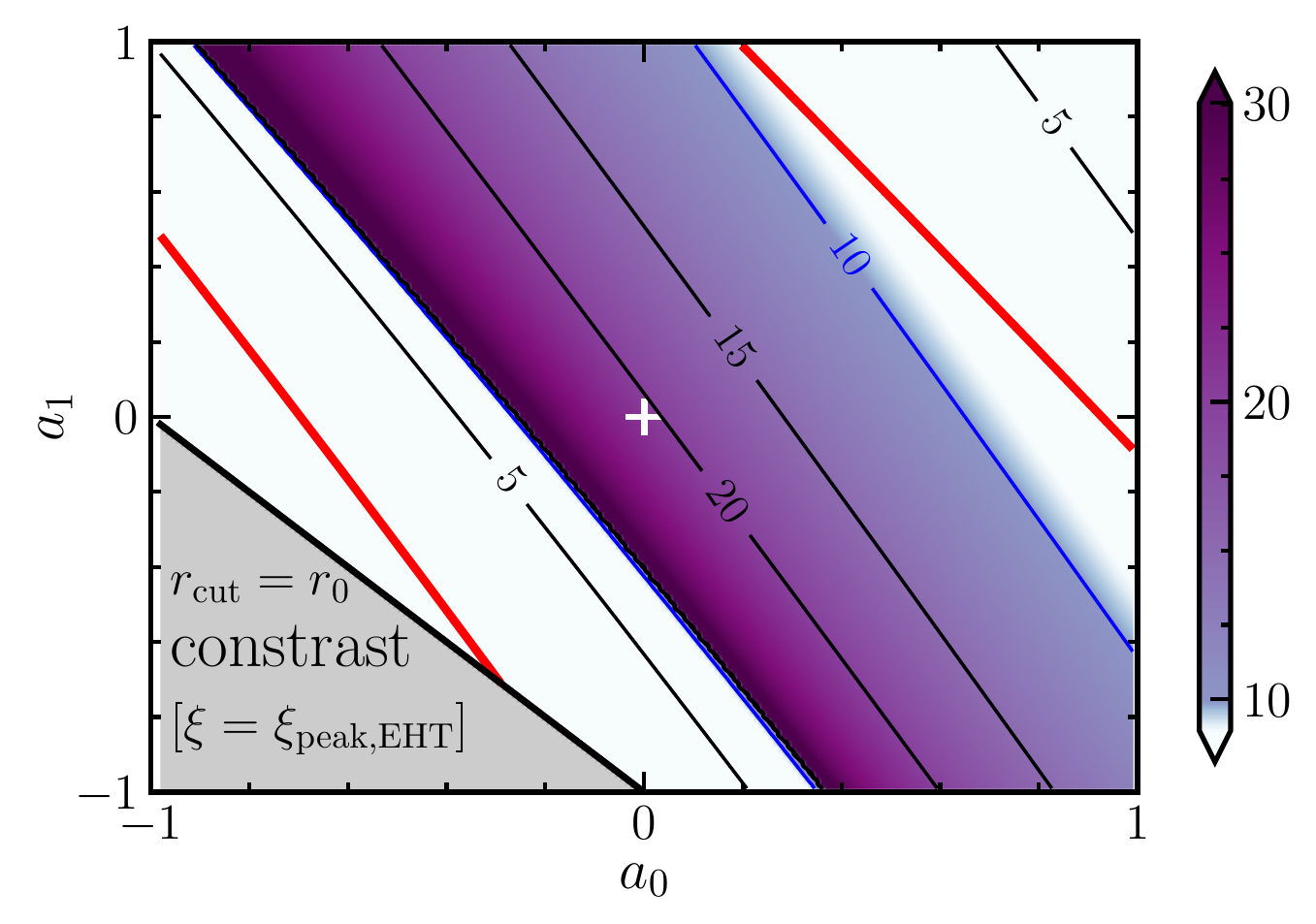}
\includegraphics[width=.68\columnwidth]{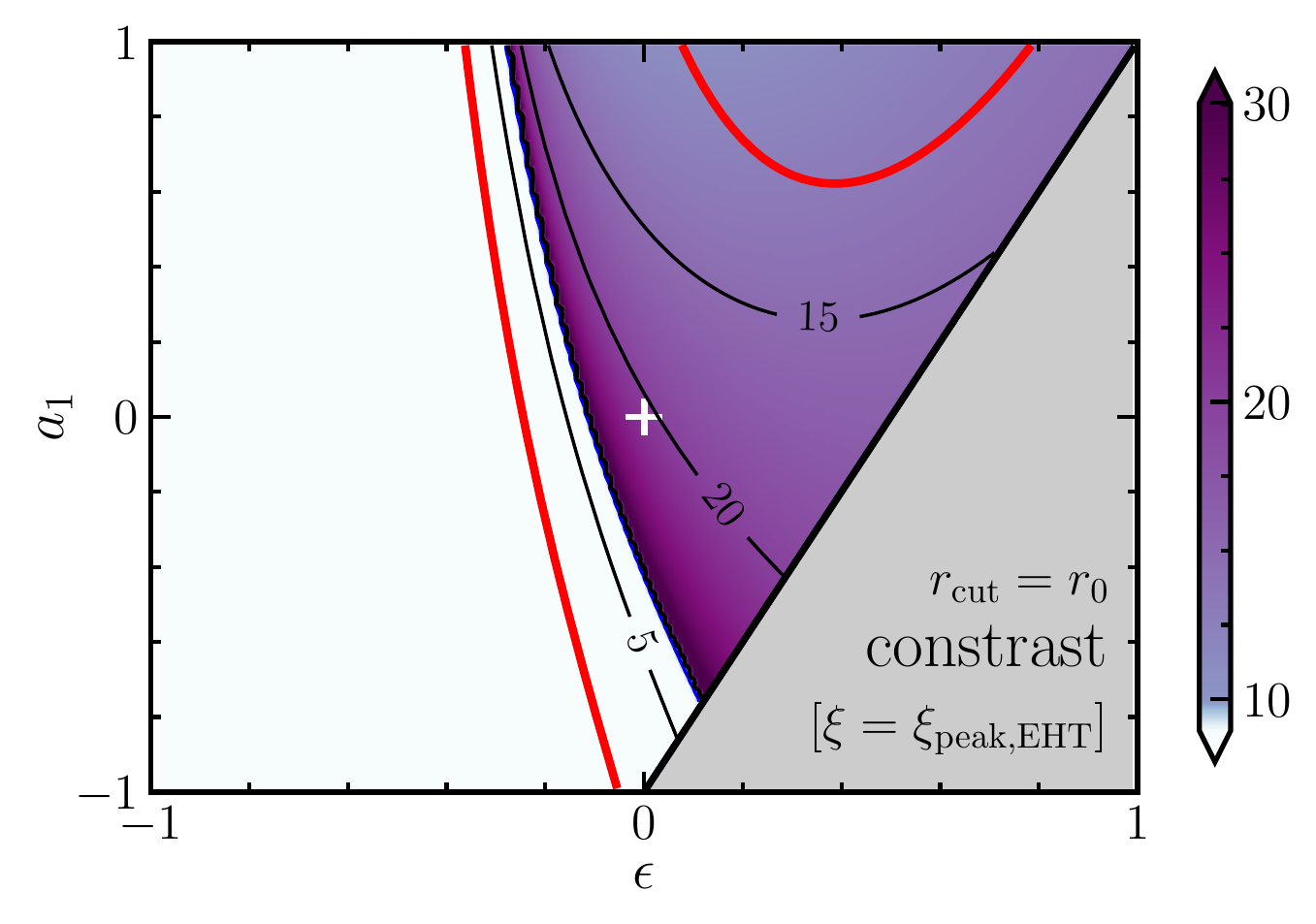}
\includegraphics[width=.68\columnwidth]{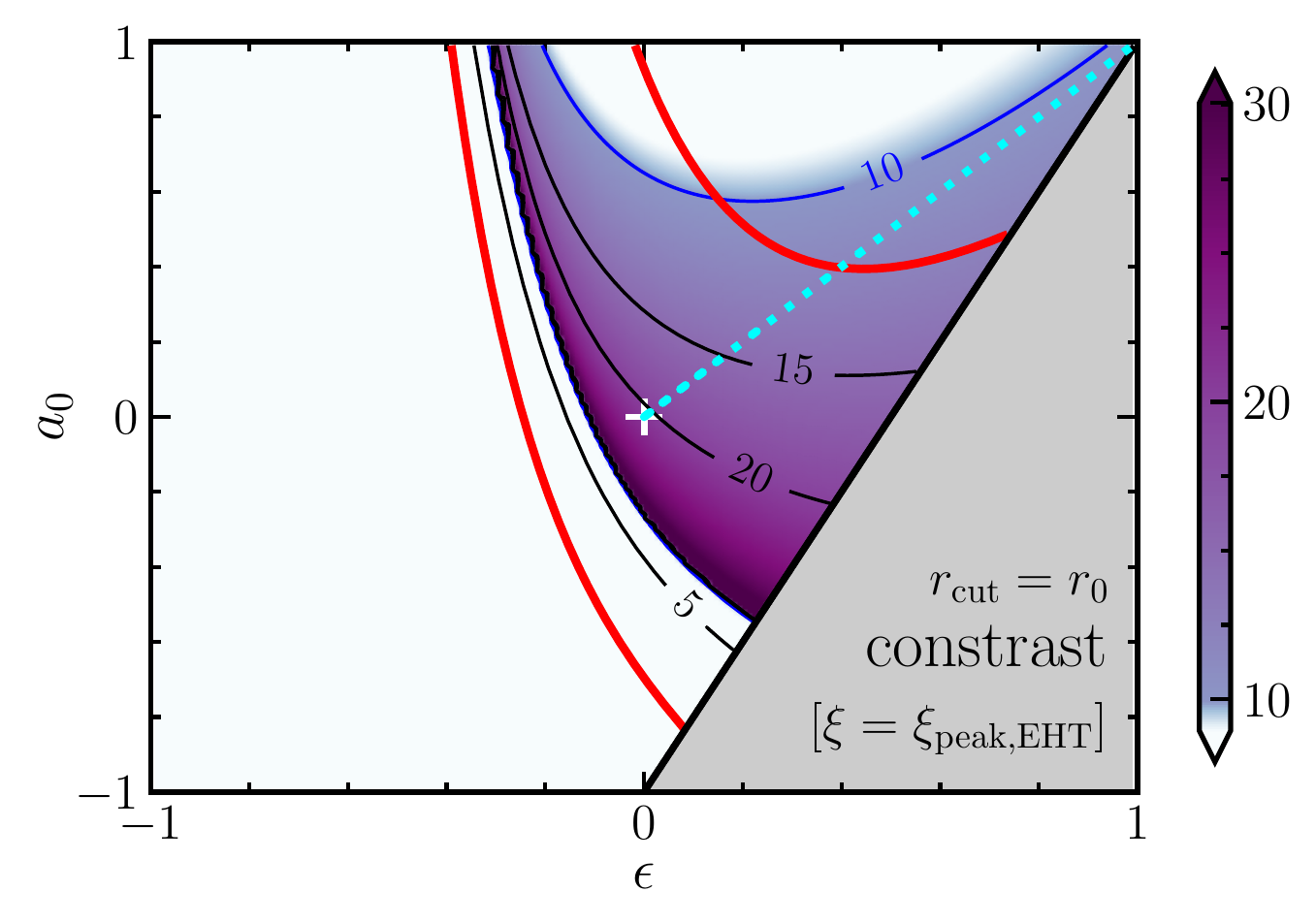}
\hspace{0.2cm}
\caption{The variation in the shadow-size (top row), and the contrast in
  the BH image at $\xi = \xi_{{\text{peak, }}{\text{\tiny{EHT}}}} = 5.5M$
  (bottom row) for the principal two-parameter RZ BH models used
  here. The red lines in all plots show $\xi = 3\sqrt{3}(1 \pm 0.17)M$,
  the approximate $1\!-\!\sigma$ bounds of allowed shadow-sizes obtained
  by the EHT for M87$^*$. When displaying the bottom row, we have allowed
  the emission region to extend all the way down to the event horizon
  (this row is the 2D analogue of the middle row of
  Fig. \ref{fig:RZ_eps_a0_a1_one_param}). Note that the peak in the
  contrast profile does not appear at $\xi = 5.5M$, except for the BHs
  with shadow-sizes of $\xi_{\text{ps}} = 5.5 M$. Thus, if the emission
  is not suppressed close to the BH, the requirement that the
  peak-contrast in the image occur at $\xi = 5.5M$ is an extremely
  stringent constraint, independent of the emission model.  The gray
  region shows the ``theoretically-disallowed'' region of the relevant RZ
  parameter space (see Table
  \ref{table:RZ_Parameter_Space_Constraints_PPN}).  The crosses (black in
  the upper panels and white in the lower ones) locate the Schwarzschild
  BH and the dotted line represents the family of Reissner-Nordstr{\"o}m
  (RN) BHs. The blue marker shows the $\bar{q} = 0.9$ RN BH (see
  \citealt{Kocherlakota2021}).}
\label{fig:RZ_eps_a0_a1_two_param}	
\end{figure*}

\begin{figure*}
\centering
\includegraphics[width=.68\columnwidth]{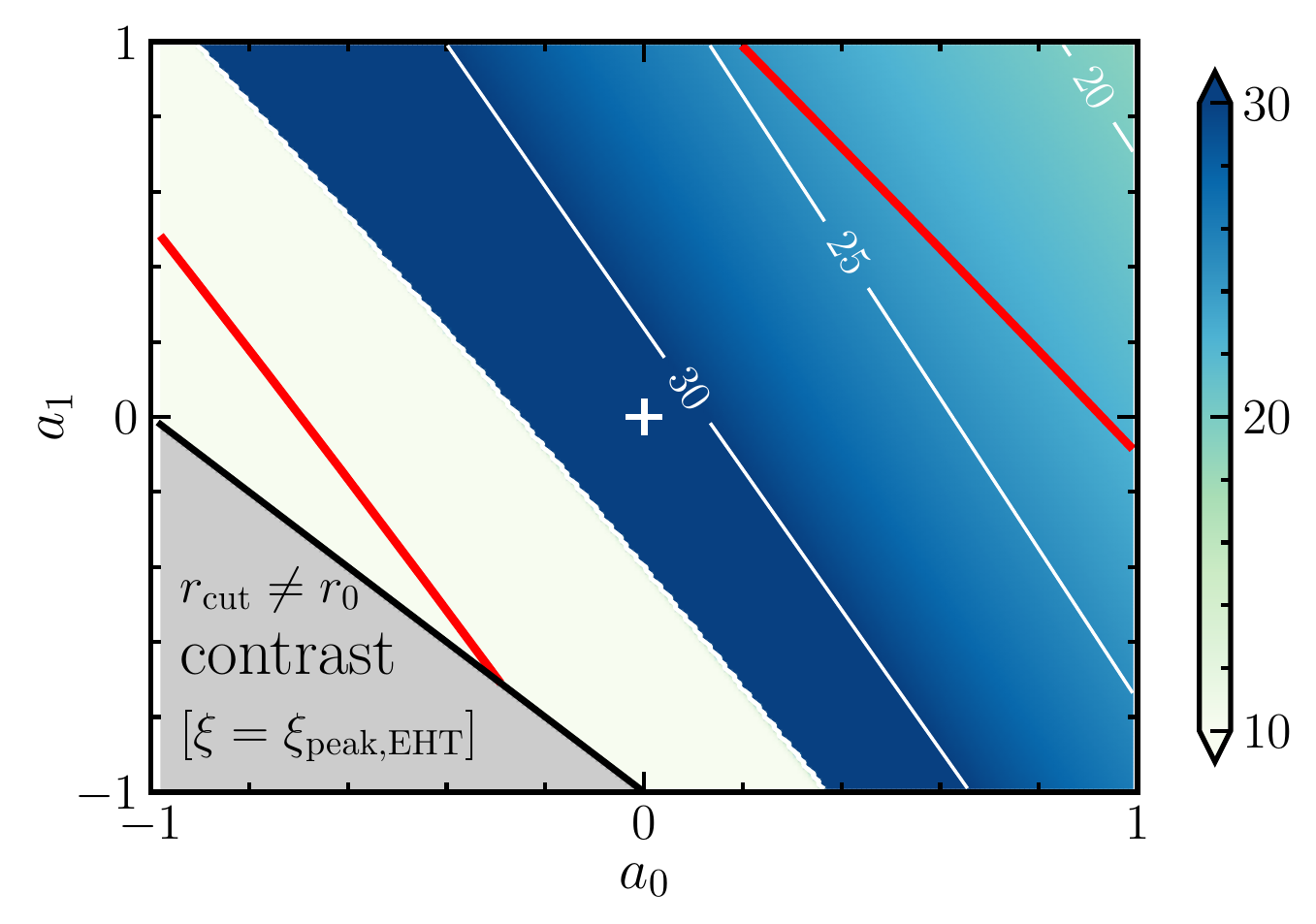}
\includegraphics[width=.68\columnwidth]{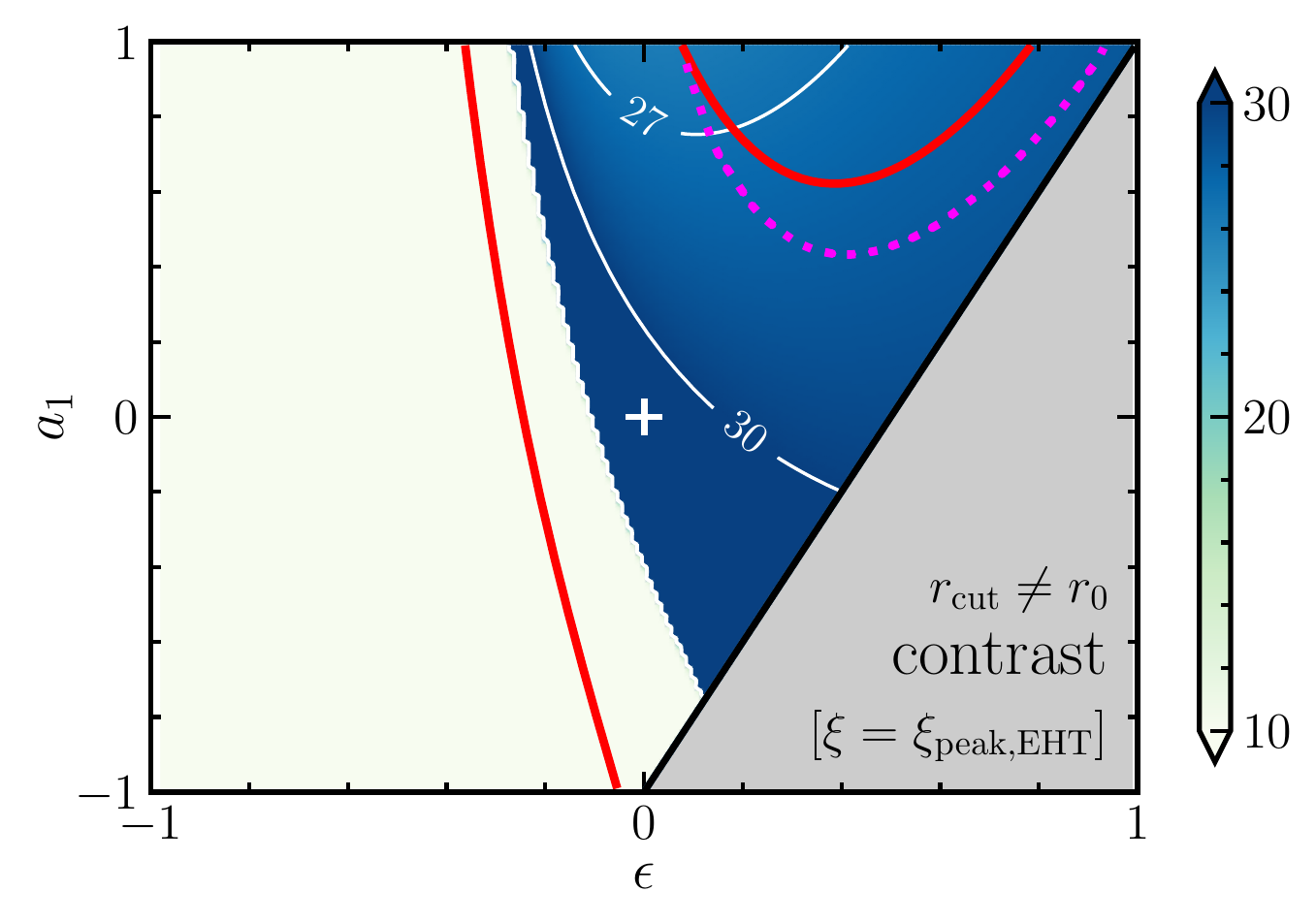}
\includegraphics[width=.68\columnwidth]{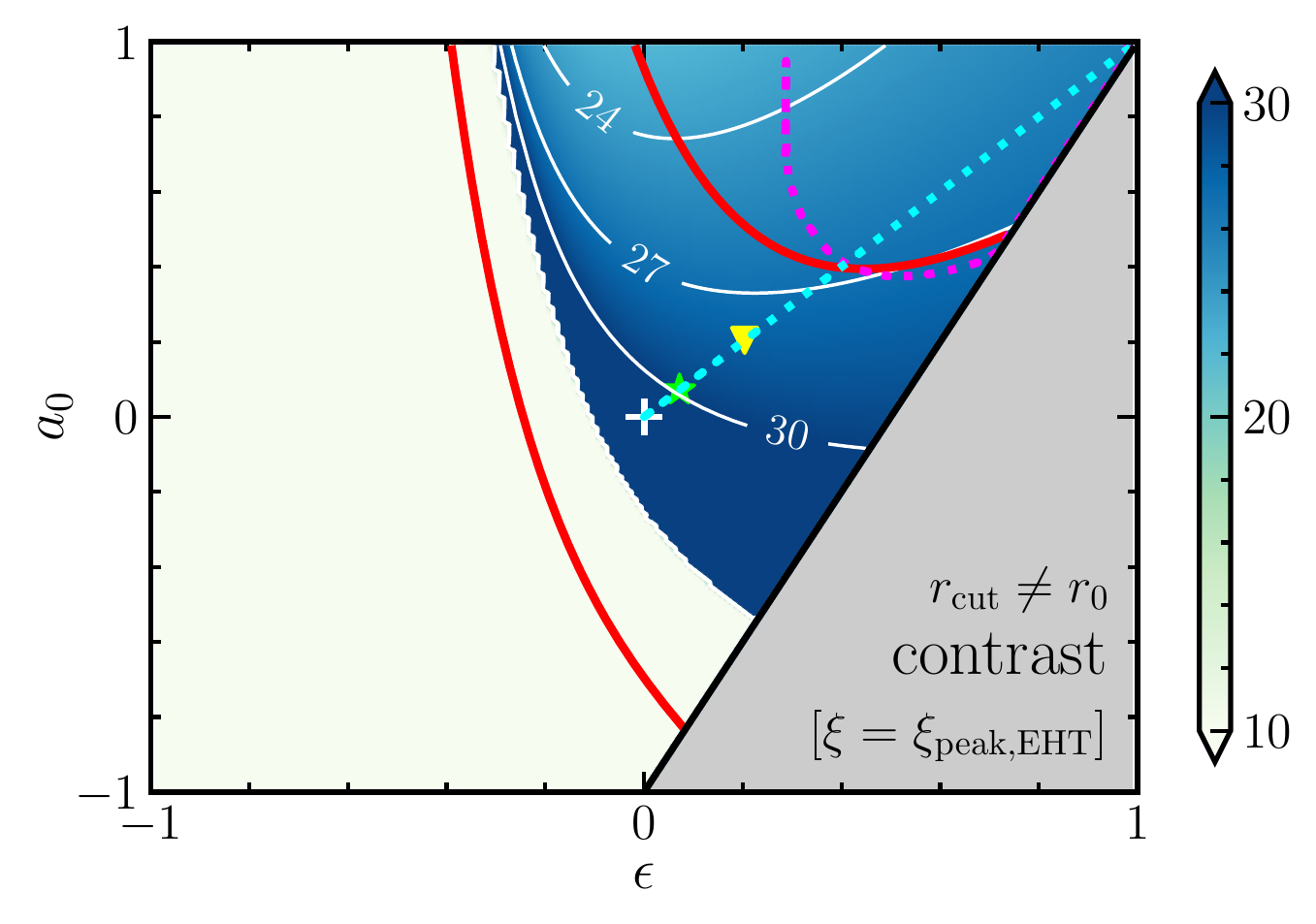}
\includegraphics[width=.68\columnwidth]{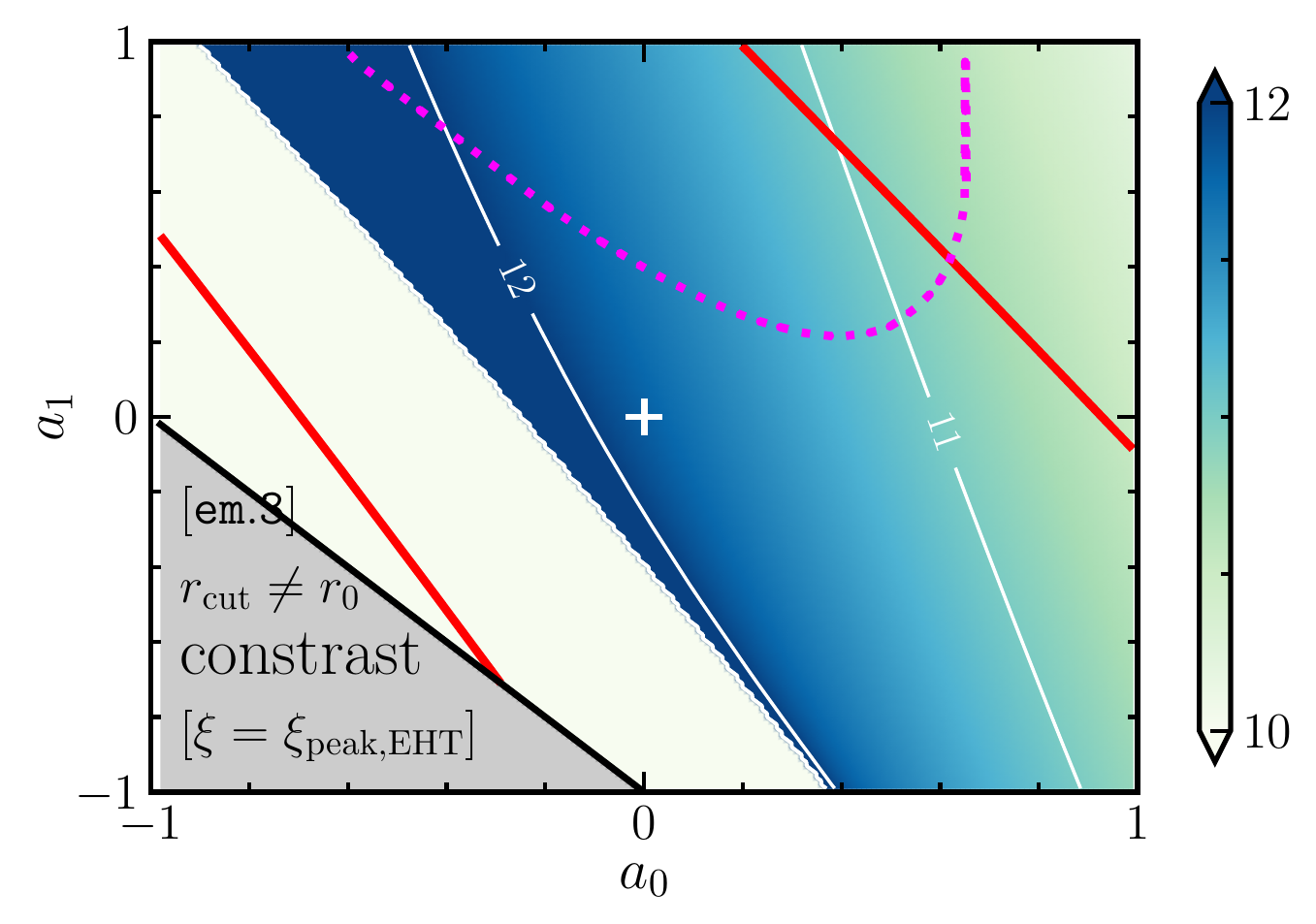}
\includegraphics[width=.68\columnwidth]{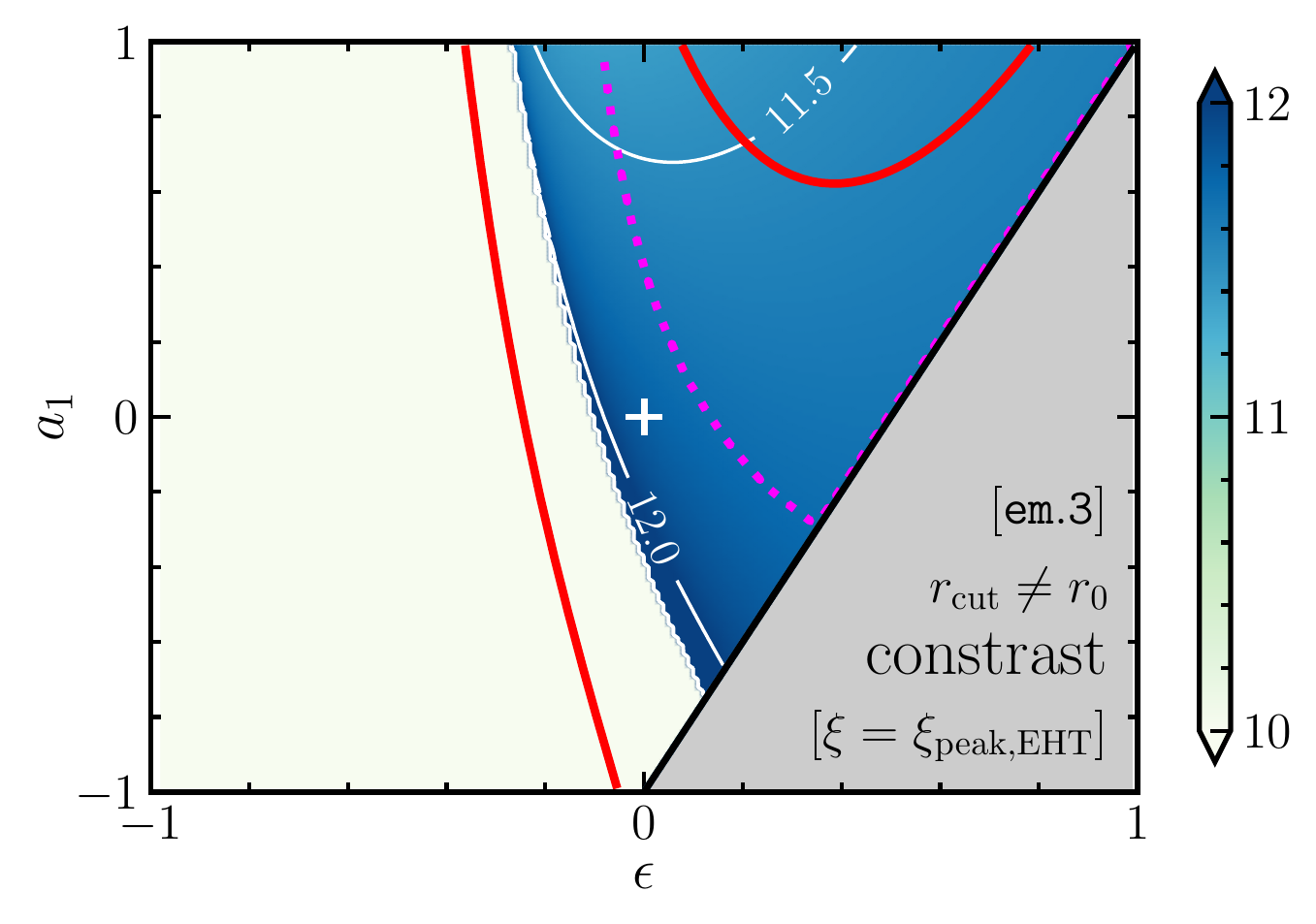}
\includegraphics[width=.68\columnwidth]{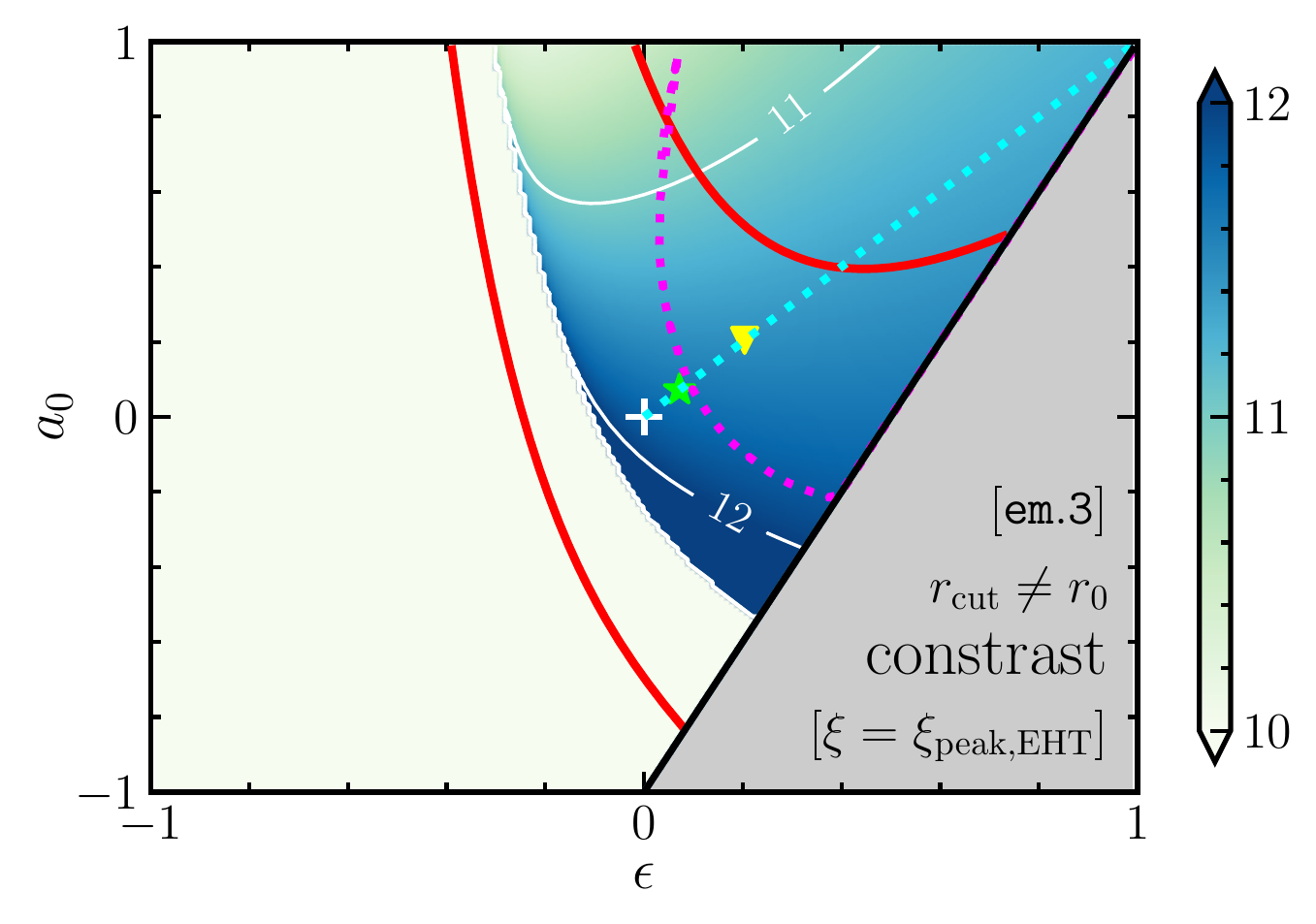}
\caption{The variation in the contrast in the BH image at $\xi =
  \xi_{{\text{peak, }}{\text{\tiny{EHT}}}} = 5.5M$ for the principal
  two-parameter RZ BH models used here. Here we impose an inner cut-off
  at $r_{\text{cut}} = r_{\text{tp}}(5.5M)$ so that one of the maxima in
  the contrast profile occurs at $\xi = 5.5 M$, i.e., at the ``correct''
  location, with the other occurring as usual at the shadow boundary $\xi
  = \xi_{\text{ps}}$. Imposing such a cut-off is only possible for BHs
  with $\xi_{\text{ps}} > 5.5M$ (i.e., not for the BHs in the light-green
  regions. 
  The magenta lines show the region in which the global maximum occurs at 
  $\xi = \xi_{\text{ps}}$ (corresponding to case B of
  Eq. \ref{eq:width_def_rcut}). Therefore, in this region, the quantity
  shown here $\mathcal{C}(5.5 M)$ only provides a lower bound for the
  peak contrast. The complementary region corresponds to the case when
  the peak occurs precisely at $\xi = 5.5 M$ (case C). From the top
  ($\mathtt{em.1}$) to the bottom row ($\mathtt{em.3}$) we have varied
  the emission coefficient prescription. The red lines denote, as usual,
  the EHT shadow-size bounds. The black cross locates the Schwarzschild
  BH and the cyan line represents the family of Reissner-Nordstr{\"o}m
  (RN) BHs. The star and triangle markers locate the RN BHs with
  specific charges of $\bar{q} = 0.50$ and $0.75$,
  respectively. }
\label{fig:RZ_eps_a0_a1_two_param_rcut}	
\end{figure*}

\section{Variation of Observed Ring Size with Metric}
\label{sec:Ring_Size_RZBHs}

In this section, we consider spherical accretion, as described in Sec.
\ref{sec:Spherical_Accretion}, onto Rezzolla-Zhidenko (RZ) BHs, and
analyse the variation in the observed image properties with RZ
parameters, which govern the deviation away from the Schwarzschild
metric. Since we have already touched on the impact of modifying the
emission model on the image properties, in Fig.
\ref{fig:RZ_eps_a0_a1_two_param}, we will fix the emission prescription
in this section to the one given in Eq. \ref{eq:Emission_Coeff}, and let
the emission zone extend all the way down to the horizon ($r_{\text{cut}}
= r_0$). In Fig. \ref{fig:RZ_eps_a0_a1_two_param_rcut}, we introduce an
inner cutoff radius in the emission region such that a peak in the
contrast profile occurs at the EHT-observed location, and also consider
how varying the emission coefficient impacts our results.

As we will discuss below, it is sufficient for our present purposes to
restrict our considerations here to the principal one-parameter and
two-parameter subfamilies of the three-parameter family
$\mathscr{M}(\epsilon; a_0; a_1)$ of RZ BHs, with $B^2 = -g_{tt}g_{rr} =
1$ and $a_2 = 0$, whose sole metric function can be written compactly as
(\citealt{Rezzolla2014}; see also \citealt{Konoplya2016a} and
\citealt{Kocherlakota2020} for the extensions to axisymmetric BH
spacetimes and to spherically-symmetric non-BH spacetimes respectively),
\begin{equation} \label{eq:N2_RZ_eps_a0_a1}
N^2(r) = 1 - \frac{2M}{r} + \frac{4a_0}{(1+\epsilon)^2}\frac{M^2}{r^2} 
+ \frac{8(\epsilon - a_0 + a_1)}{(1 + \epsilon)^3}\frac{M^3}{r^3} 
- \frac{16 a_1}{(1+\epsilon)^4}\frac{M^4}{r^4}\,, 
\end{equation}
where $M$ is the ADM mass of the BH spacetime. As in 
\cite{Rezzolla2014}, we will require that the outermost horizon 
(i.e., the largest zero of $g_{rr}^{-1} = N^2(r) = 0$) be located at 
$r = r_0$, where
\begin{equation}
  \label{eq:RZ_Horizon}
  r_0 := \frac{2M}{1+\epsilon}\,.
\end{equation}
Thus, $\epsilon > -1$ (alone) controls the size of the event horizon 
and $a_0$ measures the magnitude of violation of the solar system 
constraints ($|a_0| \lesssim 10^{-4}$; \citealt{Will:2006LRR}). These 
constraints are not applicable to theories of gravity which do not 
admit a Birkhoff-like uniqueness theorem, thus supporting the study of 
metrics with $a_0 \neq 0$. Furthermore, that no larger roots of $N^2 = 0$ 
should exist imposes non-trivial constraints on the 3D RZ parameter 
space. We tabulate these theoretical constraints for the principal axes 
and planes of this 3D space in Table 
\ref{table:RZ_Parameter_Space_Constraints_PPN}.

The $tt$-metric function above can be compared to its asymptotic
post-Newtonian (PN) expansion,
\begin{equation}
N^2 = 1 - \frac{2M}{r} + 2\sum_{n=1}^\infty \zeta_{n}\left(-\frac{M}{r}\right)^{n+1}\,,
\end{equation}
where $\zeta_1$ is typically expressed as $\zeta_1 := \beta - \gamma$,
with $\gamma$ the coefficient of the $\mathscr{O}(r^{-1})$-term in the
asymptotic expansion of $g_{rr}$. In particular, for this family of BH
metrics, the $tt$-PN expansion truncates at $\mathscr{O}(r^{-4})$, and it
is possible to find a map between the two sets of parameters as,
\begin{align}
  \label{eq:PN_Parameters}
  \zeta_1 &=\ \frac{2a_0}{(1 + \epsilon)^2}\,,\\
  \zeta_2 &=\ -\frac{4(\epsilon - a_0 + a_1)}{(1 + \epsilon)^3}\,,\\ 
  \zeta_3 &=\ -\frac{8a_1}{(1 + \epsilon)^4}\,. 
\end{align}
We indicate which PN parameters vanish for the various sections of the 
$\mathscr{M}(\epsilon; a_0; a_1)$ family we consider here in Table 
\ref{table:RZ_Parameter_Space_Constraints_PPN}. 

We also explicitly report the Ricci and Kretschmann curvature scalars, 
$\mathscr{R}$ and $\mathscr{K}$, for these RZ BHs to show that these 
spacetimes are regular everywhere except at the central singularity 
$r=0$,%
\begin{align}
\mathscr{R} =&\ 4 M^3 r^{-6}(\zeta_2 r - 3 M \zeta_3) \\
\mathscr{K} =&\ 16 M^2 r^{-12}\left[3r^6 - 12 M \zeta_1 r^5 + 2 M^2 
(7\zeta_1^2 + 10\zeta_2)r^4 \right. \nonumber \\
&\ \left. - 10 M^3 (5 \zeta_1 \zeta_2 + 3 \zeta_3)r^3
+ 2 M^4 (39 \zeta_1 \zeta_3 + 23 \zeta_2^2) r^2 \right. \nonumber \\
&\ \left. - 146 M^5 \zeta_2 \zeta_3 r + 117 M^6 \zeta_3^2\right] 
\end{align}
Considering the particular case of the $\mathscr{M}(\epsilon)-$RZ family,
it is easy to see that there is a bijective map between the metric
functions with $1/2 \leq \epsilon < \infty$ and those with $-1 < \epsilon
\leq 1/2$: the only metric-deviation parameter $\zeta_2(\epsilon)$ is
double-valued on $-1 < \epsilon < \infty$. Therefore, by setting the
``theoretical priors,'' $-1 < \epsilon \leq 1/2$, we consider (a) all
unique $\mathscr{M}(\epsilon)-$RZ BH solutions while ensuring also that
(b) $r_0(\epsilon) = 2M/(1+\epsilon)$ locates the event horizon. For
$\epsilon > 1/2$, $r_0(\epsilon)$ corresponds to the \textit{inner}
horizon with $r = \left(\epsilon + \sqrt{4\epsilon +
  \epsilon^2}\right)/(1 + \epsilon)$ then locating the outer event
horizon.  In other words, within the ranges in which the RZ parameters
are here chosen, the metric does not show any pathology.

Table \ref{table:RZ_Parameter_Space_Constraints_PPN} demonstrates how 
the theoretically-allowed range of a given parameter that characterises 
a particular structural deviation in the spacetime metric from 
Schwarzschild depends on the overall form of the metric itself. 
Notice, e.g., the striking case of the $\epsilon$ parameter: its 
theoretically-allowed domain changes character from being bounded when 
only $\epsilon$ is allowed to vary, to being unbounded otherwise. Thus, 
as a corollary, constraints from experiment on any given parameter 
(RZ or PN) will, in general, depend on the full form of the metric in 
use (see also \citealt{Volkel2020}). 
We demonstrate this below explicitly by considering both 
one-parameter and two-parameter subfamilies of the RZ metric. This does 
not \textit{a priori} rule out the existence of an observation-specific 
set of parameters or a combination thereof on which ``robust''
(independent of the overall form of the metric; even if only as a good 
approximation) constraints may be set. In the absence of knowledge of 
such a combination, it can be useful to report constraints on the 
single-parameter RZ BH metrics, as done e.g. in \cite{Abbott2021}, 
where the PN formalism was used instead. Also, since we will report our 
findings entirely in terms of RZ parameters, we note that the 
non-linear map between the two sets of coefficients implies translating 
observational constraints from the RZ- to the PN-space can yield 
drastically differing inferences, especially when $\epsilon \gtrsim -1$.

We further limit the range of our demonstrative exploration of images of 
RZ BHs to the region of ``small'' parametric deviations%
\footnote{Notice that small-$\epsilon$ deviations do \textit{not} 
correspond to small changes in the spacetime as characterized, e.g., 
by the horizon-size or the curvature invariants due to their highly 
non-linear dependence on $\epsilon$ (see also, e.g., 
\citealt{Suvorov2020}).}, %
$\Delta < 1$, where $\Delta := \max{\left(|\epsilon|, |a_0|, |a_1|
\right)}$ is the infinity norm on the space and measures the magnitude 
of the parametric deviation of an RZ BH from the Schwarzschild 
spacetime \citep{Kocherlakota2020}. This allows us to focus on BH 
spacetimes that are relatively similar to the Schwarzschild spacetime 
since, e.g., we find that there exist parameter regions for $\Delta > 1$ 
for which the (quartic) CNG equation \eqref{eq:CNG_Eq} admits multiple 
roots $r>r_0$ (cf. \citealt{Gan2021}). We relegate a systematic 
exploration of these regions to future work. 

In addition to the EHT measurements of $\xi_{{\text{peak, }}{\text{\tiny{EHT}}}}$ and
$C_{\text{peak, EHT-min}}$ that we have used above, since changes in the
spacetime metric can cause changes in the observed size of the shadow, we
will also compare how the EHT inference of the shadow-size of
M87$^*$, $3\sqrt{3}(1-0.17)M \lesssim r_{\text{sh}} :=
\xi_{\text{ps}} \lesssim 3\sqrt{3}(1+0.17)M$ \citep{EHT_M87_PaperVI,
  Kocherlakota2021}, constrains the RZ parameter space. We also comment
on the consistency between these two sets of bounds when using non-Kerr
BHs, whilst remaining within the modest ambit of the toy
accretion-emission models in use here.

We show the variation in the sizes of the horizon, photon sphere, and
shadow for the single-parameter models in the top row of Fig.
\ref{fig:RZ_eps_a0_a1_one_param}. For small-deviations from the
Schwarzschild spacetime, the variation in the shadow radius
$\xi_{\text{ps}}$ is approximately linear with respect to the relevant
metric parameter for the $\mathscr{M}(a_0)$ and $\mathscr{M}(a_1)$ BH
models, i.e., when the horizon size is held fixed to its Schwarzschild
value.  The variation in the photon sphere size is also approximately
linear with respect to $a_1$ for $\mathscr{M}(a_1)$ BHs since $a_1$
modifies the metric at a higher order in $r^{-1}$. The horizontal white
bands indicate the allowed range of these parameters, for the
single-parameter metrics, as inferred from the shadow-size bounds
discussed above. We show the corresponding contrast profiles for these
single-parameter BH metrics in the second row, when the emission region
extends all the way down to the horizon. As in
Fig. \ref{fig:Schwarzschild_Emissivity}, we show $\xi_{{\text{peak,
  }}{\text{\tiny{EHT}}}}$ and $\mathcal{C}_{\text{peak, EHT-min}}$ as
vertical orange and horizontal cyan lines respectively. It is clear then
that BHs that have shadow-sizes $\xi_{\text{ps}} \gnsim 5.5 M$ would in
general appear to have bright rings larger than the one measured by the
EHT, $\xi_{\text{peak}} \gnsim \xi_{{\text{peak,
  }}{\text{\tiny{EHT}}}}$. On the other hand, BHs with smaller
shadow-sizes could satisfy $\xi_{\text{peak}} = \xi_{{\text{peak,
  }}{\text{\tiny{EHT}}}}$ if the emission regions in their spacetimes did
not extend all the way down to the horizon but were instead cut-off at
some radius outside their respective photon spheres for some reason. In
this case however, as we have seen above in Sec.
\ref{sec:Schw_Emissivity_Results}, the maximum contrast in the image of
such BH would be reduced, from infinity to some finite value, when
$r_{\text{cut}} > r_{\text{ps}}$. To this end, we consider RZ BHs with
$\xi < \xi_{{\text{peak, }}{\text{\tiny{EHT}}}}$ and show the contrast
profiles when $r_{\text{cut}} = r_{\text{tp}}(\xi_{{\text{peak,
  }}{\text{\tiny{EHT}}}})$ in the bottom row of
Fig. \ref{fig:RZ_eps_a0_a1_one_param}. We see that since the
leading-order change in the $tt$-metric function introduced by the
$\epsilon$ or the $a_1$ parameter is at $\mathscr{O}(r^{-3})$, the impact
of small-variations ($\Delta < 1$) in these parameters on the value of
the peak contrast is relatively smaller than in the case of
small-variations in the $\mathscr{M}(a_0)$ model, where the metric is
modified already at $\mathscr{O}(r^{-2})$. We also show the impact of
varying the emission model, and demonstrate that it should be possible to
constrain the ``astrophysics,'' i.e., the exponent of the emission
coefficient, and/or the spacetime metric, with the help of additional
observables such as the maximum contrast in the image. Thus, bounds on
the RZ parameter from both sides are to be generically anticipated.
Finally, we note that if the shadow-size of a BH is ``sufficiently
smaller'' than the EHT-observed location of the contrast peak, i.e.,
$\xi_{\text{ps}} \ll \xi_{\text{peak-EHT}}$ and/or the emission
coefficient is ``sufficiently soft'', (i.e., decaying more slowly with
distance) , then the location of the absolute maximum in the contrast
profile shifts back from $\xi = \xi_{\text{tp}}(r_{\text{cut}})$ to the
shadow boundary $\xi = \xi_{\text{ps}}$. We discuss these aspects in more
detail in Sec.  \ref{sec:Fractional_Width}. This indicates that the
two-sided bounds discussed above will generically exist even when the
restriction on the size of the deviations $\Delta < 1$ imposed here is
lifted.

We shift our focus now to the question of whether multi-parameter
parametrized metrics can also be meaningfully constrained when using the
same set of observables. For our present demonstrative purposes, we use
here the two-parameter RZ models to explore the change in the bounds
inferred on the individual parameters when changing the overall structure
of the metric. First, with a cartoon, we visualise the impact, on the
parameter space of the $\mathscr{M}(\epsilon, a_0)$ RZ BHs, of the EHT
shadow-size constraints (red bands) and the reported minimum bound of the
peak-contrast in the image (blue bands) in Fig. \ref{fig:cartoon}. The
red lines (solid and dashed) represent shadow-size isocontours whereas
the blue lines show two sample peak-contrast isocontours, when the peak
occurs at $\xi = 5.5M$. If in the future we are able to constrain the
shadow-sizes of supermassive compact objects such as M87$^*$ or Sgr~A$^*$
with increasing precision, the width of the red band would shrink and
thus constrain the RZ parameter space more severely. A similar statement
holds for the blue band if the flux-sensitivity of the EHT can be
increased. As will become evident from the figures below, while these
bands overlap significantly, these constraints are not in general
degenerate. In the limit of infinite precision, picking out a single
shadow radius yields at best a family of two-parameter RZ BHs (solid- or
dashed-red lines in this figure), while a precise peak contrast
measurement would pick out a different family of RZ BHs (solid- or
dashed-blue lines) in general. Thus, with two observables, it should be
possible to uniquely determine the underlying metric parameters of a 2D
family of BHs.\footnote{We note that this is not always the case: there
are regions of the parameter space, e.g., close to the extremal limit,
where the contour a families of BHs can have precisely the same shadow
radius and peak contrast values. In these cases, additional measurements,
such as those corresponding to other contrast values, would be needed to
break the degeneracy.} Interestingly, as can be seen from this figure, it
is possible for two BHs with differing horizon sizes to have identical
shadow-sizes (filled circle and empty circle, both lying on the same
red-line).

More concretely, the variation in the shadow-sizes, as well in the value
of the contrast in the image at $\xi = \xi_{{\text{peak, }}{\text{\tiny{EHT}}}}$, when the
emission zone extends all the way to the horizon, with the relevant RZ
parameters for the 2D RZ BHs is reported in Fig.
\ref{fig:RZ_eps_a0_a1_two_param}. The red lines in all figures bound the
region in which the EHT shadow-size bounds are respected. As above, the
(blue) contours $\mathcal{C}(\xi = \xi_{{\text{peak, }}{\text{\tiny{EHT}}}}) = 10$ in
Fig. \ref{fig:RZ_eps_a0_a1_two_param} represent the bounds
obtained from the requirement that $\mathcal{C} (\xi_{{\text{peak, }}{\text{\tiny{EHT}}}})
> \mathcal{C}_{\text{peak, EHT-min}}$. Thus, the first takeaway is that BHs 
with shadow-sizes larger than $\xi_{{\text{peak, }}{\text{\tiny{EHT}}}}$ fail to 
meet the contrast constraint, whereas those that have significantly smaller 
shadow-sizes are also expected to fail, for the reasons discussed above. 
Notice however that since we do not cut off the emission outside the
horizon, the peak in the contrast profile must necessarily occur at $\xi
= \xi_{\text{ps}}$. Therefore, if $\xi_{\text{ps}} \neq 
\xi_{{\text{peak, }}{\text{\tiny{EHT}}}}$, 
the contrast peak in the image does not occur at the EHT-observed
location. 

Furthermore, the observed lower bound on the peak contrast imposes no
additional constraints on the RZ parameter space in this scenario
($r_{\text{cut}} = r_0$), and no insight into the emission coefficient is
possible (see Sec.~\ref{sec:Fractional_Width}).  $\alpha$, which we have
used to locate the observed peak in the contrast, is of finite precision
(limited in part by the angular resolution of the EHT array in 2017), we
expect a small spread in this ``allowed'' region around $\xi_{\text{ps}}
= \xi_{{\text{peak, }}{\text{\tiny{EHT}}}}$. The consistency of the two
bands of constraints -- location of the contrast-peak and shadow-size --
demonstrates a null-test of the calibration procedure employed by the EHT
to obtain the latter from the former. While the test here is limited due
to the restriction to spherical symmetry and a fixed BH ADM mass, our
approach involving analytical models is very efficient as a path-finding
step for further exploration of the parameter space with numerical
simulations.  We explore the impact of using the observed ring width,
another useful observable that is sensitive to the gradient of the
emission coefficient, in the following section for single-parameter BH
models.

Finally, we note that the EHT-allowed range of a particular
parameter depends on the underlying metric. We see that while for the
$\mathscr{M}(\epsilon)$ BH models, we find $-0.22 \lesssim \epsilon \leq
0.5$ to be the EHT-allowed range, it is immediately evident from
Fig. \ref{fig:RZ_eps_a0_a1_two_param}, that the allowed range for
$\epsilon$ is quite significantly different for the
$\mathscr{M}(\epsilon; a_1)$ and $\mathscr{M}(\epsilon; a_0)$ models.

\begin{figure*}
\includegraphics[width=\columnwidth]{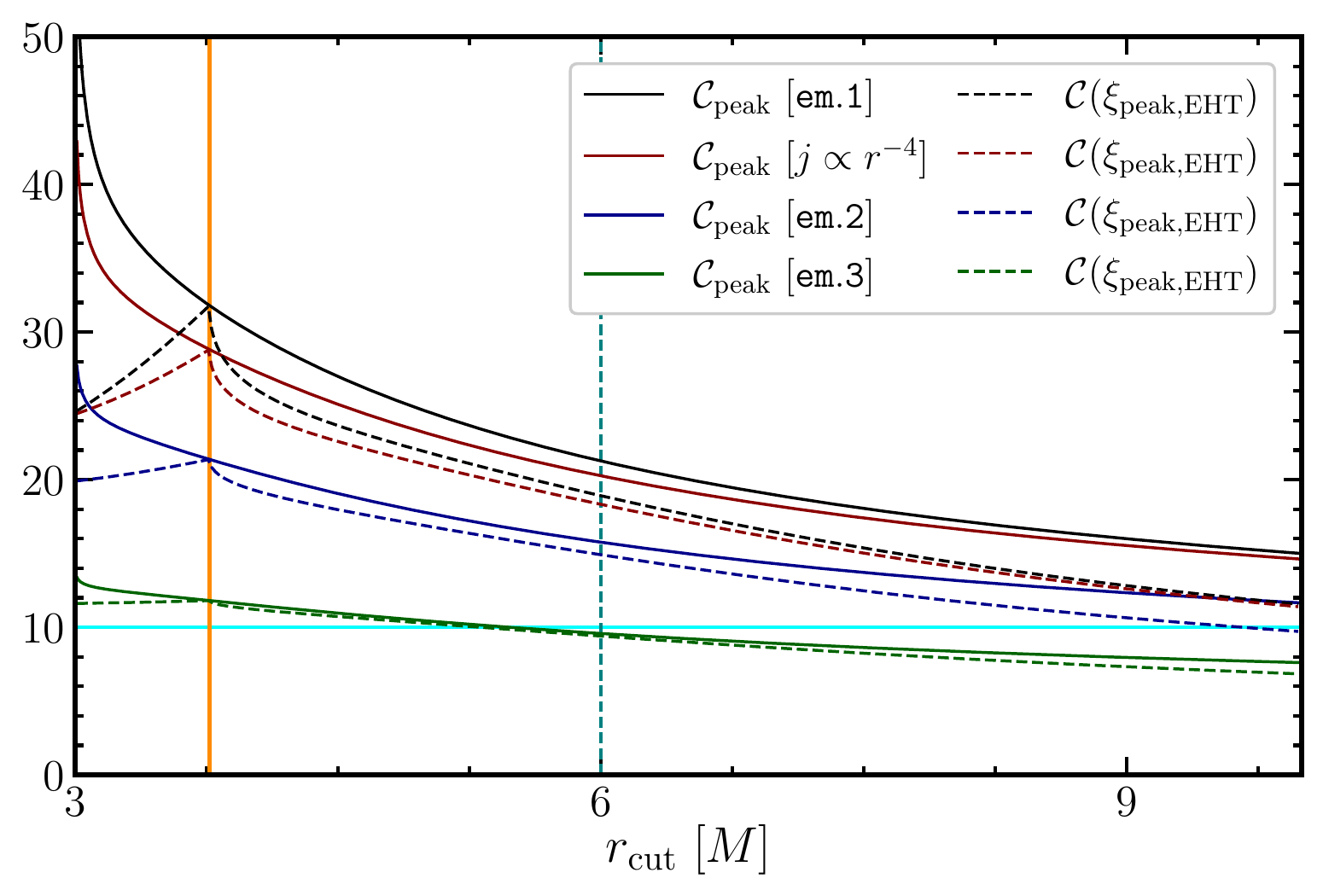}
\hspace{0.2cm}
\includegraphics[width=\columnwidth]{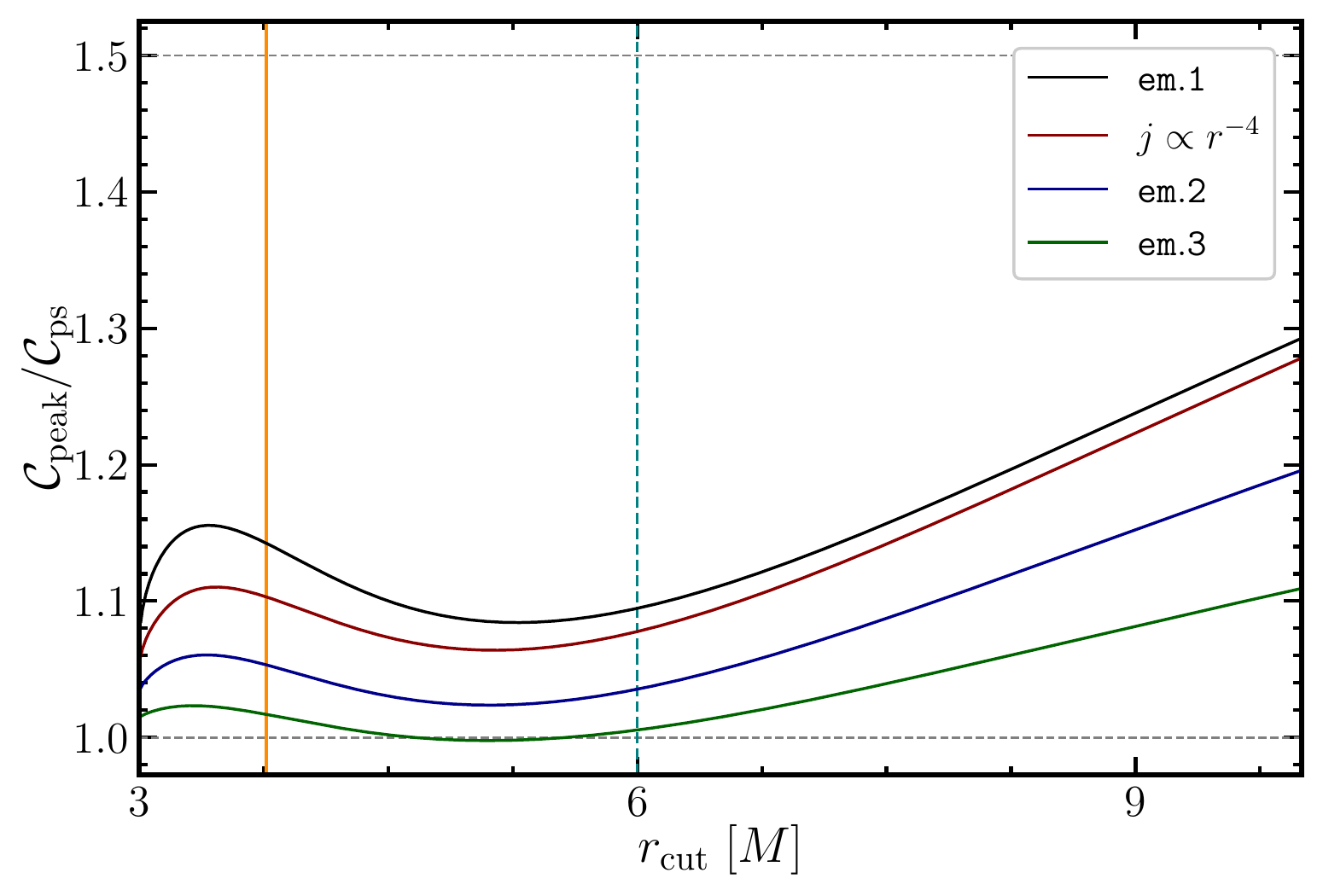}
\caption{In the left panel, we show the variation in the peak value of the
  contrast in the image of a Schwarzschild BH with varying
  $r_{\text{cut}}$. In the right panel, the relative difference in the
  peak contrast and the contrast at the shadow boundary is displayed.
  The vertical orange and green lines correspond to $r =
  r_{\text{tp}}(\xi_{{\text{peak, }}{\text{\tiny{EHT}}}}) \approx 3.77 M$ and $r = 6 M$,
  respectively.}
\label{fig:Schw_Images_Contrast_rcut}
\end{figure*}

We show in Fig. \ref{fig:RZ_eps_a0_a1_two_param_rcut} the contrast at
$\xi_{{\text{peak, }}{\text{\tiny{EHT}}}}$ when one of the contrast maxima also
occurs at this location, i.e., for $r_{\text{cut}} = r_{\text{tp}}(
\xi_{{\text{peak, }}{\text{\tiny{EHT}}}})$. To explore the impact of varying the emission
prescription, we show in the top and bottom rows the two emission models
$\mathtt{em.1}$ and $\mathtt{em.3}$ respectively. As expected, on
introducing this cutoff, for BHs with $\xi_{\text{ps}} < 
\xi_{{\text{peak, }}{\text{\tiny{EHT}}}}$, the value of the contrast at 
$\xi_{{\text{peak, }}{\text{\tiny{EHT}}}}$ increases (compare the bottom row 
of Fig. \ref{fig:RZ_eps_a0_a1_two_param} with the top row of Fig.
\ref{fig:RZ_eps_a0_a1_two_param_rcut}), thus making weak the bounds
imposed by the lower bound of the peak-contrast obtained by the EHT. At
the same time, unsurprisingly, for the ``softer'' (fall-off) emission
prescription, we find tighter constraints on the RZ parameter space from
the lower bound on the peak-contrast. Thus, even in the most conservative
case ($\mathtt{em.1}$), while the contrast-constraints do not appear in
the $\Delta < 1$ region of the parameter space, it is evident that with
increasingly better observations, metric tests will only get better and
increasingly more robust.

The figure also hints at the possibility that there
exist regions of total degeneracy, where two-parameter families of BHs
may have the same shadow-size and approximately the same $\mathcal{C}
(\xi_{{\text{peak, }}{\text{\tiny{EHT}}}})$ values (and possibly the same contrast
profiles), e.g., in the top right regions in the $\epsilon\!-\!a_0$ and
$\epsilon\!-\!a_1$ parameter spaces. This degeneracy could in principle be
broken by introducing new observables such as the fractional-width of the
ring in the observed image (see the discussion in Sec.
\ref{sec:Fractional_Width}) or the locations of the photon sub-ring and
the Lyapunov exponent, the latter of which we explore in upcoming work
\citep{Kocherlakota2022}.

\section{Fractional Width of the Observed Ring as a Discriminator of 
Emission Model}
\label{sec:Fractional_Width}

In this section we consider the definition of the width of the 
observed circular ring, and explore its use in obtaining insights into local 
emission physics.

For a Schwarzschild BH, we show in the left panel of
Fig. \ref{fig:Schw_Images_Contrast_rcut} the variation in the peak
contrast with changing emission region cut-off location
$r_{\text{cut}}$. This figure demonstrates that independently of the
emission model, requiring a large cut-off radius, to push the peak in the
observed image to larger sizes in order to match the EHT reported value
of $\xi_{{\text{peak, }}{\text{\tiny{EHT}}}} = 5.5 M$ can be
effective. However, for significantly larger $r_{\text{cut}}$, the
peak-contrast value falls below the EHT reported value of
$\mathcal{C}_{\text{peak, EHT-min}} = 10$. In the right panel, we show
the fractional difference between the peak contrast and the contrast at
the photon sphere for different $r_{\text{cut}}$, to measure the drop in
the contrast from the observed peak to the shadow boundary. We note in
particular that $\mathcal{C}_{\text{ps}} > 0.5
\mathcal{C}_{\text{peak}}$, at least for $3 < r_{\text{cut}}/M < 9$.
Thus, a truly sharp drop in the contrast occurs always close to the
shadow boundary, $\xi = \xi_{\text{ps}}$. The non-monotonic behaviour of
the curves shown here appears to stem from the non-monotonic behaviour of
the redshift function for photons emitted radially-inwards, $\gamma_-$
(see the left panel of Fig. \ref{fig:Schw_Redshifts}).

We denote by $w$ the width of the observed ring defined, within the
context of the spherical models we use here, using the ``half-peak''
location(s), $\xi^\pm_{\text{hp}} :=
\xi^\pm(\mathcal{C}_{\text{peak}}/2)$, where the contrast becomes half of
the peak contrast value, with the superscripts $\pm$ denoting that
$\xi^+_{\text{hp}} > \xi_{\text{peak}}$ and vice-versa.  For the case
when $r_{\text{cut}} = r_0$, since the peak contrast value necessarily
occurs at the shadow boundary and is mathematically infinite,
$\mathcal{C}_{\text{ps}} := \mathcal{C}(\xi = \xi_{\text{ps}})
\rightarrow \infty$, we use $w = 0$. When $\xi_{\text{ps}} <
\xi_{{\text{peak, }}{\text{\tiny{EHT}}}}$, introducing an ad hoc interior
cut-off radius for the emission region becomes possible. As discussed
above, the only way to find the location of the bright emission-ring to
correspond to the one measured in the M87* image in the present set up is
to set $r_{\text{cut}} = r_{\text{tp}}(\xi_{{\text{peak,
  }}{\text{\tiny{EHT}}}})$.  The contrast values of the two (finite)
maxima in this case, i.e., $\mathcal{C}_{\text{ps}}$ and
$\mathcal{C}_{{\text{peak, }}{\text{\tiny{EHT}}}} :=
\mathcal{C}(\xi_{{\text{peak, }}{\text{\tiny{EHT}}}})$, outline four
possibilities
\begin{equation}
  \label{eq:width_def_rcut}
w =
\left\{
\begin{alignedat}{3}
&  \xi^+_{\text{hp}} - \xi_{\text{ps}}\,,\ \  
&& \mathcal{C}_{\text{ps}} > 2\mathcal{C}_{{\text{peak, }}{\text{\tiny{EHT}}}}
&& \quad \quad {\rm [case\ A]} \\
&  \xi^+_{\text{hp}} - \xi_{\text{ps}}\,,\ \  
&& 2\mathcal{C}_{{\text{peak, }}{\text{\tiny{EHT}}}} \geq \mathcal{C}_{\text{ps}} > \mathcal{C}_{{\text{peak, }}{\text{\tiny{EHT}}}}
&& \quad \quad {\rm [case\ B]} \\
&  \xi^+_{\text{hp}} - \xi_{\text{ps}} \,,\ \ 
&&  \mathcal{C}_{{\text{peak, }}{\text{\tiny{EHT}}}} \geq \mathcal{C}_{\text{ps}} \geq \mathcal{C}_{{\text{peak, }}{\text{\tiny{EHT}}}}/2
&& \quad \quad {\rm [case\ C]} \\
&  \xi^+_{\text{hp}}  - \xi^-_{\text{hp}} \,,\ \ 
&& \mathcal{C}_{{\text{peak, }}{\text{\tiny{EHT}}}}/2 > \mathcal{C}_{\text{ps}}
&& \quad \quad {\rm [case\ D]} \,,
\end{alignedat}
\right.
\end{equation}
with $\xi_{\text{peak}} = \xi_{\text{ps}}$ in cases A and B and 
$\xi_{\text{peak}} = \xi_{{\text{peak, }}{\text{\tiny{EHT}}}}$ otherwise.

In the first two cases, the global maximum of the contrast profile occurs
at the shadow boundary even when an inner cut-off radius for the emission
is invoked. However, in these cases, the width of the ring $w$ is clearly
non-zero.  Furthermore, in case A, $\xi = \xi_{{\text{peak,
  }}{\text{\tiny{EHT}}}}$ lies outside the ring width. In case C, the
peak of emission occurs at $\xi = \xi_{{\text{peak,
  }}{\text{\tiny{EHT}}}}$ with the shadow boundary forming the inner edge
of the emission-ring. Finally, case D represents a scenario where the
emission ring would appear to be disjoint from the shadow boundary. For
all of the metric deviations and all of the emission models we consider
here, we find that cases A and D are never realised. We direct the reader
to see Fig. \ref{fig:RZ_eps_a0_a1_two_param_rcut} where the only
scenarios that appear to be possible, when $\xi_{\text{ps}} <
\xi_{{\text{peak, }}{\text{\tiny{EHT}}}}$ and $r_{\text{cut}} =
r_{\text{tp}}( \xi_{{\text{peak, }}{\text{\tiny{EHT}}}})$, are case B
(demarcated by the blue lines) and case C. Thus, the inner edge of the
observed emission ring is always bounded by the shadow boundary
independently of whether or not there is an inner cut-off to the region
of emission. This is a central result of the current work which has
recently been subject to debate (see, e.g., \citealt{Gralla2021}).

\begin{figure}
\includegraphics[width=\columnwidth]
{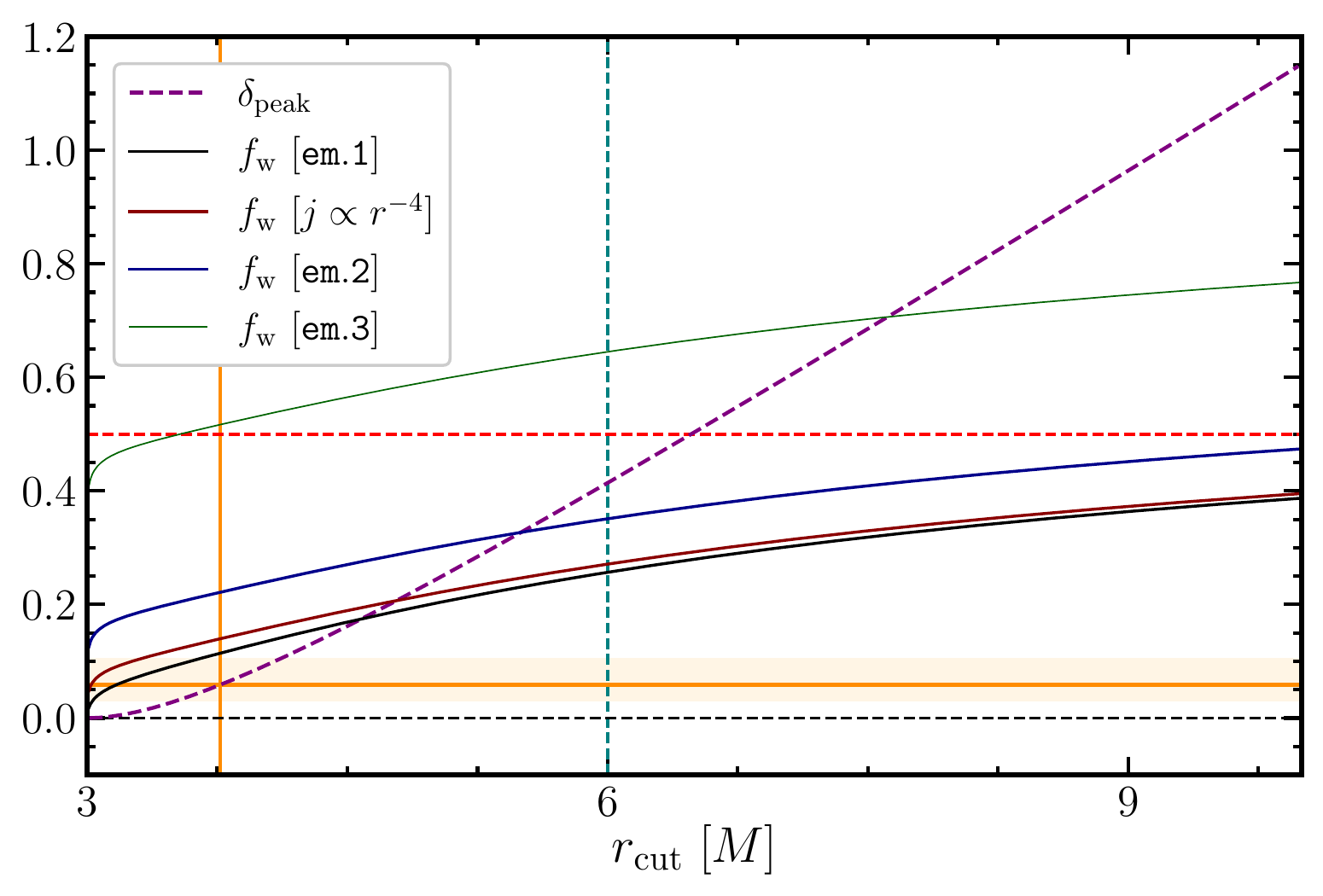}
\hspace{0.2cm}
\caption{We show here the variation, for the Schwarzschild BH, in the 
location of the peak in the contrast profile relative to the shadow-size, 
$\delta_{\text{peak}} := \xi_{\text{peak}}/\xi_{\text{ps}} - 1$, when the 
emission region has an artificial interior cutoff at $r = r_{\text{cut}}$. 
We also display the change in the fractional-width of the observed ring 
in the image, $f_{\text{w}} := w/2\xi_{\text{peak}}$ with $r_{\text{cut}}$. 
The fractional-width of the ring could be a useful observable to restrict 
the emission coefficient $j$.  The vertical orange and green lines have 
the same meaning as in Fig. \ref{fig:Schw_Images_Contrast_rcut}. The 
dashed-red line shows the EHT bound, $f_{\text{w}} \leq 0.5$.}
\label{fig:Schw_Ring_Characteristics_rcut}
\end{figure}

\begin{figure}
\includegraphics[width=\columnwidth]
{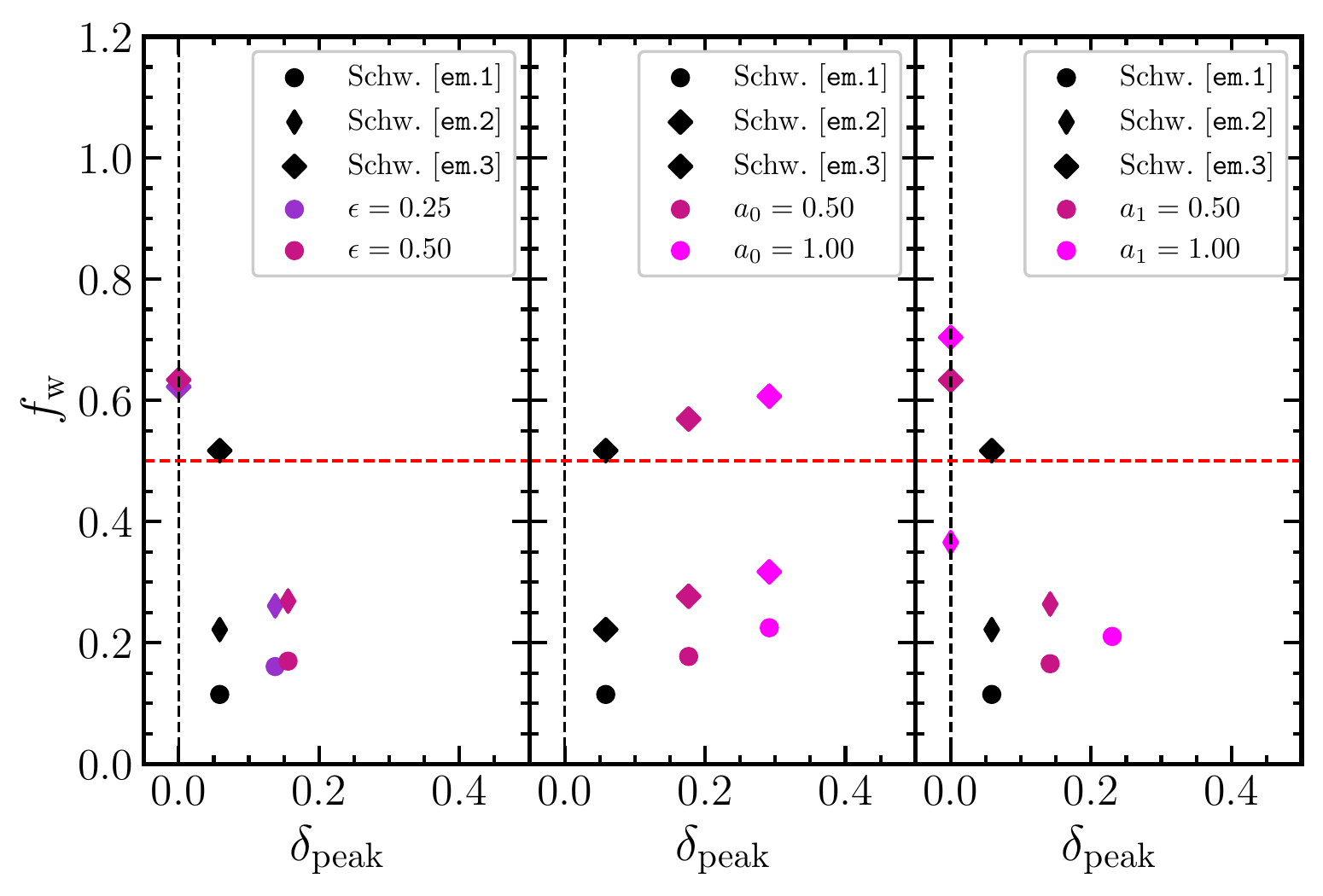}
\caption{We show here the variation, for the single-parameter RZ BHs, in 
$\delta_{\text{peak}}$ and $f_{\text{w}}$ with the relevant deviation 
parameter, for $r_{\text{cut}} = r_{\text{tp}}(\xi_{{\text{peak, }}{\text{\tiny{EHT}}}})$. 
We see that the soft emission prescription $j = k_2 r^{-2}~[\mathtt{em.3}]$ 
typically fails the $f_{\text{w}} \lesssim 0.5$ bound [dashed-red line], 
demonstrating how 
this observable can be used to constrain emission physics. The clear 
correlation with the metric deviation parameter also indicates how 
tests of gravity may be possible with this observable. See also the bottom 
row of Fig. \ref{fig:RZ_eps_a0_a1_one_param}}
\label{fig:RZ_eps_a0_a1_one_param_delpeak_fw}
\end{figure}

\begin{figure*}
\includegraphics[width=\columnwidth]
{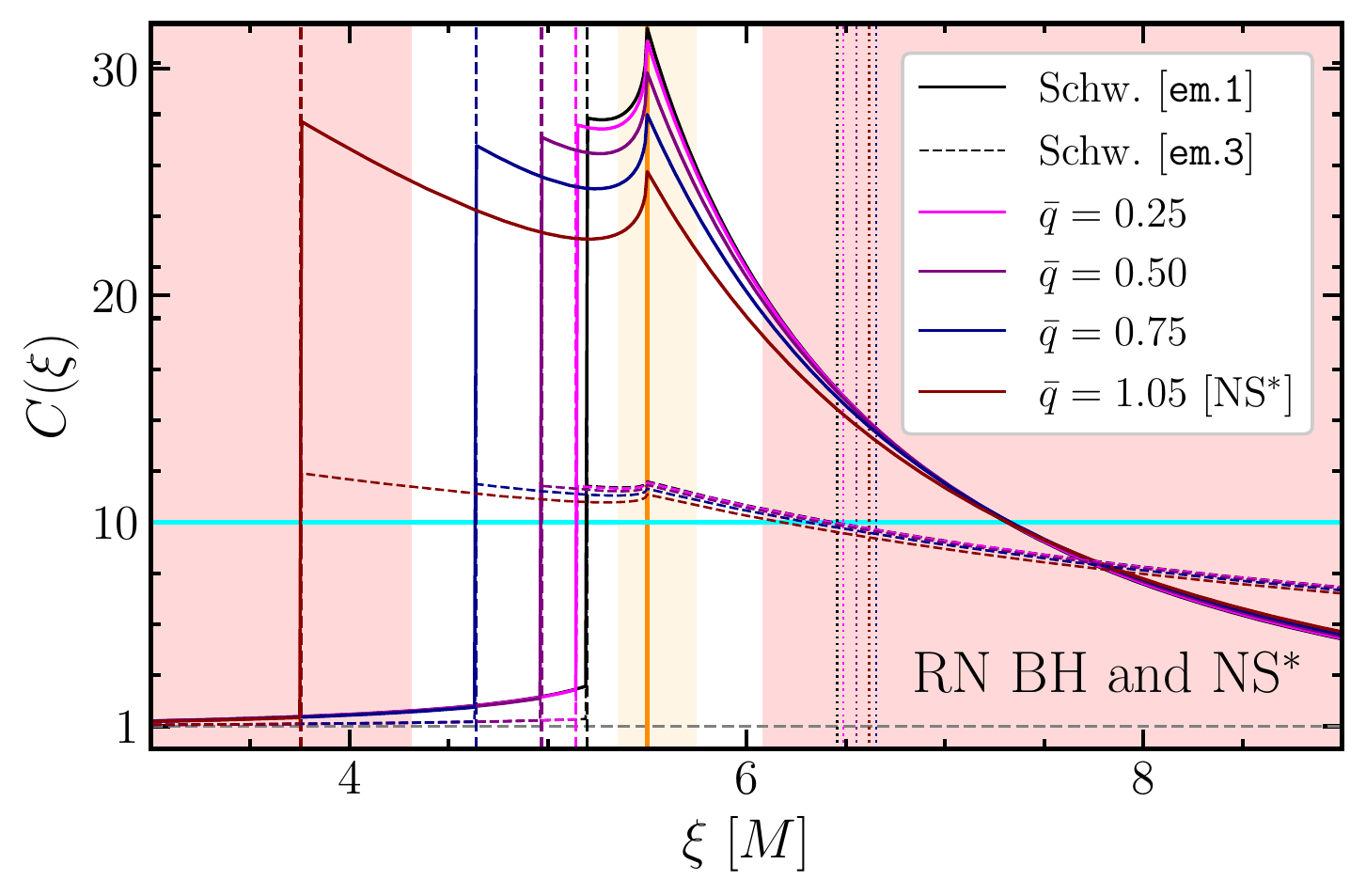}
\hspace{0.2cm}
\includegraphics[width=\columnwidth]
{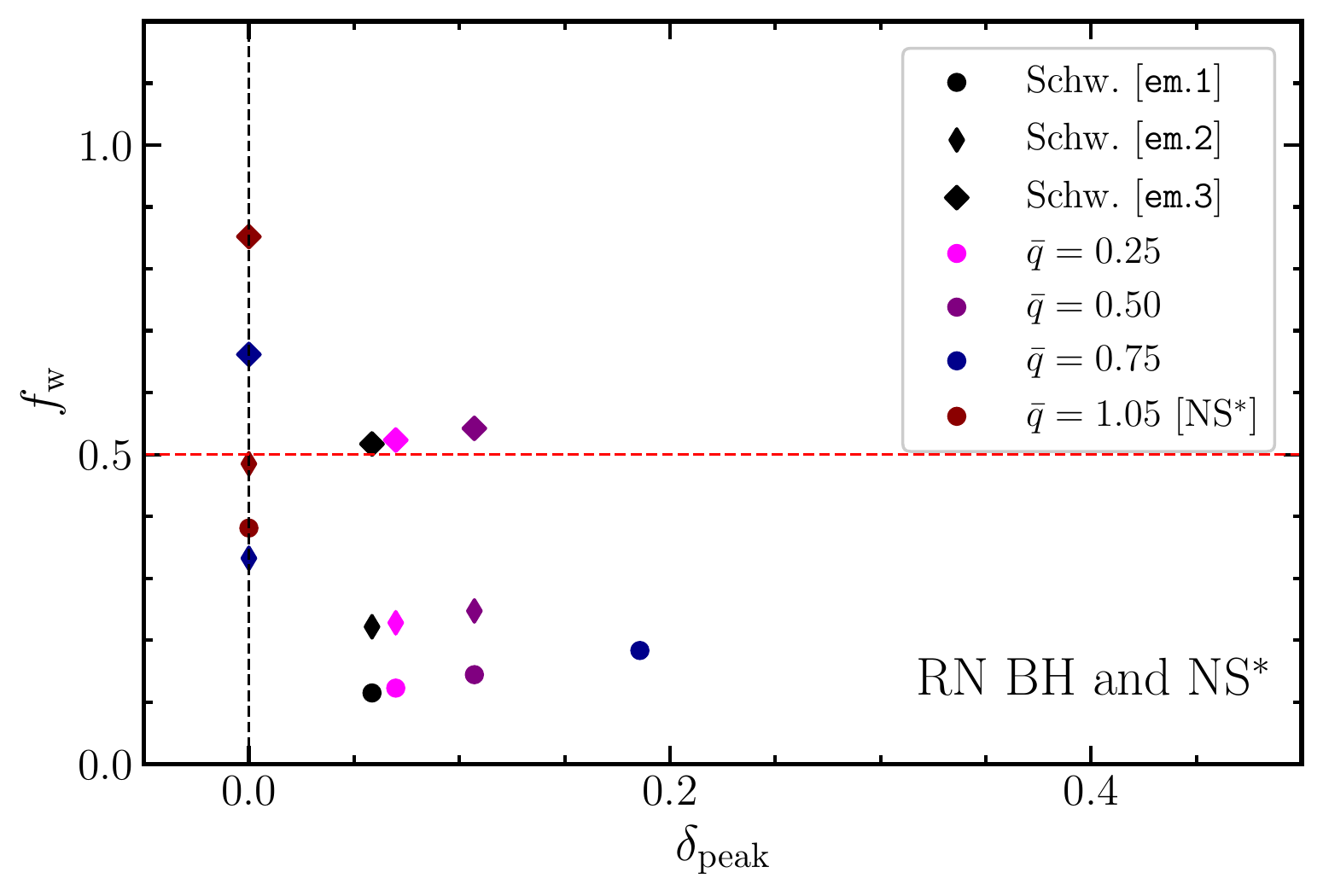}
\includegraphics[width=\columnwidth]
{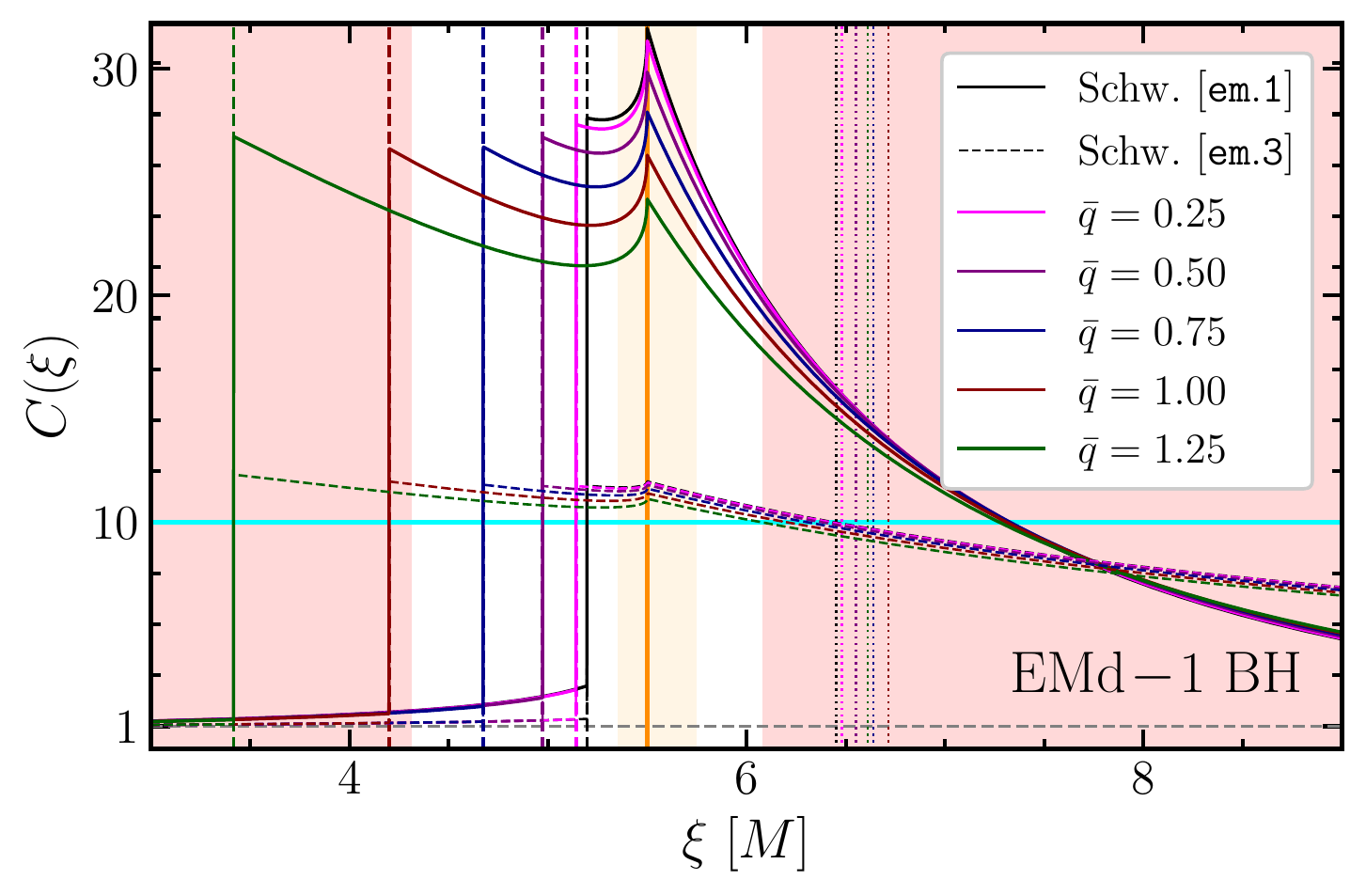}
\hspace{0.2cm}
\includegraphics[width=\columnwidth]
{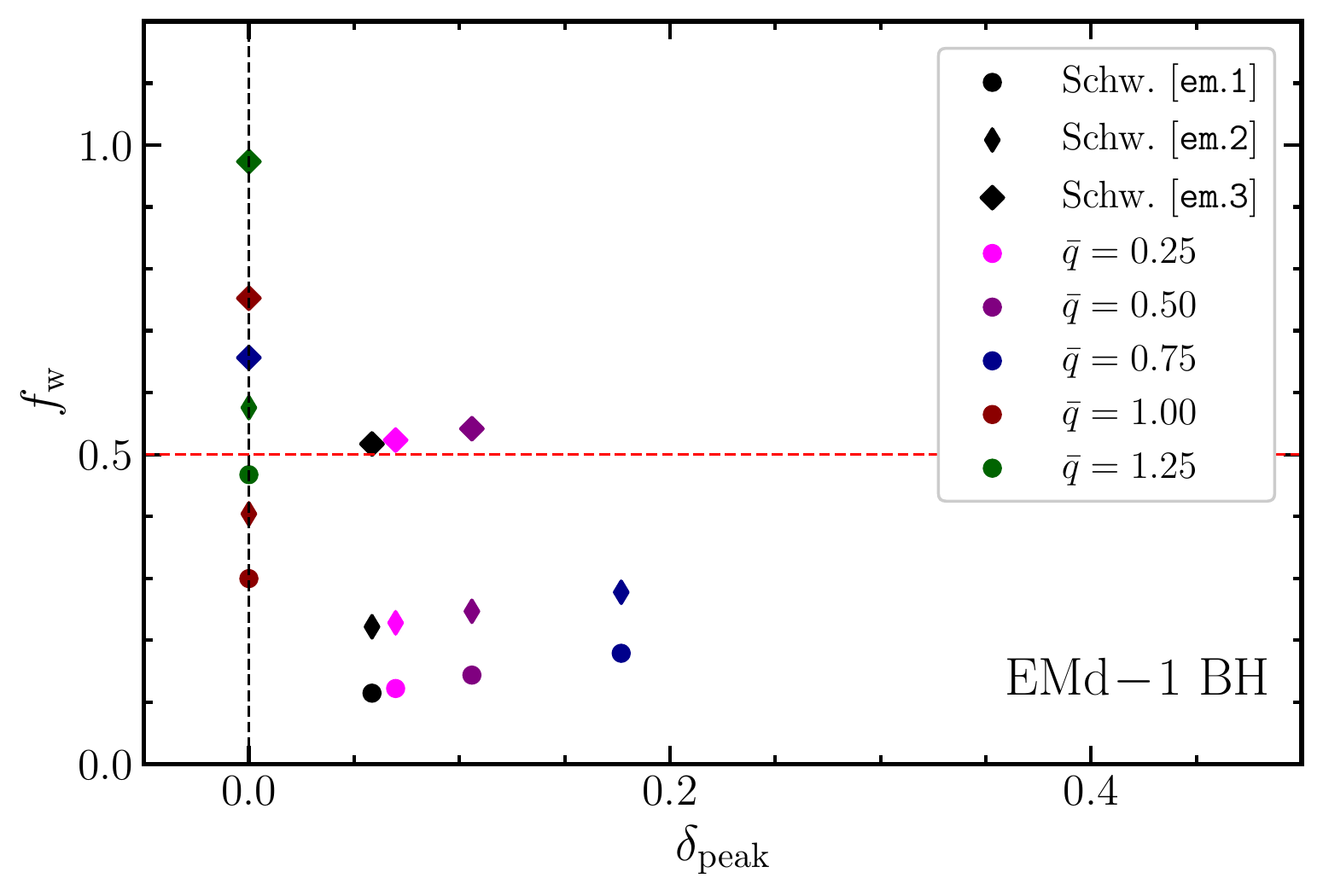}
\caption{In the left panels, we display the variation in 
the observed contrast $\mathcal{C}$ for a series of Reissner-Nordstr{\"o}m 
(RN) and Einstein-Maxwell-dilaton-1 (EMd-1) BHs, with increasing specific 
charges, $\bar{q} = q/M$. We also discuss the case of a single RN naked 
singularity (NS$^*$) The vertical dashed correspond to the size of the 
shadow cast by the compact object, whereas the vertical dotted lines 
denote the half-peak location for the emission model $[\mathtt{em.1}]$. 
The remaining lines and shaded-regions retain their meaning from Fig. 
\ref{fig:Schw_Images_Contrast_rcut} and Fig. 
\ref{fig:RZ_eps_a0_a1_one_param}. In the right panels, we show the 
corresponding fractional deviation of the contrast peak relative to the 
shadow-size $\delta_{\text{peak}}$ and the fractional width of the ring 
relative to the shadow diameter $f_{\text{w}}$ (compare with left panels). 
This figure demonstrates how the maximum observed contrast in the image is 
sensitive to $j$, whereas the width of the ring can potentially be used to 
test gravity. The dashed-red line shows the EHT reported bound, 
$f_{\text{w}} \leq 0.5$, for comparison.}
\label{fig:RN_EMd1_delpeak_fw}
\end{figure*} 

In Fig. \ref{fig:Schw_Ring_Characteristics_rcut}, for the Schwarzschild BH, 
we show the fractional-deviation of the location of the observed peak from 
the shadow boundary $\delta_{\text{peak}} := \xi_{\text{peak}}/
\xi_{\text{ps}} - 1$ and the fractional-width of the observed ring relative 
to the observed ring diameter $f_{\text{w}} := w/2\xi_{\text{peak}}$ (see, 
e.g., Sec. 7.2 of \citealt{EHT_M87_PaperVI}). The EHT reports a maximum 
value of $f_{\text{w}} \lesssim 0.5$. It is interesting to note that the 
range of $r_{\text{cut}}$ for which the observed contrast peak appears at 
the location seen by the EHT also satisfies the fractional-width 
constraint, despite the underlying accretion-emission model being highly 
simplistic, only for sufficiently steep (large-$n$) emission coefficients. 
Thus, the fractional-width of the ring can be yet another promising 
discriminator of emission physics. 

In Fig. \ref{fig:RZ_eps_a0_a1_one_param_delpeak_fw}, we show the variation 
of $\delta_{\text{peak}}$ and $f_{\text{w}}$ for the various one-parameter 
RZ BH models, with varying emission coefficient and spacetime metric. 
Here we use $r_{\text{cut}} = r_{\text{tp}}(\xi_{{\text{peak, }}{\text{\tiny{EHT}}}})$ to 
force the peak in the observed contrast to appear at the EHT observed 
location, $\xi = \xi_{{\text{peak, }}{\text{\tiny{EHT}}}}$. It becomes clear from this figure 
that this is not always possible for BHs with sufficiently small shadow 
sizes, $\xi_{\text{ps}} \ll \xi_{{\text{peak, }}{\text{\tiny{EHT}}}}$. Note how either a 
soft emission coefficient or a small shadow-size can cause the peak at 
$\xi = \xi_{{\text{peak, }}{\text{\tiny{EHT}}}}$ to become a secondary local maximum.

Finally, to ascertain which features and possibilities discussed here can
appear when using BH and non-BH spacetimes that arise as solutions in
various known theories of gravity, we consider the image properties of
the Reissner-Nordstr{\"o}m (RN) BHs and of the
Gibbons-Maeda-Garfinkle-Strominger-Horowitz \citep{Gibbons1988,
  Garfinkle1991}, which we shall refer to as the EMd-1 BHs. These BHs
arise as static, charged solutions of the Einstein-Maxwell equations and
of the Einstein-Maxwell-dilaton low-energy effective action of the
heterotic string respectively. We also consider a RN naked singularity
(NS$^*$) spacetime for completeness. In Fig. \ref{fig:RN_EMd1_delpeak_fw}
we show the contrast profiles of the RN and EMd-1 BHs such that a maximum
occurs at the EHT observed value $\xi=\xi_{{\text{peak,
  }}{\text{\tiny{EHT}}}}$. Notice however that since the shadow-size
decreases monotonically with specific charge in both cases, for
sufficiently small shadow-sizes the peak at $\xi=\xi_{{\text{peak,
  }}{\text{\tiny{EHT}}}}$ is only a secondary local maximum. Thus, such
BH solutions would be ruled out, at least within the accretion and
emission models we use here.  Finally, we note also that the RN BH is
exactly described by the $\mathscr{M}( \epsilon, a_0)$ RZ BH metric with
$\epsilon(\bar{q}) = a_0(\bar{q}) = 2/(1 + \sqrt{1 - \bar{q}^2}) - 1$,
where $\bar{q} = q/M$ is the specific electromagnetic charge of the BH
\citep[see, e.g.,][]{Kocherlakota2020}. We mark with a red dotted line
this family of BHs in the $(\epsilon, a_0)$ parameter space in
Figs. \ref{fig:RZ_eps_a0_a1_two_param} and
\ref{fig:RZ_eps_a0_a1_two_param_rcut} for ready comparison.

As can be seen from this figures in this section the fractional-width of 
the ring always increases when the emission model becomes softer (i.e., 
smaller-$n$ for $j = k_n r^{-n}$), as anticipated. We expect that when 
using more realistic accretion and emission models, this observable will 
continue to remain an excellent observable with which the EHT can 
constrain both emission physics and spacetime geometry.

\section{Conclusions and Discussion}
\label{sec:Conclusions}

We have studied here radially freely falling accretion onto
spherically-symmetric BHs, whose spacetime geometries are characterized
in general by three-parameters $\{\epsilon, a_0, a_1\}$. To study
features of the image formed on the observer's sky we employ various
isotropic, monochromatic emission models that are characterized roughly
by the radial profile of the emission coefficient $j \sim r^{-n}$.
Further, to study the impact of artificially suppressed emission close to
the event horizon, we introduce the parameter $r_{\text{cut}}$, the
radius inside which we set $j = 0$.

For this set of accretion-emission models, we find that the observed size
of the bright ring, characterized by the location of the peak of the
intensity contrast $\xi = \xi_{\text{peak}}$, is independent of the
emission coefficient $j$. However, $\xi_{\text{peak}}$ increases
monotonically with $r_{\text{cut}}$. Since it is possible to push out the
location of the peak in the contrast profile by modulating
$r_{\text{cut}}$, for BHs with shadow-sizes smaller than the observed
size of the peak of emission in the EHT image of M87$^*$,
$\xi_{\text{ps}} < \xi_{{\text{peak, }}{\text{\tiny{EHT}}}}$, one can
find a cutoff radius, $r_{\text{cut}} = r_{\text{tp}}(\xi_{{\text{peak,
  }}{\text{\tiny{EHT}}}})$, such that the contrast peak occurs exactly at
the EHT observed location.  BHs with shadow-sizes larger than
$\xi_{{\text{peak, }}{\text{\tiny{EHT}}}}$ are ruled out since the
location of the peak contrast in their images can never occur at $\xi =
\xi_{{\text{peak, }}{\text{\tiny{EHT}}}}$. It is also possible in
principle to constrain BHs with very small shadow-sizes when they violate
the condition $\mathcal{C}_{\text{peak}} = \mathcal{C}(\xi_{{\text{peak,
  }}{\text{\tiny{EHT}}}}) > \mathcal{C}_{\text{peak,
    EHT-min}}$. Additionally, for BHs with very small shadow-sizes
($\xi_{\text{ps}} \ll \xi_{{\text{peak, }}{\text{\tiny{EHT}}}}$), the
contrast peak shifts back to the shadow boundary, with $\xi =
\xi_{{\text{peak, }}{\text{\tiny{EHT}}}}$ locating only a secondary local
maximum, when $r_{\text{cut}} = r_{\text{tp}}(\xi_{{\text{peak,
  }}{\text{\tiny{EHT}}}})$. Thus, using the measurement of the location
of the peak contrast along with its measured lower bound imposes
constraints on the space of metric-deviation parameters. Note that since
the emission coefficient is primarily responsible in setting the scale of
the contrast in the image, and ``softer'' emission coefficients
(small-$n$) imply tighter constraints, it is indeed possible to set
conservative bounds using the emission model we have proposed here
[$\mathtt{em.1}$; see Sec.~\ref{sec:Emission_Coeff}]; this model accounts
for the entire binding energy of cold infalling fluid. Finally, we can
also use simultaneously the shadow-size bounds reported by the EHT to
constrain further the RZ parameter spaces. While the ``constraint bands''
we report from these two sets of measurements (location and value of the
image contrast peak on the one side and the shadow bounds on the other)
show overlap, as they should, they are typically not degenerate, as
evidenced by the shape of the contour lines.

There is no \textit{a priori} reason for emission to cease at some inner
radius outside the event horizon $r_{\text{cut}} > r_0$ at least within
the present context of spherical accretion. Thus, if we set
$r_{\text{cut}} = r_0$ and consider the various two-parameter family of
RZ BHs, we then find that the constraints on the relevant parameter
spaces that we obtain by imposing the requirement that the location of
the peak in the observed contrast profile coincide with the EHT-reported
value, $\xi_{\text{peak}} = \xi_{{\text{peak, }}{\text{\tiny{EHT}}}} =
\alpha M/2 \approx 5.5 M$, are far more severe than the EHT shadow-size
bounds. Essentially, the former imposes the condition that
$\xi_{\text{ps}} \approx 5.5 M$, and is equivalent to an extremely
precise shadow-size measurement. Even in this case, for
higher-dimensional BH parameter spaces, it is not impossible to constrain
exactly the spacetime metric of M87$^*$. To break these types of
inevitable degeneracies, we require additional observables, such as the
fractional size of the ring width $f_{\text{w}}$ that the EHT has
measured (\citealt{EHT_M87_PaperI}; see, e.g., Fig.
\ref{fig:Schw_Ring_Characteristics_rcut}) or the sizes of photon subrings
that the ngEHT hopes to observe soon. At the same time, it is also clear
that the constraints we are able to establish on general properties of
the spacetime near M87$^*$, such as the size of the photon sphere or the
event horizon ($\epsilon$), using presently available EHT measurements
are nevertheless substantial. This is particularly relevant for various
well-known solutions of alternative theories of gravity, where the
arbitrary multi-pole moments of the BH metric are controlled by a small
number of parameters or ``charges'' (see, e.g.,
\citealt{Kocherlakota2021} and compare with \citealt{Kocherlakota2020}).

A significant limitation of the present work involves our use of a
spherical accretion model, which implies that the redshifts and effective
path-lengths shown in Sec.~\ref{sec:Spherical_Accretion} may be modified
significantly when working, e.g., even with emission sourced from
stationary geometrically-thick tori. While in the latter case naturally
the circular symmetry of the image on the celestial plane is lost, more
importantly, the sharp drop in contrast at the shadow boundary may change
due effectively to reduced Doppler beaming. At the same time, we do not
expect significant qualitative changes in the results we have discussed
above, with the advantage that these have been obtained without the use
of GRMHD or ray-tracing codes: when using about $10^4$ grid points (BH
metrics) to generate the contrast contour plots shown in
Figs. \ref{fig:RZ_eps_a0_a1_two_param} and
\ref{fig:RZ_eps_a0_a1_two_param_rcut}, our code runs for less than a day
on a standard laptop. In the same spirit, it would be useful to
characterize in future the percentage changes in the various image-domain
features we have explored here when employing more realistic
semi-analytic accretion-emission models as well as ray-traced images of
GRMHD simulations involving hot, magnetized accretion flows around
Rezzolla-Zhidenko BHs.

While this work was in preparation, some of these issues have recently
been addressed in \citet{Bauer2021}, where the $\mathscr{M}(a_1)-$RZ
metric is also used, as well as in \citet{Ozel2021} and
\citet{Younsi2021}, where analytic models away from spherical symmetry
are considered. In particular, it is worth noting that the range of the
single-parameter $|a_1| < 0.2$ adopted by \citet{Bauer2021} is
considerably smaller than what is investigated here (see Table
\ref{table:RZ_Parameter_Space_Constraints_PPN}), leading to a suppression
of the variation of the characteristic size of the emission ring around
an RZ black hole with spacetime geometry (see Fig.
\ref{fig:RZ_eps_a0_a1_one_param_delpeak_fw}). This is highly restrictive
since the bounds on the parameter $a_1$ that one can infer from the
shadow-size measurement of M87$^*$ (see the top right panel of
Fig. \ref{fig:RZ_eps_a0_a1_one_param}) are approximately an
order-of-magnitude larger. Our results are also in agreement with those
in Sec. 2 of \citet{Ozel2021}, where spherical accretion is also
considered.  While \citet{Ozel2021} and \citet{Younsi2021} argue that the
size of the bright emission ring is not disjoint from the shadow boundary
using semi-analytic non-spherical accretion and emission models around
Kerr and non-Kerr black holes respectively, we establish the same here by
introducing an agnostic inner cut-off for the emission region, albeit
within spherical symmetry, which has been a point of recent debate (see,
e.g., \citealt{Gralla2021}).

\textit{Acknowledgements}: It is a pleasure to thank Ramesh Narayan for
insightful discussions on emission coefficients, and Enrico Barausse,
Guillermo Delgado, and Sebastian Voelkel for comments. We also thank our
EHT colleagues for useful suggestions. Support comes from the ERC Synergy
Grant ``BlackHoleCam: Imaging the Event Horizon of black holes'' (Grant
No. 610058) and from the ERC Advanced Grant ``JETSET: Launching,
propagation and emission of relativistic jets from binary mergers and
across mass scales'' (Grant No. 884631).


\bibliographystyle{mnras}


\begin{thebibliography}{}
\makeatletter
\relax
\def\mn@urlcharsother{\let\do\@makeother \do\$\do\&\do\#\do\^\do\_\do\%\do\~}
\def\mn@doi{\begingroup\mn@urlcharsother \@ifnextchar [ {\mn@doi@}
  {\mn@doi@[]}}
\def\mn@doi@[#1]#2{\def\@tempa{#1}\ifx\@tempa\@empty \href
  {http://dx.doi.org/#2} {doi:#2}\else \href {http://dx.doi.org/#2} {#1}\fi
  \endgroup}
\def\mn@eprint#1#2{\mn@eprint@#1:#2::\@nil}
\def\mn@eprint@arXiv#1{\href {http://arxiv.org/abs/#1} {{\tt arXiv:#1}}}
\def\mn@eprint@dblp#1{\href {http://dblp.uni-trier.de/rec/bibtex/#1.xml}
  {dblp:#1}}
\def\mn@eprint@#1:#2:#3:#4\@nil{\def\@tempa {#1}\def\@tempb {#2}\def\@tempc
  {#3}\ifx \@tempc \@empty \let \@tempc \@tempb \let \@tempb \@tempa \fi \ifx
  \@tempb \@empty \def\@tempb {arXiv}\fi \@ifundefined
  {mn@eprint@\@tempb}{\@tempb:\@tempc}{\expandafter \expandafter \csname
  mn@eprint@\@tempb\endcsname \expandafter{\@tempc}}}

\bibitem[\protect\citeauthoryear{{Abdujabbarov}, {Rezzolla}  \&
  {Ahmedov}}{{Abdujabbarov} et~al.}{2015}]{Abdujabbarov2015}
{Abdujabbarov} A.~A.,  {Rezzolla} L.,   {Ahmedov} B.~J.,  2015, \mn@doi [MNRAS] {10.1093/mnras/stv2079}, \href
  {http://adsabs.harvard.edu/abs/2015MNRAS.454.2423A} {454, 2423}

\bibitem[\protect\citeauthoryear{{Bambi}}{{Bambi}}{2013}]{Bambi2013a}
{Bambi} C.,  2013, \mn@doi [Phys. Rev. D] {10.1103/PhysRevD.87.107501}, \href
  {https://ui.adsabs.harvard.edu/abs/2013PhRvD..87j7501B} {87, 107501}

\bibitem[\protect\citeauthoryear{{Bambi} \& {Modesto}}{{Bambi} \&
  {Modesto}}{2013}]{Bambi2013}
{Bambi} C.,  {Modesto} L.,  2013, \mn@doi [Phys. Lett. B]
  {10.1016/j.physletb.2013.03.025}, \href
  {http://adsabs.harvard.edu/abs/2013PhLB..721..329B} {721, 329}

\bibitem[\protect\citeauthoryear{{Bardeen}}{{Bardeen}}{1974}]{Bardeen1974}
{Bardeen} J.~M.,  1974, in {Dewitt-Morette} C.,  ed.,  Vol. 64, Gravitational
  Radiation and Gravitational Collapse. p.~132

\bibitem[\protect\citeauthoryear{{Bauer}, {C{\'a}rdenas-Avenda{\~n}o}, {Gammie}
   \& {Yunes}}{{Bauer} et~al.}{2021}]{Bauer2021}
{Bauer} A.,  {C{\'a}rdenas-Avenda{\~n}o} A.,  {Gammie} C.~F.,   {Yunes} N.,
  2021, arXiv e-prints, \href
  {https://ui.adsabs.harvard.edu/abs/2021arXiv211102178B} {p. arXiv:2111.02178}

\bibitem[\protect\citeauthoryear{{Broderick}, {Johannsen}, {Loeb}  \&
  {Psaltis}}{{Broderick} et~al.}{2014}]{Broderick2014}
{Broderick} A.~E.,  {Johannsen} T.,  {Loeb} A.,   {Psaltis} D.,  2014, \mn@doi
  [ApJ] {10.1088/0004-637X/784/1/7}, \href
  {http://adsabs.harvard.edu/abs/2014ApJ...784....7B} {784, 7}

\bibitem[\protect\citeauthoryear{{Broderick}, {Tiede}, {Pesce}  \&
  {Gold}}{{Broderick} et~al.}{2021}]{Broderick2021}
{Broderick} A.~E.,  {Tiede} P.,  {Pesce} D.~W.,   {Gold} R.,  2021, arXiv
  e-prints, \href {https://ui.adsabs.harvard.edu/abs/2021arXiv210509962B} {p.
  arXiv:2105.09962}

\bibitem[\protect\citeauthoryear{{Event Horizon Telescope Collaboration}
  et~al.,}{{Event Horizon Telescope Collaboration}
  et~al.}{2019a}]{EHT_M87_PaperI}
{Event Horizon Telescope Collaboration} et~al., 2019a, \mn@doi [ApJ
  Lett.] {10.3847/2041-8213/ab0ec7}, \href
  {http://adsabs.harvard.edu/abs/2019ApJ...875L...1E} {875, L1}

\bibitem[\protect\citeauthoryear{{Event Horizon Telescope Collaboration}
  et~al.,}{{Event Horizon Telescope Collaboration}
  et~al.}{2019b}]{EHT_M87_PaperII}
{Event Horizon Telescope Collaboration} et~al., 2019b, \mn@doi [ApJ
  Lett.] {10.3847/2041-8213/ab0c96}, \href
  {http://adsabs.harvard.edu/abs/2019ApJ...875L...2E} {875, L2}

\bibitem[\protect\citeauthoryear{{Event Horizon Telescope Collaboration}
  et~al.,}{{Event Horizon Telescope Collaboration}
  et~al.}{2019c}]{EHT_M87_PaperIII}
{Event Horizon Telescope Collaboration} et~al., 2019c, \mn@doi [ApJ
  Lett.] {10.3847/2041-8213/ab0c57}, \href
  {http://adsabs.harvard.edu/abs/2019ApJ...875L...3E} {875, L3}

\bibitem[\protect\citeauthoryear{{Event Horizon Telescope Collaboration}
  et~al.,}{{Event Horizon Telescope Collaboration}
  et~al.}{2019d}]{EHT_M87_PaperIV}
{Event Horizon Telescope Collaboration} et~al., 2019d, \mn@doi [ApJ
  Lett.] {10.3847/2041-8213/ab0e85}, \href
  {http://adsabs.harvard.edu/abs/2019ApJ...875L...4E} {875, L4}

\bibitem[\protect\citeauthoryear{{Event Horizon Telescope Collaboration}
  et~al.,}{{Event Horizon Telescope Collaboration}
  et~al.}{2019e}]{EHT_M87_PaperV}
{Event Horizon Telescope Collaboration} et~al., 2019e, \mn@doi [ApJ
  Lett.] {10.3847/2041-8213/ab0f43}, \href
  {http://adsabs.harvard.edu/abs/2019ApJ...875L...5E} {875, L5}

\bibitem[\protect\citeauthoryear{{Event Horizon Telescope Collaboration}
  et~al.,}{{Event Horizon Telescope Collaboration}
  et~al.}{2019f}]{EHT_M87_PaperVI}
{Event Horizon Telescope Collaboration} et~al., 2019f, \mn@doi [ApJ
  Lett.] {10.3847/2041-8213/ab1141}, \href
  {http://adsabs.harvard.edu/abs/2019ApJ...875L...6E} {875, L6}

\bibitem[\protect\citeauthoryear{Gan, Wang, Wu  \& Yang}{Gan
  et~al.}{2021}]{Gan2021}
Gan Q.,  Wang P.,  Wu H.,   Yang H.,  2021, \mn@doi [Phys. Rev. D]
  {10.1103/PhysRevD.104.024003}, 104, 024003

\bibitem[\protect\citeauthoryear{{Garfinkle}, {Horowitz}  \&
  {Strominger}}{{Garfinkle} et~al.}{1991}]{Garfinkle1991}
{Garfinkle} D.,  {Horowitz} G.~T.,   {Strominger} A.,  1991, \mn@doi [Phys.
  Rev. D] {10.1103/PhysRevD.43.3140}, \href
  {http://adsabs.harvard.edu/abs/1991PhRvD..43.3140G} {43, 3140}

\bibitem[\protect\citeauthoryear{{Gebhardt}, {Adams}, {Richstone}, {Lauer},
  {Faber}, {G{\"u}ltekin}, {Murphy}  \& {Tremaine}}{{Gebhardt}
  et~al.}{2011}]{Gebhardt11}
{Gebhardt} K.,  {Adams} J.,  {Richstone} D.,  {Lauer} T.~R.,  {Faber} S.~M.,
  {G{\"u}ltekin} K.,  {Murphy} J.,   {Tremaine} S.,  2011, \mn@doi
  [ApJ] {10.1088/0004-637X/729/2/119}, \href
  {http://adsabs.harvard.edu/abs/2011ApJ...729..119G} {729, 119}

\bibitem[\protect\citeauthoryear{{Gibbons} \& {Maeda}}{{Gibbons} \&
  {Maeda}}{1988}]{Gibbons1988}
{Gibbons} G.~W.,  {Maeda} K.-I.,  1988, \mn@doi [Nuc. Phys. B]
  {10.1016/0550-3213(88)90006-5}, \href
  {http://adsabs.harvard.edu/abs/1988NuPhB.298..741G} {298, 741}

\bibitem[\protect\citeauthoryear{Gimeno-Soler, Font, Herdeiro  \&
  Radu}{Gimeno-Soler et~al.}{2019}]{Gimeno-Soler2019}
Gimeno-Soler S.,  Font J.~A.,  Herdeiro C.,   Radu E.,  2019, \mn@doi [Phys.
  Rev. D] {10.1103/PhysRevD.99.043002}, 99, 043002

\bibitem[\protect\citeauthoryear{{Gralla}}{{Gralla}}{2021}]{Gralla2021}
{Gralla} S.~E.,  2021, \mn@doi [Phys. Rev. D] {10.1103/PhysRevD.103.024023}, \href
  {https://ui.adsabs.harvard.edu/abs/2021PhRvD.103b4023G} {103, 024023}

\bibitem[\protect\citeauthoryear{{Gralla}, {Holz}  \& {Wald}}{{Gralla}
  et~al.}{2019}]{Gralla2019}
{Gralla} S.~E.,  {Holz} D.~E.,   {Wald} R.~M.,  2019, \mn@doi [Phys. Rev. D]
  {10.1103/PhysRevD.100.024018}, \href
  {https://ui.adsabs.harvard.edu/abs/2019PhRvD.100b4018G} {100, 024018}

\bibitem[\protect\citeauthoryear{{Hioki} \& {Maeda}}{{Hioki} \&
  {Maeda}}{2009}]{Hioki2009}
{Hioki} K.,  {Maeda} K.-I.,  2009, \mn@doi [Phys. Rev. D]
  {10.1103/PhysRevD.80.024042}, \href
  {https://ui.adsabs.harvard.edu/abs/2009PhRvD..80b4042H} {80, 024042}

\bibitem[\protect\citeauthoryear{{Jaroszynski} \& {Kurpiewski}}{{Jaroszynski}
  \& {Kurpiewski}}{1997}]{Jaroszynski1997}
{Jaroszynski} M.,  {Kurpiewski} A.,  1997, \aap, \href
  {https://ui.adsabs.harvard.edu/abs/1997A&A...326..419J} {326, 419}

\bibitem[\protect\citeauthoryear{{Johannsen}, {Wang}, {Broderick}, {Doeleman},
  {Fish}, {Loeb}  \& {Psaltis}}{{Johannsen} et~al.}{2016}]{Johannsen2016c}
{Johannsen} T.,  {Wang} C.,  {Broderick} A.~E.,  {Doeleman} S.~S.,  {Fish}
  V.~L.,  {Loeb} A.,   {Psaltis} D.,  2016, \mn@doi [Phys. Rev. Lett.]
  {10.1103/PhysRevLett.117.091101}, \href
  {http://adsabs.harvard.edu/abs/2016PhRvL.117i1101J} {117, 091101}

\bibitem[\protect\citeauthoryear{{Johnson} et~al.,}{{Johnson}
  et~al.}{2020}]{Johnson2020}
{Johnson} M.~D.,  et~al., 2020, \mn@doi [Science Adv.]
  {10.1126/sciadv.aaz1310}, \href
  {https://ui.adsabs.harvard.edu/abs/2020SciA....6.1310J} {6, eaaz1310}

\bibitem[\protect\citeauthoryear{{Jusufi}, {Saurabh}, {Azreg-A{\"\i}nou},
  {Jamil}, {Wu}  \& {Bambi}}{{Jusufi} et~al.}{2021}]{Jusufi2021}
{Jusufi} K.,  {Saurabh} K.,  {Azreg-A{\"\i}nou} M.,  {Jamil} M.,  {Wu} Q.,
  {Bambi} C.,  2021, arXiv e-prints, \href
  {https://ui.adsabs.harvard.edu/abs/2021arXiv210608070J} {p. arXiv:2106.08070}

\bibitem[\protect\citeauthoryear{{Kocherlakota} \& {Rezzolla}}{{Kocherlakota}
  \& {Rezzolla}}{2020}]{Kocherlakota2020}
{Kocherlakota} P.,  {Rezzolla} L.,  2020, \mn@doi [Phys. Rev. D]
  {10.1103/PhysRevD.102.064058}, \href
  {https://ui.adsabs.harvard.edu/abs/2020PhRvD.102f4058K} {102, 064058}

\bibitem[\protect\citeauthoryear{Kocherlakota et~al.}{Kocherlakota
  et~al.}{2021}]{Kocherlakota2021}
Kocherlakota P.,  et~al., 2021, \mn@doi [Phys. Rev. D]
  {10.1103/PhysRevD.103.104047}, 103, 104047

\bibitem[\protect\citeauthoryear{{Kocherlakota}, {Rezzolla}  \&
  {Wielgus}}{{Kocherlakota} et~al.}{2022}]{Kocherlakota2022}
{Kocherlakota} P.,  {Rezzolla} L.,   {Wielgus} M.,  2022, in preparation

\bibitem[\protect\citeauthoryear{{Konoplya}, {Rezzolla}  \&
  {Zhidenko}}{{Konoplya} et~al.}{2016}]{Konoplya2016a}
{Konoplya} R.,  {Rezzolla} L.,   {Zhidenko} A.,  2016, \mn@doi [Phys. Rev. D]
  {10.1103/PhysRevD.93.064015}, \href
  {http://adsabs.harvard.edu/abs/2016PhRvD..93f4015K} {93, 064015}

\bibitem[\protect\citeauthoryear{{Lara}, {V{\"o}lkel}  \& {Barausse}}{{Lara}
  et~al.}{2021}]{Lara2021}
{Lara} G.,  {V{\"o}lkel} S.~H.,   {Barausse} E.,  2021, arXiv e-prints, \href
  {https://ui.adsabs.harvard.edu/abs/2021arXiv211000026L} {p. arXiv:2110.00026}

\bibitem[\protect\citeauthoryear{{Mazur} \& {Mottola}}{{Mazur} \&
  {Mottola}}{2004}]{Mazur2004}
{Mazur} P.~O.,  {Mottola} E.,  2004, \mn@doi [PNAS] {10.1073/pnas.0402717101}, \href
  {http://adsabs.harvard.edu/abs/2004PNAS..101.9545M} {101, 9545}

\bibitem[\protect\citeauthoryear{{Medeiros}, {Psaltis}  \&
  {{\"O}zel}}{{Medeiros} et~al.}{2020}]{Medeiros2020}
{Medeiros} L.,  {Psaltis} D.,   {{\"O}zel} F.,  2020, \mn@doi [ApJ]
  {10.3847/1538-4357/ab8bd1}, \href
  {https://ui.adsabs.harvard.edu/abs/2020ApJ...896....7M} {896, 7}

\bibitem[\protect\citeauthoryear{{Mizuno} et~al.,}{{Mizuno}
  et~al.}{2018}]{Mizuno2018}
{Mizuno} Y.,  et~al., 2018, \mn@doi [Nature Astronomy]
  {10.1038/s41550-018-0449-5}, \href
  {https://ui.adsabs.harvard.edu/abs/2018NatAs...2..585M} {2, 585}

\bibitem[\protect\citeauthoryear{{Narayan}, {Johnson}  \& {Gammie}}{{Narayan}
  et~al.}{2019}]{Narayan2019}
{Narayan} R.,  {Johnson} M.~D.,   {Gammie} C.~F.,  2019, \mn@doi [ApJ
  Lett.] {10.3847/2041-8213/ab518c}, \href
  {https://ui.adsabs.harvard.edu/abs/2019ApJ...885L..33N} {885, L33}

\bibitem[\protect\citeauthoryear{{Olivares} et~al.,}{{Olivares}
  et~al.}{2020}]{Olivares2020}
{Olivares} H.,  et~al., 2020, \mn@doi [MNRAS] {10.1093/mnras/staa1878}, \href
  {https://ui.adsabs.harvard.edu/abs/2020MNRAS.497..521O} {497, 521}

\bibitem[\protect\citeauthoryear{{Ozel}, {Psaltis}  \& {Younsi}}{{Ozel}
  et~al.}{2021}]{Ozel2021}
{Ozel} F.,  {Psaltis} D.,   {Younsi} Z.,  2021, arXiv e-prints, \href
  {https://ui.adsabs.harvard.edu/abs/2021arXiv211101123O} {p. arXiv:2111.01123}

\bibitem[\protect\citeauthoryear{{Psaltis}, {Medeiros}, {Christian}, {Ozel}  \&
  {the EHT Collaboration}}{{Psaltis} et~al.}{2020}]{Psaltis2020_EHT}
{Psaltis} D.,  {Medeiros} L.,  {Christian} P.,  {Ozel} F.,   {the EHT
  Collaboration} 2020, \mn@doi [Phys. Rev. Lett.]
  {10.1103/PhysRevLett.125.141104}, \href
  {https://ui.adsabs.harvard.edu/abs/2020arXiv201001055P} {125, 141104}

\bibitem[\protect\citeauthoryear{{Pu} \& {Broderick}}{{Pu} \&
  {Broderick}}{2018}]{Pu2018}
{Pu} H.-Y.,  {Broderick} A.~E.,  2018, \mn@doi [ApJ]
  {10.3847/1538-4357/aad086}, \href
  {https://ui.adsabs.harvard.edu/abs/2018ApJ...863..148P} {863, 148}

\bibitem[\protect\citeauthoryear{{Rezzolla} \& {Zhidenko}}{{Rezzolla} \&
  {Zhidenko}}{2014}]{Rezzolla2014}
{Rezzolla} L.,  {Zhidenko} A.,  2014, \mn@doi [Phys. Rev. D]
  {10.1103/PhysRevD.90.084009}, \href
  {http://adsabs.harvard.edu/abs/2014PhRvD..90h4009R} {90, 084009}

\bibitem[\protect\citeauthoryear{{Sen}}{{Sen}}{1992}]{Sen1992}
{Sen} A.,  1992, \mn@doi [Phys. Rev. Lett.] {10.1103/PhysRevLett.69.1006},
  \href {http://adsabs.harvard.edu/abs/1992PhRvL..69.1006S} {69, 1006}

\bibitem[\protect\citeauthoryear{Shaikh \& Joshi}{Shaikh \&
  Joshi}{2019}]{Shaikh2019b}
Shaikh R.,  Joshi P.~S.,  2019, \mn@doi [JCAP] {10.1088/1475-7516/2019/10/064},
  10, 064

\bibitem[\protect\citeauthoryear{{Shaikh}, {Kocherlakota}, {Narayan}  \&
  {Joshi}}{{Shaikh} et~al.}{2019}]{Shaikh2019}
{Shaikh} R.,  {Kocherlakota} P.,  {Narayan} R.,   {Joshi} P.~S.,  2019, \mn@doi
  [Mon. Not. R. Astron. Soc.] {10.1093/mnras/sty2624}, \href
  {https://ui.adsabs.harvard.edu/abs/2019MNRAS.482...52S} {482, 52}

\bibitem[\protect\citeauthoryear{Suvorov}{Suvorov}{2020}]{Suvorov2020}
Suvorov A.~G.,  2020, \mn@doi [Class. Quant. Grav.] {10.1088/1361-6382/aba6a8},
  37, 185001

\bibitem[\protect\citeauthoryear{{The Ligo Scientific Collaboration}, {the
  VIRGO Collaboration}  \& {the KAGRA Collaboration}}{{The Ligo Scientific
  Collaboration} et~al.}{2021}]{Abbott2021}
{The Ligo Scientific Collaboration} {the VIRGO Collaboration}  {the KAGRA
  Collaboration} 2021, \mn@doi [\apjl] {10.3847/2041-8213/ac082e}, \href
  {https://ui.adsabs.harvard.edu/abs/2021ApJ...915L...5A} {915, L5}

\bibitem[\protect\citeauthoryear{{V{\"o}lkel} \& {Barausse}}{{V{\"o}lkel} \&
  {Barausse}}{2020}]{Volkel2020}
{V{\"o}lkel} S.~H.,  {Barausse} E.,  2020, \mn@doi [Phys. Rev. D]
  {10.1103/PhysRevD.102.084025}, \href
  {https://ui.adsabs.harvard.edu/abs/2020PhRvD.102h4025V} {102, 084025}

\bibitem[\protect\citeauthoryear{{V{\"o}lkel}, {Barausse}, {Franchini}  \&
  {Broderick}}{{V{\"o}lkel} et~al.}{2021}]{Volkel2021}
{V{\"o}lkel} S.~H.,  {Barausse} E.,  {Franchini} N.,   {Broderick} A.~E.,
  2021, \mn@doi [Class. Quantum Grav.] {10.1088/1361-6382/ac27ed},
  \href {https://ui.adsabs.harvard.edu/abs/2021CQGra..38uLT01V} {38, 21LT01}

\bibitem[\protect\citeauthoryear{{Will}}{{Will}}{2006}]{Will:2006LRR}
{Will} C.~M.,  2006, \mn@doi [Liv. Rev. Relativ.] {10.12942/lrr-2006-3},
  \href {http://ads.nao.ac.jp/abs/2006LRR.....9....3W} {9, 3}

\bibitem[\protect\citeauthoryear{Yang, Liu, Zhu, Zhao, Wu, Yang  \& Jamil}{Yang
  et~al.}{2021}]{Yang2021}
Yang S.,  Liu C.,  Zhu T.,  Zhao L.,  Wu Q.,  Yang K.,   Jamil M.,  2021,
  \mn@doi [Chin. Phys. C] {10.1088/1674-1137/abc066}, 45, 015102

\bibitem[\protect\citeauthoryear{{Younsi}, {Wu}  \& {Fuerst}}{{Younsi}
  et~al.}{2012}]{Younsi2012}
{Younsi} Z.,  {Wu} K.,   {Fuerst} S.~V.,  2012, \mn@doi [A&A]
  {10.1051/0004-6361/201219599}, \href
  {http://adsabs.harvard.edu/abs/2012A%26A...545A..13Y} {545, A13}

\bibitem[\protect\citeauthoryear{{Younsi}, {Zhidenko}, {Rezzolla}, {Konoplya}
  \& {Mizuno}}{{Younsi} et~al.}{2016}]{Younsi2016}
{Younsi} Z.,  {Zhidenko} A.,  {Rezzolla} L.,  {Konoplya} R.,   {Mizuno} Y.,
  2016, \mn@doi [Phys. Rev. D] {10.1103/PhysRevD.94.084025}, \href
  {http://adsabs.harvard.edu/abs/2016PhRvD..94h4025Y} {94, 084025}

\bibitem[\protect\citeauthoryear{{Younsi}, {Psaltis}  \& {{\"O}zel}}{{Younsi}
  et~al.}{2021}]{Younsi2021}
{Younsi} Z.,  {Psaltis} D.,   {{\"O}zel} F.,  2021, arXiv e-prints, \href
  {https://ui.adsabs.harvard.edu/abs/2021arXiv211101752Y} {p. arXiv:2111.01752}

\bibitem[\protect\citeauthoryear{{Yuan} \& {Narayan}}{{Yuan} \&
  {Narayan}}{2014}]{Yuan2014}
{Yuan} F.,  {Narayan} R.,  2014, \mn@doi [ARAA] {10.1146/annurev-astro-082812-141003}, \href
  {http://adsabs.harvard.edu/abs/2014ARA%26A..52..529Y} {52, 529}

\makeatother
\end{thebibliography}
\input{emring.bbl}


\appendix

\section{Schwarzschild Black Hole: Photon Radial Motion and Redshift}
\label{app:Schw_Images_Addl}

\begin{figure*}
\includegraphics[width=\columnwidth]{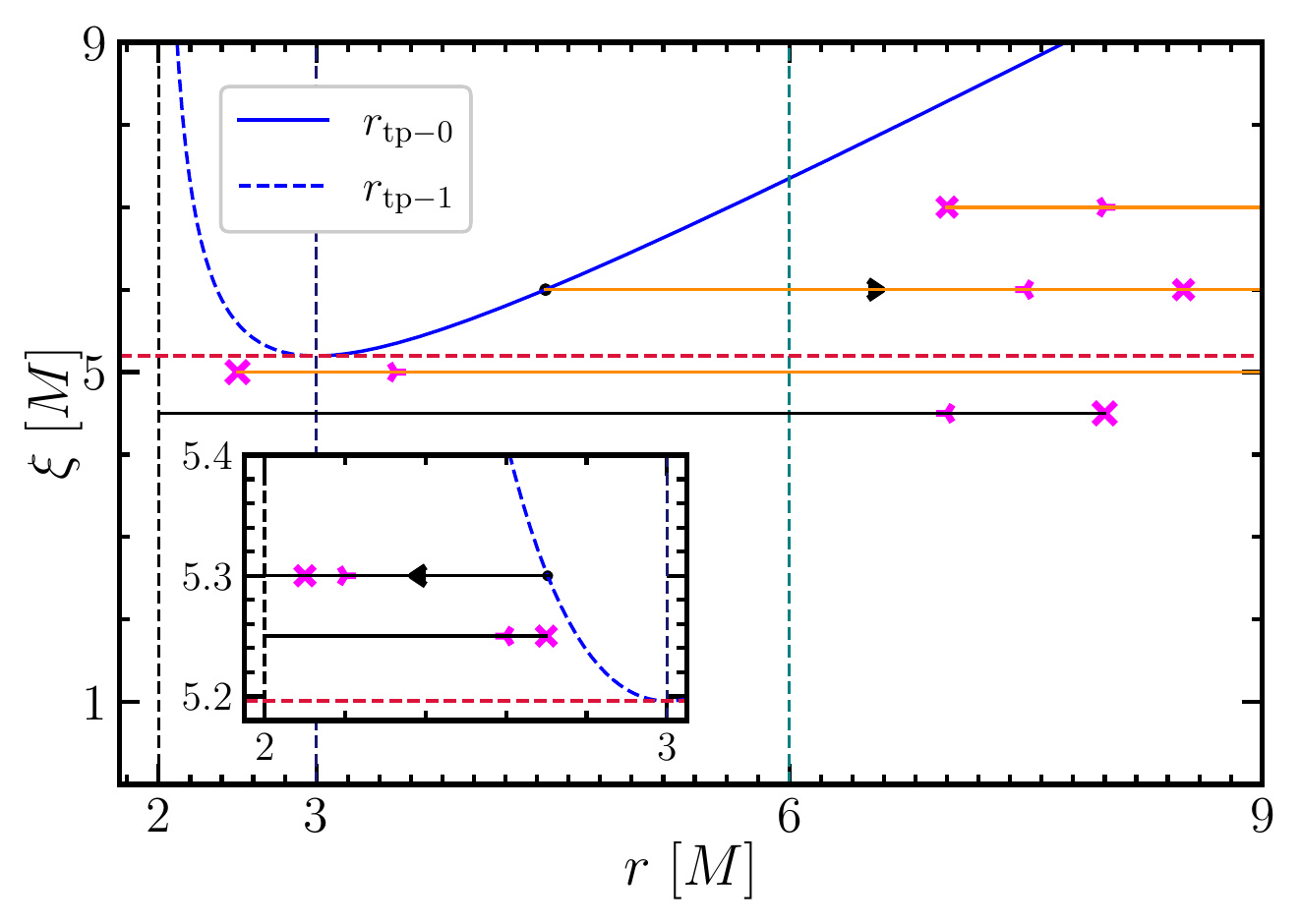}
\vspace{0.1cm}
\includegraphics[width=\columnwidth]{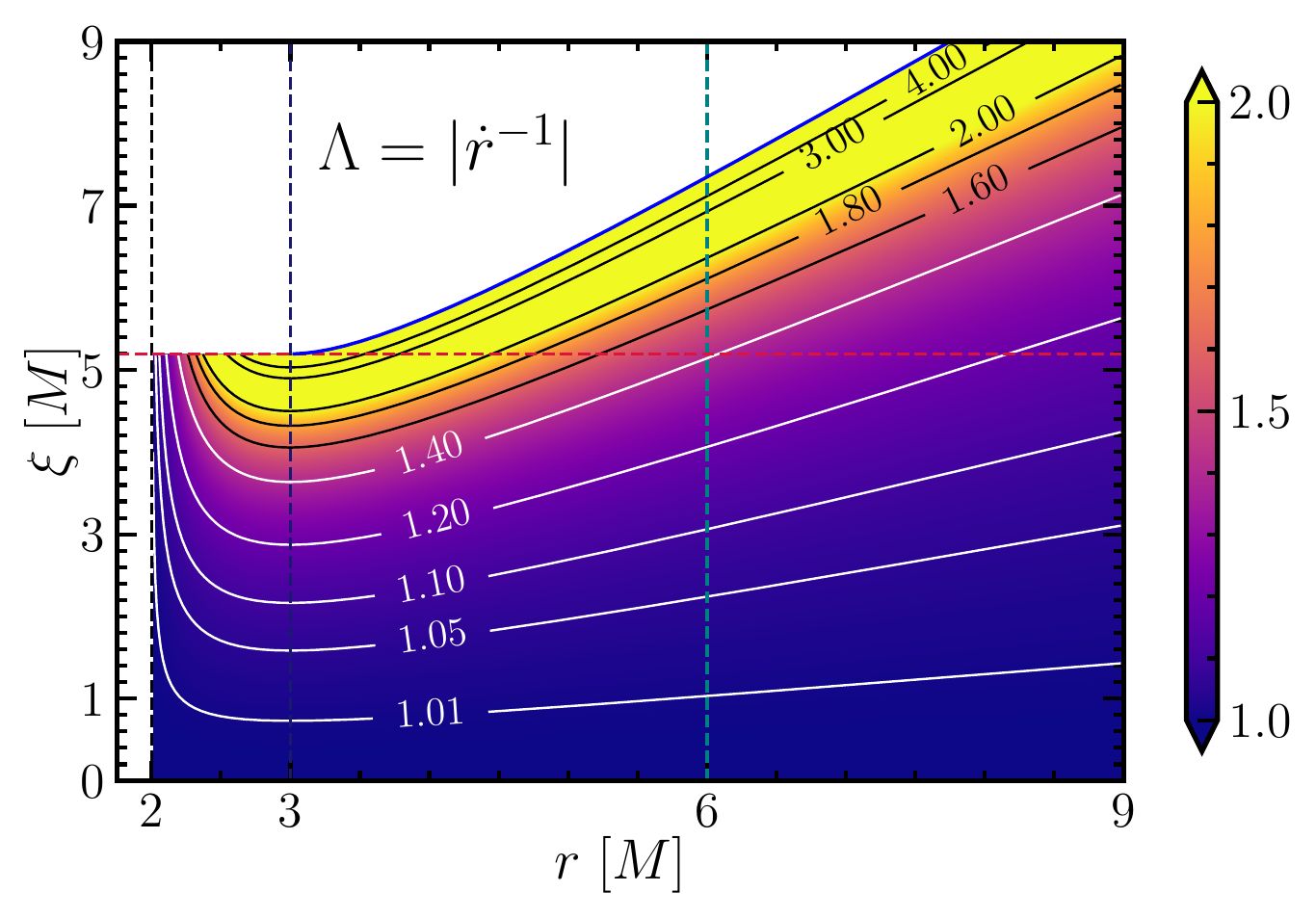}
\caption{In the left panel, we present a schematic that shows sample
  starting positions and velocities for photons, represented by magenta
  crosses and arrows respectively, and the radial evolution of their
  orbits. The black arrows show the ``final'' direction of the radial
  velocity after reflection at the relevant turning point. Photons which
  reach asymptotic observers, thus contributing to the observed
  intensity, are shown in orange lines. In the right panel, we show the
  modulus of the inverse radial velocity of photons that reach asymptotic
  observers, $\Lambda$. This quantity a key ingredient in the intensity
  integral \eqref{eq:Intensity_Integral_Inner_Boundary} we have used
  above, and measures the ``effective path-length'' of the null geodesic
  through the emitting fluid. The vertical blue and green lines correspond
  to the locations of the Schwarzschild photon sphere and the
  Schwarzschild innermost stable circular orbit (ISCO), respectively,
  i.e., $r = 3 M$ and $r = 6M$, respectively. The horizontal red line
  corresponds to the size of the Schwarzschild shadow, $\xi =
  \xi_{\text{ps}} = 3\sqrt{3} M$.}
\label{fig:Schw_Photon_Radial_Motion}
\end{figure*}

\begin{figure*}
\includegraphics[width=\columnwidth]{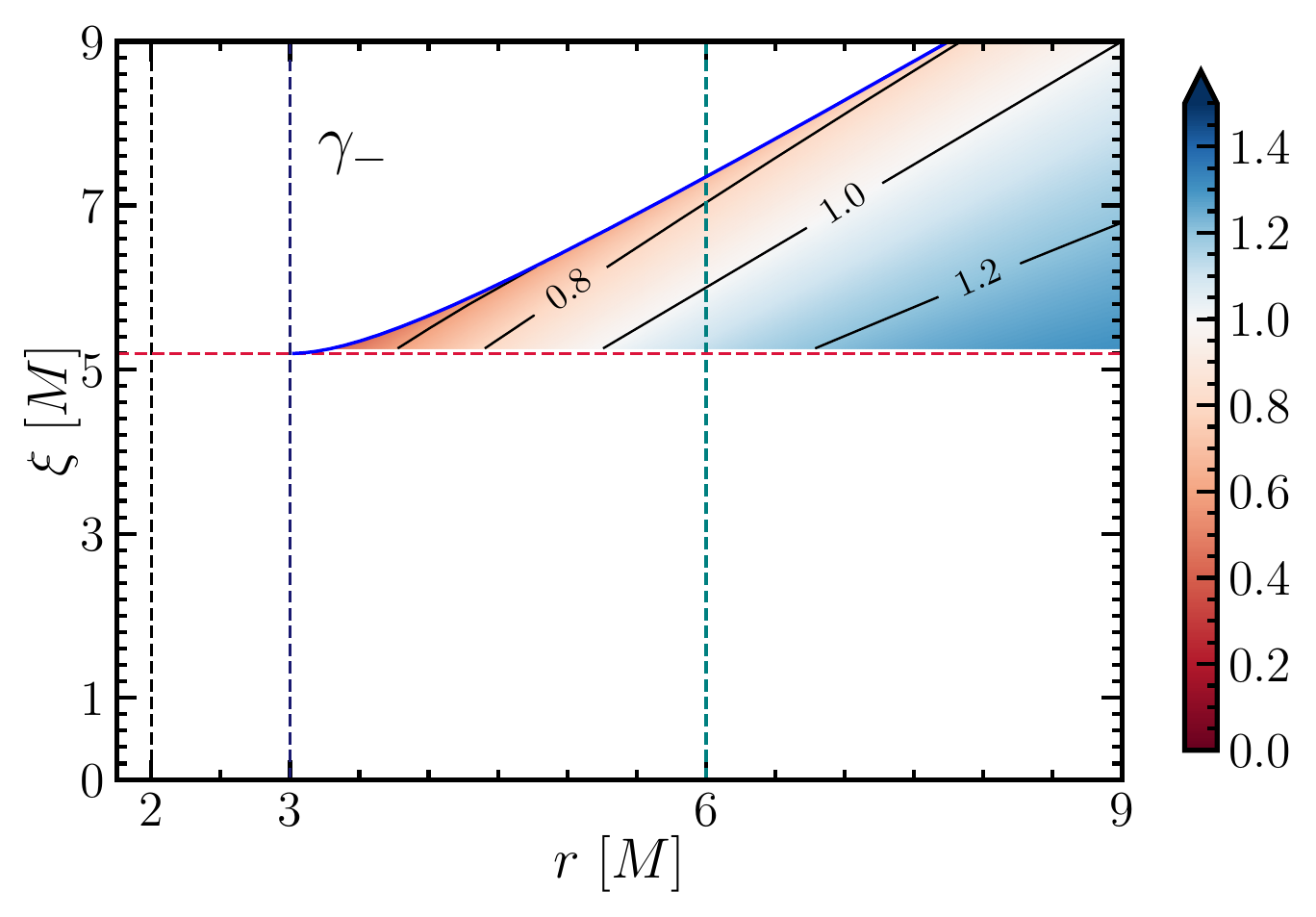}
\vspace{0.1cm}
\includegraphics[width=\columnwidth]{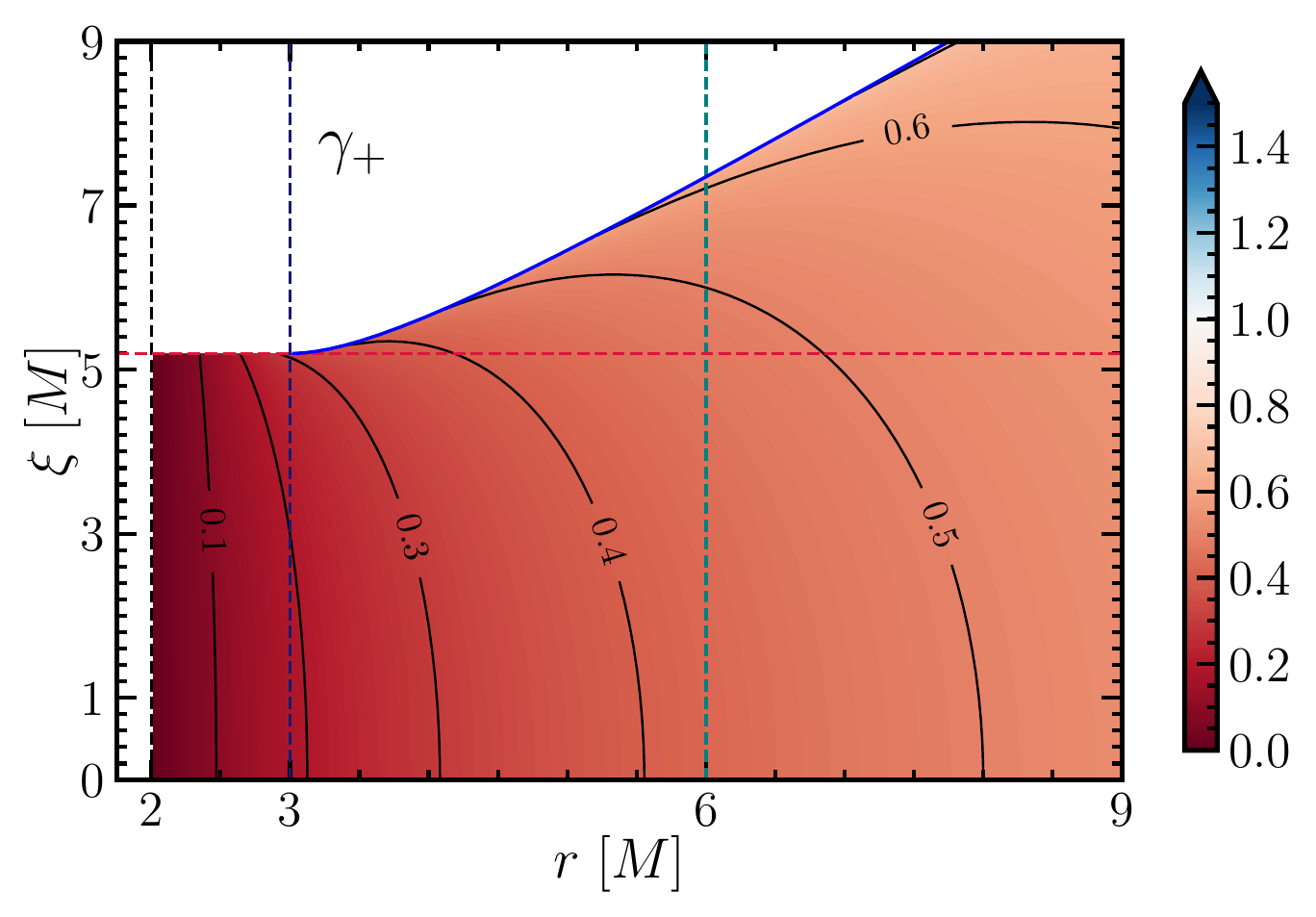}
\caption{Here we show the redshifts $\gamma_\pm$ experienced by photons
  that reach asymptotic observers, depending on where they were emitted
  from ($x$-axis) and in which way they were moving initially, i.e.,
  whether they were emitted toward the BH ($-$) or away from it ($+$). As
  expected, only radially-ingoing photons experience blueshifts, and
  photons emitted close to the horizon experience infinite redshifts.
  The vertical and horizontal lines have the same meaning as in
  Fig. \ref{fig:Schw_Photon_Radial_Motion}.}
\label{fig:Schw_Redshifts}
\end{figure*}

The left panel of Fig. \ref{fig:Schw_Photon_Radial_Motion} shows
characteristic sample values of conserved quantities $\xi$ and starting
positions $r$ from which emitted photons can reach asymptotic observers,
and thus participate in image formation (shown in orange lines). These
can be distributed into two sets based whether they have a
radially-ingoing leg ($\xi > \xi_{\text{ps}}$) or not ($\xi <
\xi_{\text{ps}}$). This simple schematic then immediately makes clear the
logic behind writing the key equation we have used here, Eq.
\ref{eq:Intensity_Integral_Inner_Boundary}. Ingoing photons with impact
parameters larger than that of the photon on the CNG, $\xi >
\xi_{\text{ps}}$, and emitted far away from the BH ``turn'' at the outer
turning point $r_{\text{tp-0}}$ and start moving radially-outwards
whereas those that are initially outgoing and emitted near the horizon
fall back onto the BH since these are reflected at the inner turning
point $r_{\text{tp-1}}$. The expressions for both these quantities (shown
in blue in the figure) are given as,
\begin{equation}
r_{\text{tp-k}}(\xi) = \frac{2\xi}{\sqrt{3}}\cos{\left[\frac{1}{3}
\cos^{-1}{\left(-\frac{\xi_{\text{ps}}}{\xi}\right)} - \frac{2\pi k}{3}
\right]}\,.
\end{equation}

The right panel of Fig. \ref{fig:Schw_Photon_Radial_Motion} shows the
modulus of the inverse radial velocity $\Lambda = |1/k_{\pm, \pm}^r|$ of
photons moving through the Schwarzschild spacetime. This factor is one of
the two purely emission-independent pieces in the intensity integral
\eqref{eq:Intensity_Integral_Inner_Boundary}, and the bright yellow
region shows where this term contributes the most, i.e., close to the
orbit's turning point- where its radial velocity is the smallest.
Equivalently, this is where the affine-time along the null geodesic per
unit radius (and thus the amount of intensity contributed to the null
geodesic, when a source of emission is present) is the largest.  

In Fig. \ref{fig:Schw_Redshifts}, we display the variation in the redshifts
experienced by ingoing (-) and outgoing photons emitted with an impact
parameter $\xi$ at a radius $r$. We highlight the change in $\gamma_-$
from representing a blueshift ($>1$) to a redshift close to the turning
point of a photon orbit (due to reduced radial velocity).

\bsp	
\label{lastpage}
\end{document}